\documentclass[aps,amssymb,amsmath,notitlepage,superscriptaddress,prx,twocolumn,nofootinbib,longbibliography]{revtex4-1}

\usepackage{amsfonts}
\usepackage{graphicx,graphics,epsfig,times,bm,bbm,amssymb,amsmath,amsfonts,mathrsfs}
\usepackage[normalem]{ulem}
\usepackage{wrapfig}
\usepackage{setspace}
\usepackage{subfigure}
\usepackage{dsfont}
\usepackage{braket}
\usepackage[pdfstartview=FitH]{hyperref}
\hypersetup{
    colorlinks=true,       
    linkcolor=cyan,          
    citecolor=magenta,        
    filecolor=magenta,      
    urlcolor=cyan,           
    runcolor=cyan
}
\usepackage[pdftex]{color}
\usepackage{upgreek }
\usepackage[x11names]{xcolor}
\usepackage{tikz}
\usepackage{natbib}
\usepackage{chngcntr}

\newcommand{\beq}{\begin{equation}}
\newcommand{\eeq}{\end{equation}}
\newcommand{\beqnn}{\begin{equation*}}
\newcommand{\eeqnn}{\end{equation*}}
\newcommand{\beann}{\begin{eqnarray*}}
\newcommand{\eeann}{\end{eqnarray*}}

\newcommand{\mc}{\mathcal}

\newcommand{\bes} {\begin{subequations}}
\newcommand{\ees} {\end{subequations}}
\newcommand{\bea} {\begin{eqnarray}}
\newcommand{\eea} {\end{eqnarray}}

\newcommand{\ignore}[1]{}

\begin{document}

\title{Simulated Quantum Annealing Comparison between All-to-All Connectivity Schemes}

\author{Tameem Albash}
\thanks{These authors contributed equally to this work.}
\affiliation{Department of Physics and Astronomy, University of Southern California, Los Angeles, California 90089, USA}
\affiliation{Center for Quantum Information Science \& Technology, University of Southern California, Los Angeles, California 90089, USA}
\affiliation{Information Sciences Institute, University of Southern California, Marina del Rey, California 90292, USA}
\author{Walter Vinci}
\thanks{These authors contributed equally to this work.}
\affiliation{Department of Physics and Astronomy, University of Southern California, Los Angeles, California 90089, USA}
\affiliation{Center for Quantum Information Science \& Technology, University of Southern California, Los Angeles, California 90089, USA}
\affiliation{Department of Electrical Engineering, University of Southern California, Los Angeles, California 90089, USA}
\author{Daniel A. Lidar}
\affiliation{Department of Physics and Astronomy, University of Southern California, Los Angeles, California 90089, USA}
\affiliation{Center for Quantum Information Science \& Technology, University of Southern California, Los Angeles, California 90089, USA}
\affiliation{Department of Electrical Engineering, University of Southern California, Los Angeles, California 90089, USA}
\affiliation{Department of Chemistry, University of Southern California, Los Angeles, California 90089, USA}
%
\begin{abstract}
%
Quantum annealing aims to exploit quantum mechanics to speed up the search for the solution to optimization problems. Most problems exhibit complete connectivity between the logical spin variables after they are mapped to the Ising spin Hamiltonian of quantum annealing. To account for hardware constraints of current and future physical quantum annealers, methods enabling the embedding of fully connected graphs of logical spins into a constant-degree graph of physical spins are therefore essential.  Here, we compare the recently proposed embedding scheme for quantum annealing with all-to-all connectivity due to Lechner, Hauke and Zoller (LHZ) [Science Advances 1 (2015)] to the commonly used minor embedding (ME) scheme. Using both simulated quantum annealing and parallel tempering simulations, we find that for a set of instances randomly chosen from a class of fully connected, random Ising problems, the ME scheme outperforms the LHZ scheme when using identical simulation parameters, despite the fault tolerance of the latter to weakly correlated spin-flip noise. This result persists even after we introduce several decoding strategies for the LHZ scheme, including a minimum-weight decoding algorithm that results in substantially improved performance over the original LHZ scheme. We explain the better performance of the ME scheme in terms of more efficient spin updates, which allows it to better tolerate the correlated spin-flip errors that arise in our  model of quantum annealing.  Our results leave open the question of whether the performance of the two embedding schemes can be improved using scheme-specific parameters and new error correction approaches.

\end{abstract}
\maketitle

\section{Introduction}
Many important and hard optimization problems can be mapped to finding the ground state of classical Ising models \cite{Barahona1982,2013arXiv1302.5843L}. The observation that such models can also be solved via quantum annealing \cite{kadowaki_quantum_1998,farhi_quantum_2001,Santoro} has spurred a great deal of recent interest in  building special-purpose analog quantum devices \cite{2002quant.ph.11152K,Dwave}
with the hope of realizing quantum speedups \cite{speedup}. In general, all pairs of spins in the Ising Hamiltonian can be coupled, resulting in long-range interactions, e.g., as in the Sherrington-Kirkpatrick model \cite{Sherrington75,Kirkpatrick:1978dn,PhysRevB.39.11828}.  A direct physical implementation of such spin systems would require all-to-all connectivity, an impossibility for analog devices with local hardware connectivity.  Instead one must embed the logical problem defined by the given Ising model into the available physical device connectivity. This embedding represents long-range logical interactions in terms of short-range physical interactions, a process which introduces unavoidable tradeoffs between energy scales, hardware resources, and accuracy \cite{Choi1,Choi2,klymko_adiabatic_2012,Cai:2014nx,Vinci:2015jt,Lechner:2015,Boothby2015a}. 

The first embedding technique used for quantum annealing is due to Choi and involves a graph-theoretical construction known as minor embedding (ME)  \cite{Choi1,Choi2}.  In ME, the spins of the given Ising problem with long-range interactions, which we refer to as the logical spins, are replaced by chains of physical spins with short-range interactions implemented on the device hardware. Physical spins within a chain are induced to behave as a single logical spin using strong ferromagnetic interactions acting as energy penalties. The hardware connectivity graph should allow for the minor embedding of complete graphs, i.e., the latter should be obtainable from the former via edge contractions. A well known example is the ``Chimera" graph used in the D-Wave quantum annealing devices~\cite{Bunyk:2014hb}.

Recently, Lechner, Hauke and Zoller (LHZ) proposed an elegant alternative embedding technique, realized on a two-dimensional triangular-shaped grid \cite{Lechner:2015}. In the LHZ scheme, the relative alignment (parallel or antiparallel) of pairs of logical spins is mapped to a physical hardware spin. Both the logical local fields and the logical couplings are mapped to local fields acting on the physical spins. This is appealing, since it places the burden of implementing the logical problem entirely on the implementation of local fields, thus obviating the need for simultaneous control of local fields and couplings. Couplings between the physical spins (acting as energy penalties) must still be introduced  to ensure that the mapping is consistent.

Since in the LHZ scheme the logical spins are encoded redundantly and non-locally in the physical degrees of freedom, it was suggested in Ref.~\cite{Lechner:2015} that the scheme includes an intrinsic fault tolerance, with some similarities to the error robustness of topological quantum memories \cite{Dennis:02}. This has been confirmed by Pastawski and Preskill (PP) under the assumption of a noise model of random spin-flips \cite{Pastawski:2015}. 
More specifically, PP pointed out that the LHZ scheme may be viewed as a classical low-density parity-check code that makes the scheme highly robust against weakly correlated spin-flip noise.

While allowing for a rigorous analytical treatment, weakly correlated random spin-flips are unfortunately not the most relevant errors in quantum annealing, the algorithm for which the LHZ scheme was proposed. In adiabatic quantum computing and quantum annealing, a time-dependent Hamiltonian evolves a system from an easily prepared and trivial ground state to the non-trivial ground state of the Hamiltonian of interest. In a closed system, the adiabatic theorem provides a guarantee that the probability of dynamical excitations can be made arbitrarily small if the evolution is sufficiently slow relative to the inverse of the minimum gap \cite{Kato:50,Jansen:07,lidar:102106}. Under the favorable assumption of weak system-environment coupling, decoherence in open-system, finite temperature quantum annealing takes place in the instantaneous energy eigenbasis \cite{childs_robustness_2001,amin_decoherence_2009,Albash:2015nx}. In both the closed and open-system cases, errors may occur throughout the evolution, resulting in the generation of final-time excited states that differ from the ground state by a large number of spin-flips, that are neither random nor weakly correlated \cite{Young:13,q108,PAL:13,PAL:14}, as also recognized in Ref.~\cite{Pastawski:2015}. 

Motivated by these considerations, here we study whether the ME or the LHZ scheme is  preferable under a realistic noise model for quantum annealing. This is particularly pertinent since the LHZ scheme would require the development of a new quantum annealing architecture, while the ME scheme has already been experimentally implemented in numerous studies, e.g., Refs.~\cite{proteins-dwave,Bian:2013zp,Ortiz:2015,Trummer:2015,Nazareth:2015,Rosenberg:2015,Vinci:2014fk,Adachi:2015qe}.
We employ simulated quantum annealing (SQA) \cite{sqa1,Heim:2014jf}, a quantum Monte Carlo method that iteratively updates an approximation to the instantaneous quantum Gibbs state governed by the time-dependent system Hamiltonian [Eq.~\eqref{eq:adiabatic} below]. 
While it is not a completely faithful description of the annealing process since it neglects unitary dynamics, it is the method of choice for large-scale simulations of stoquastic (sign-problem free) quantum annealing \cite{Bravyi:2014bf}, and has been used to successfully describe experiments on the D-Wave devices \cite{q-sig,DWave-16q,DWave-entanglement,Boixo:2014yu,q-sig2}.  Our SQA simulations aim to model the behavior of two quantum annealing devices implementing the same annealing schedule with identical physical parameters, but with different Hamiltonians, representing the ME and LHZ embedded problem Hamiltonians, respectively. 
In addition to SQA, we also use parallel tempering (PT) \cite{Geyer:91,Hukushima:1996}, a highly efficient variant of classical Monte Carlo that enables us to focus purely on the classical Hamiltonian, independently of the quantum evolution that takes place during the annealing.  

For the range of parameters and problem sizes studied, we find that the ME scheme outperforms the LHZ scheme for the majority of instances studied.
Using PT simulations on the final (classical) Hamiltonian, we show that the two schemes perform similarly when the number of updates is sufficiently large. We interpret this to mean that the difference between the two schemes arises from more efficient SQA updates for the ME Hamiltonian, arising during the evolution governed by the intermediate (quantum) Hamiltonian. Our results imply that the LHZ scheme does not necessarily exhibit an intrinsic fault tolerance against a noise model that is  appropriate for quantum annealing. 

\begin{figure*}[t]
\hspace{-0.2in} 
\subfigure[] {\includegraphics[width=0.5\columnwidth]{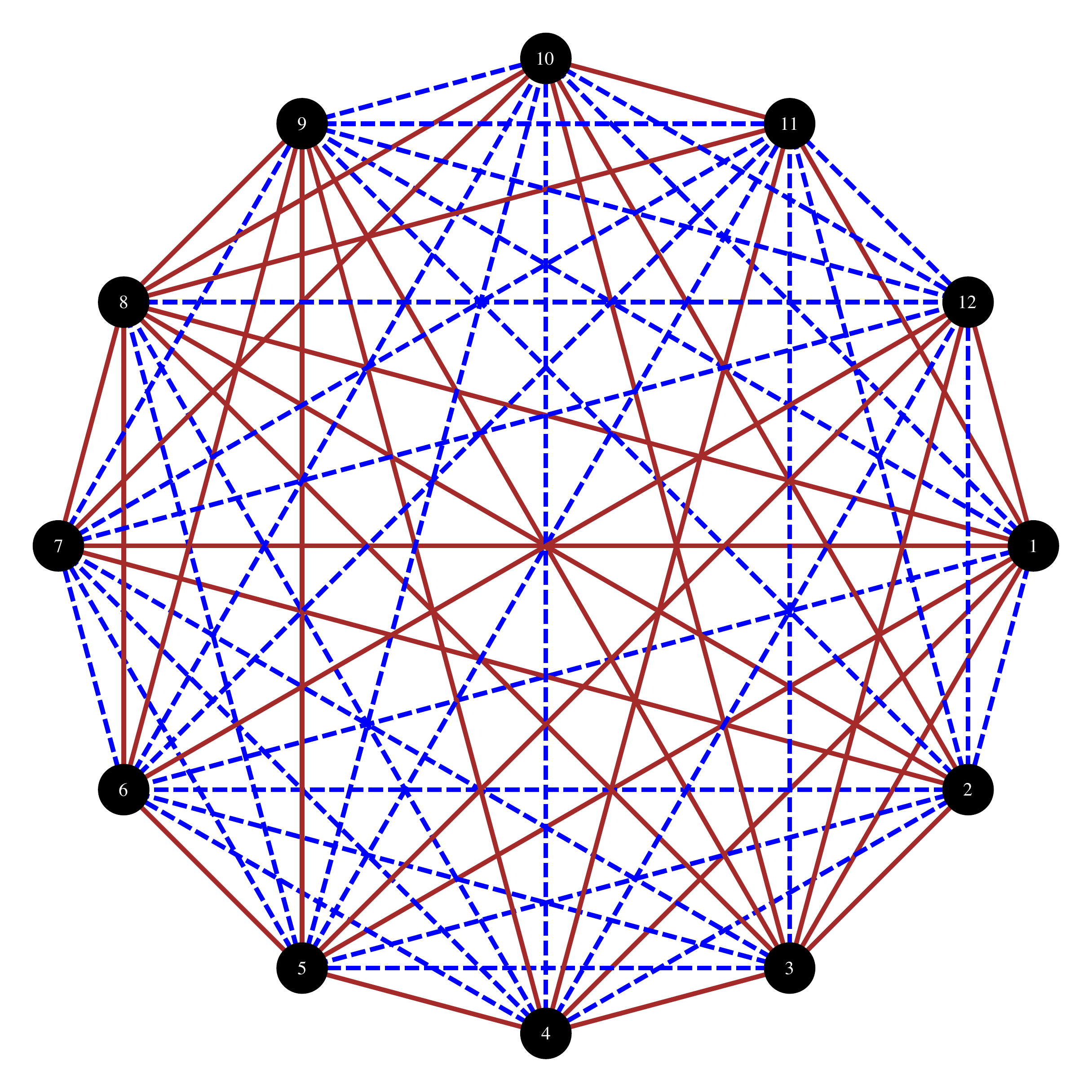}\label{fig:embs1}} \hspace{-0.1in} 
\subfigure[]{\includegraphics[width=0.7\columnwidth]{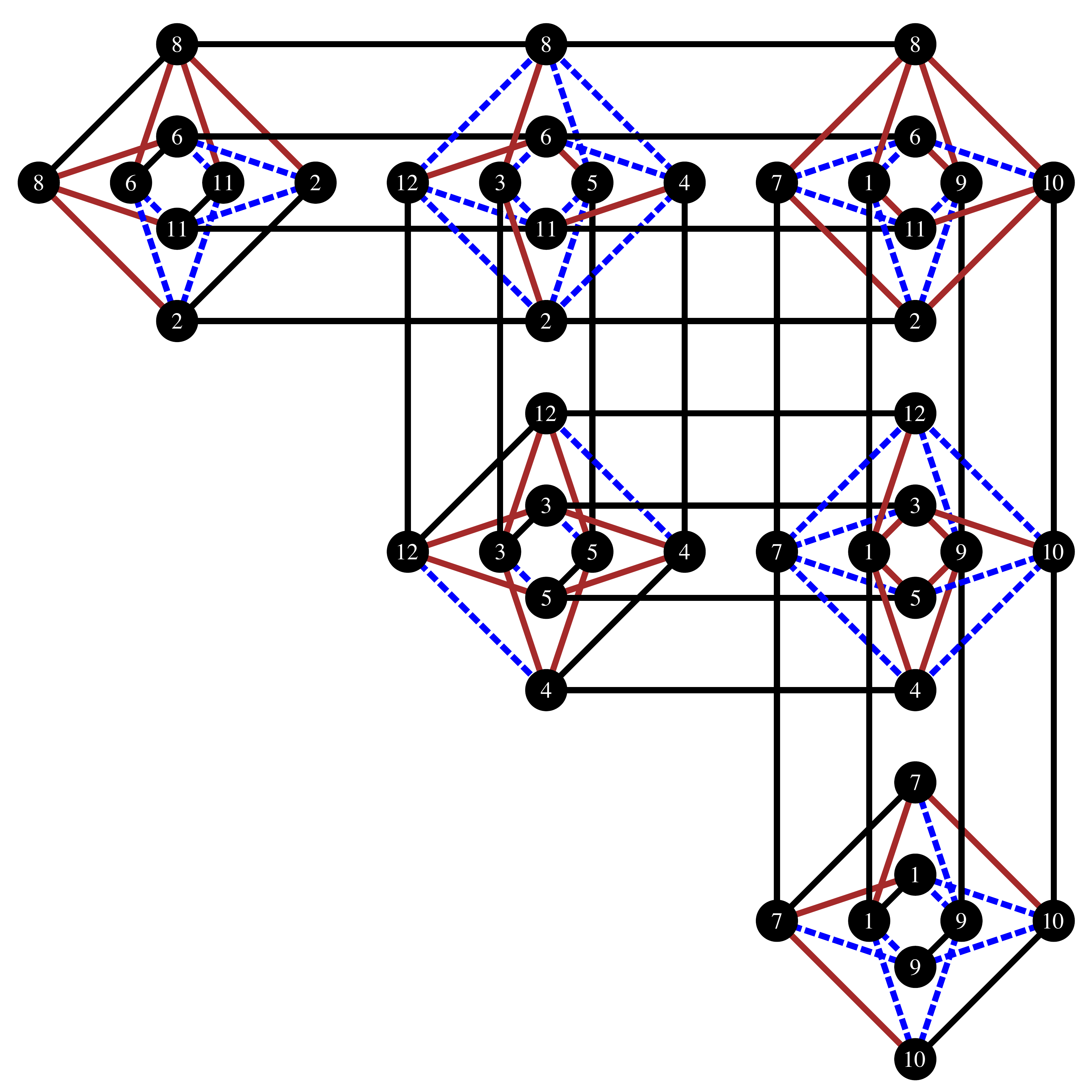}\label{fig:embs2}} \hspace{0.3in} 
\subfigure[] {\includegraphics[width=0.8\columnwidth]{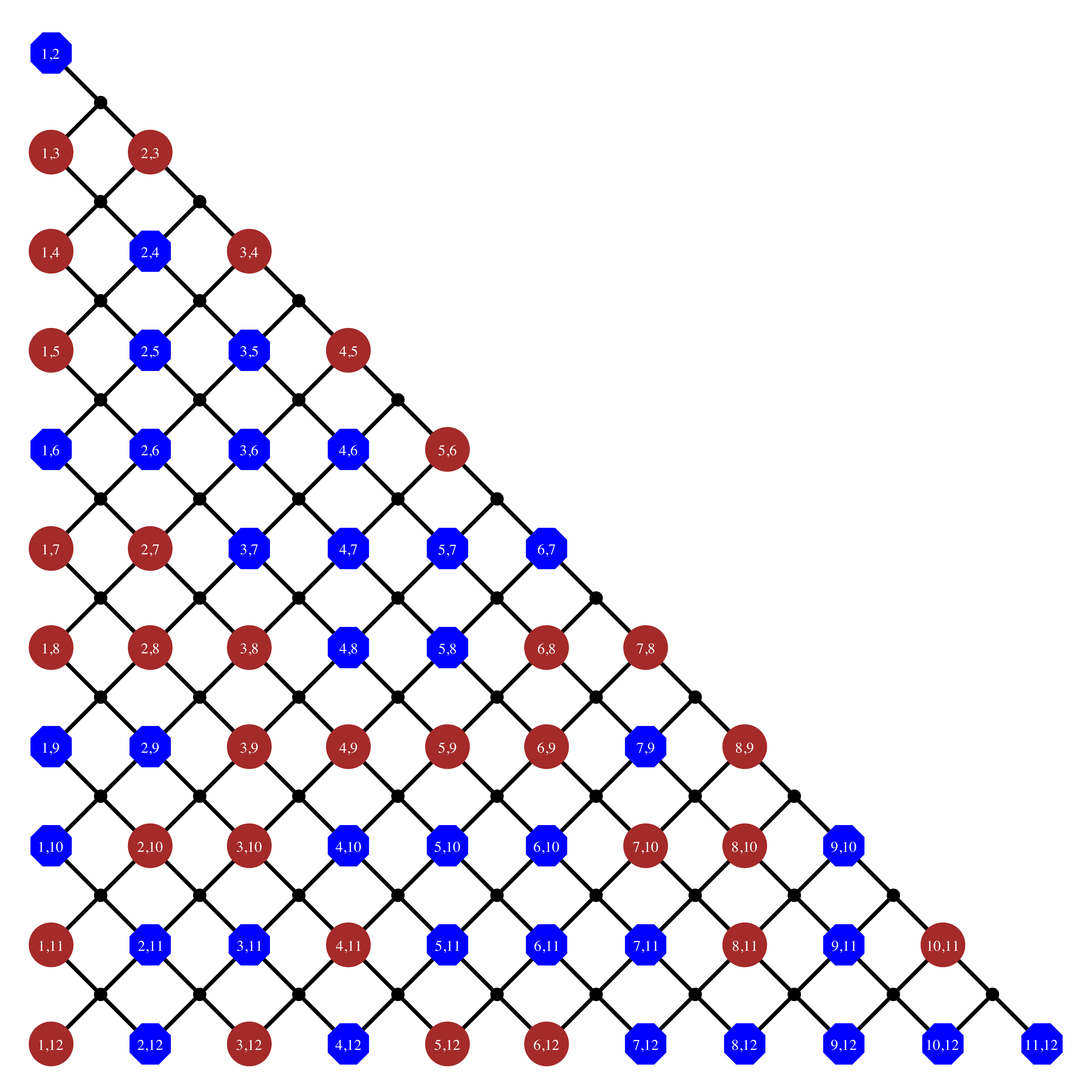}\label{fig:embs3}}
\caption{\textbf{All-to-all connectivity and its representation using the ME and LHZ schemes.} (Color online) (a) A complete graph of $12$ logical qubits. Brown (solid) and blue (dashed) lines correspond to types of couplings (e.g.,  ferromagnetic and antiferromagnetic). (b)  ME of the all-to-all Hamiltonian shown in (a). Brown (solid) and blue (dashed) lines are the physical realizations of the logical couplings. Black lines are the energy penalties within the logical chains. (c)  LHZ implementation of the same Hamiltonian. Red circles and blue octagons represent physical qubits with either negative (ferromagnetic) or positive (antiferromagnetic) local fields. Small black dots and lines represent three and four-body energy penalties.} 
\label{fig:embs}
\end{figure*}

The rest of this paper is organized as follows. In Sec.~\ref{sec:EQA} we describe the ME and LHZ schemes for embedded quantum annealing in more detail, as well as errors and majority vote decoding. We present our SQA and PT results in Sec.~\ref{sec:NR}, where we compare the ME and LHZ schemes subject to majority vote decoding. In Sec.~\ref{sec:OP} we 
consider several other decoding strategies for the LHZ scheme. In particular we describe how to map the decoding of the LHZ scheme to the decoding of a fully-connected, quadratic Sourlas code \cite{Sourlas:1989,Sourlas:1994}.This allows us to define minimum-weight (MWD) and maximum-likelihood (MLD) decoding strategies, the latter being optimal in the case of errors generated by random spin-flips. We also consider the belief propagation strategy proposed in Ref.~\cite{Pastawski:2015}.
We conclude and provide a broader discussion of our results in Sec.~\ref{sec:Conc}. Additional details are provided in the Appendix.
%
\section{Embedded Quantum Annealing}
\label{sec:EQA}
%
In quantum annealing one encodes the solution of a hard combinatorial optimization problem into the ground state of a classical Ising Hamiltonian:
\beq 
\label{eq:HP}
H_{\mathrm{P}} = \sum_{i } h_i \sigma^z_i + \sum_{i<j } J_{ij}\sigma^z_i\sigma^z_j\ .
\eeq
Hard problems  typically  result in highly frustrated spin systems with a spin-glass phase at low temperatures \cite{parisi2003constraint}. A quantum annealing device attempts to solve for the ground state of Eq.~\eqref{eq:HP} by implementing the following time-dependent Hamiltonian
\beq
H(t) = A(t) H_X + B(t)  H_{\mathrm{P}}\ , \qquad t\in[0,t_f] \ ,
\label{eq:adiabatic}
\eeq
where 
$H_{\mathrm{X}} = -\sum_{i } \sigma^x_i$ 
is the ``driver" term whose ground state is a uniform superposition in the computational basis that initializes the computation. 
The annealing schedule is specified by the functions $A(t)$ and $B(t)$, where typically $A(0)\gg B(0)$ and $A(t_f)\ll B(t_f)$. The most general optimization problem on $N$ binary variables is defined on an all-to-all connectivity problem Hamiltonian $H_{\mathrm{P}}$ graphically represented in Fig.~\ref{fig:embs1}. 

The construction of a physical  device that implements the system described by Eq.~\eqref{eq:adiabatic} with all-to-all connectivity can be technically demanding because it requires the implementation of long-range interactions. While in trapped ions long-range interactions are unproblematic \cite{Islam:2013mi}, not all physical implementations support such interactions; e.g., in most superconducting and semiconducting devices interactions are hard-wired and quasi-local \cite{Bunyk:2014hb,Barends:2015kl,Eng:2015if,Casparis:2015fe}, and for ultracold gases in optical lattices interactions are determined by the geometry \cite{Bloch:2008qa}. Schemes that approximate long-range interactions with short-range interactions are thus desirable for certain implementations. The ME and LHZ schemes are both designed to address this challenge, as we briefly review next.

%
\subsection{Minor Embedding Scheme}
%

The ME scheme replaces long-range interactions between logical qubits by short-range interactions between chains (or clusters) of physical qubits, where each chain represents a logical qubit, and correspondingly replaces the logical problem Hamiltonian $H_{\mathrm{P}}$ [Eq.~\eqref{eq:HP}] by a minor-embedded Hamiltonian $H^{\mathrm{ME}}_{\mathrm{P}}$. Physical qubits belonging to the same logical-qubit chain are connected by strong ferromagnetic interactions (acting as energy penalties), which lowers the energy of configurations where these qubits are aligned at the end of the anneal. 
``Broken" chains (where not all members of a chain are aligned) do not correspond to logical states and must be corrected, e.g., by majority vote. For a more formal and complete exposition of ME see, e.g., Ref.~\cite{Vinci:2015jt}.

We consider here a particular ME on ``Chimera" graphs. Chimera is a degree-six graph formed from the tiling of an $L\times L$ grid of $K_{4, 4}$ cells. Despite its low degree and near planarity, an $L \times L$ Chimera graph allows for the minor embedding of an all-to-all logical problem $H_{\mathrm{P}}$  defined on $N = 4L$ logical qubits where each logical qubit is represented by a chain of $\lceil N/4 \rceil+1$ physical qubits.  Figure~\ref{fig:embs2} depicts a minor embedded Hamiltonian $H^{\mathrm{ME}}_{\mathrm{P}}$ of the all-to-all Hamiltonian $H_{\mathrm{P}}$ of Fig.~\ref{fig:embs1}, on a Chimera subgraph. Blue and red thin links correspond to physical couplers representing the logical interactions of $H_{\mathrm{P}}$, while black couplers represent strong ferromagnetic interactions that couple physical qubits belonging to the same logical chain.  Note in particular that while the full Chimera graph would have $8L^2=N^2/2$ spins, the embedding of the complete graph of size $N$ only uses $N(\lceil N/4 \rceil+1)$, i.e., about half of these spins.  For a more detailed description of the ME scheme on Chimera graphs see, e.g., Refs.~\cite{Choi1,Choi2}.

\begin{figure*}[t]
   \subfigure[]{\includegraphics[width=1\columnwidth]{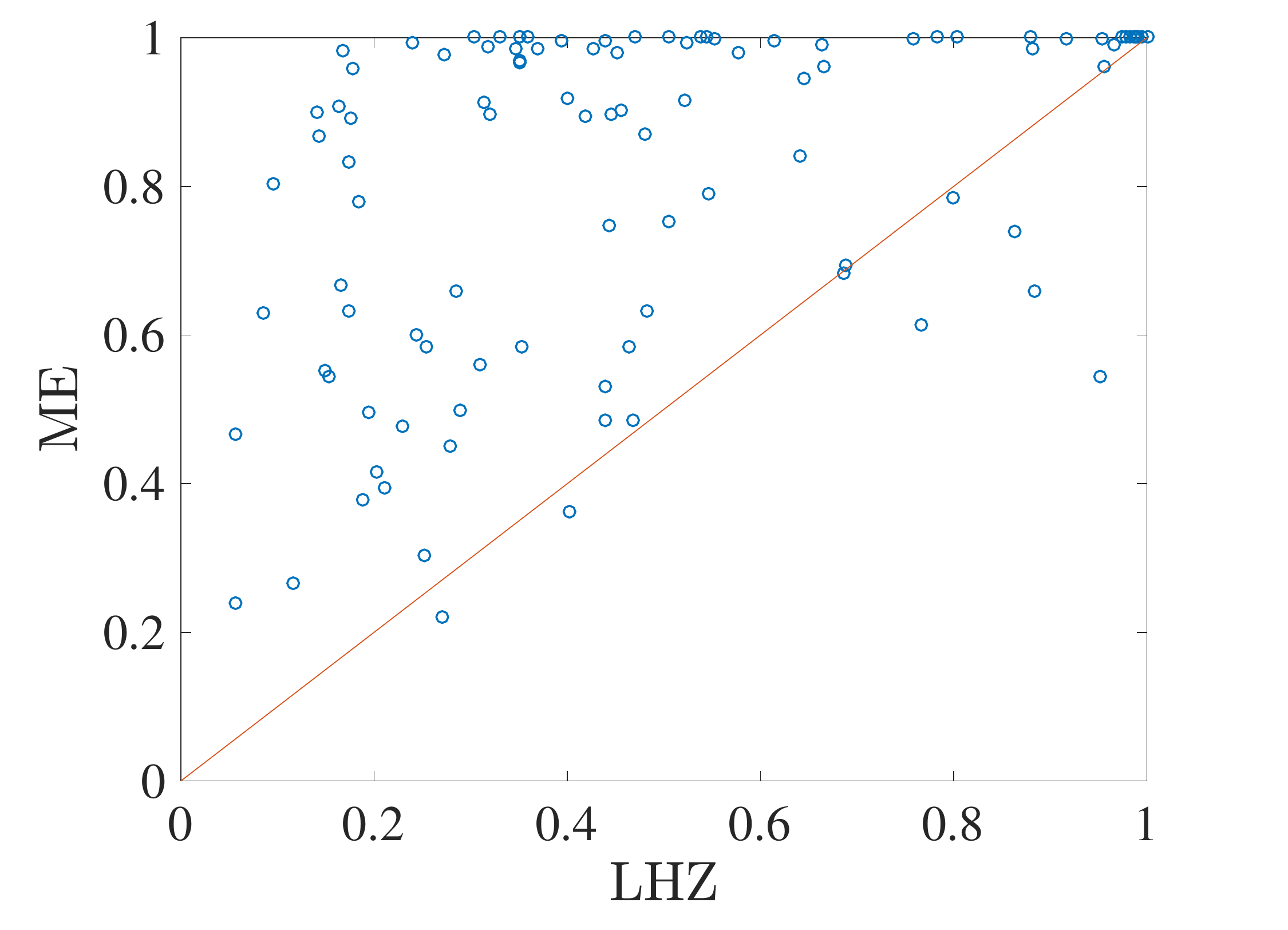}\label{fig:SQAComp1}} \hspace{-0in}
   \subfigure[]{\includegraphics[width=1\columnwidth]{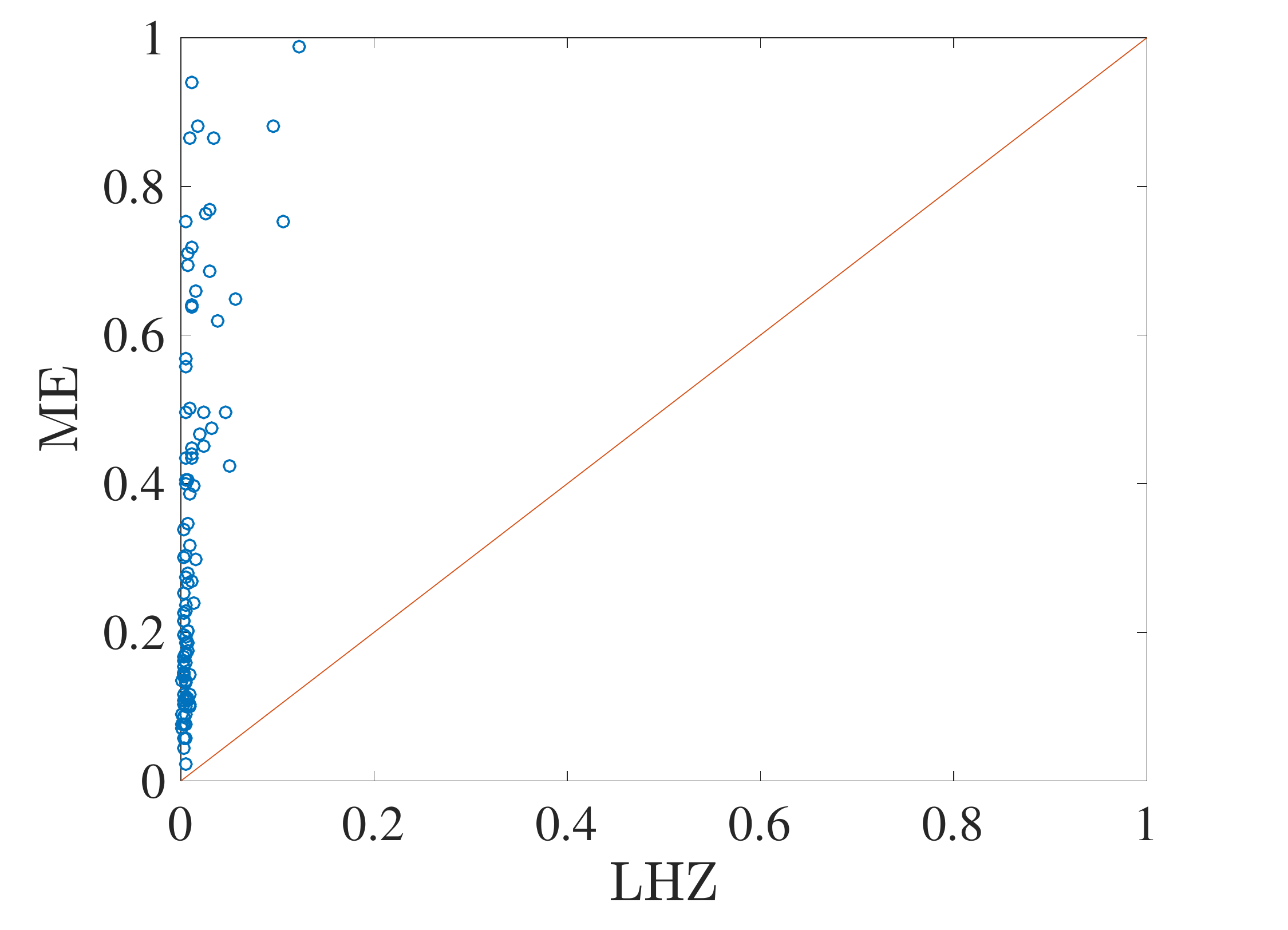}\label{fig:SQAComp2}} \hspace{-0.3in}
   \caption{\textbf{Relative performance of the ME and LHZ schemes.} (Color online) Shown are scatter plots of success probabilities for $100$ random (a) $K_8$ and (b) $K_{16}$ instances generated via SQA and majority vote decoding. The ME scheme exhibits substantially better performance. Majority vote decoding with the same number of votes ($3$ and $5$ for $K_8$ and $K_{16}$, respectively) was used for both schemes, and the optimal energy penalty was used for each instance.  SQA parameters: $10^4$ sweeps, $\beta = 1$.  Note that the effective temperature of SQA is modulated through the anneal by the annealing schedule (Fig.~\ref{fig:AnnealingSchedule} in Appendix~\ref{sec:NS}).  By the end of the anneal, the effective (dimensionless) temperature is $\approx 0.05$, smaller than the absolute value of the smallest coupling used in our simulations, $0.1$.}
\label{fig:SQAComp}
\end{figure*}

\subsection{Lechner-Hauke-Zoller Scheme}
\label{sec:LHZ}
%

In the LHZ scheme, the embedded Hamiltonian $H^{\mathrm{LHZ}}_{\mathrm{P}}$ is defined on  $K = \binom{N}{2}$ physical qubits corresponding to the $K$ logical interactions of the original problem Hamiltonian $H_{\mathrm{P}}$. The logical couplings $J_{ij} \equiv J_k$ are mapped to local fields applied to the physical qubits (we denote the physical qubit's value in the computational basis by $q_k \equiv q_{i,j} = \pm 1$ ).  The value of a physical qubit in the LHZ scheme encodes the relative alignment of the corresponding pair of logical qubits: a physical qubit pointing up (down) corresponds to an aligned (anti-aligned) logical pair. Since $K>N$, the physical system includes physical states that do not correspond to logical states, similarly to what happens in the ME case. The appearance of these spurious states is suppressed by imposing a sufficient number of constraints (energy penalties). These constraints are designed to favor an even number of spin-flips around any closed loop in the logical spins, in order to induce a consistent mapping. The form of the LHZ Hamiltonian we consider here is the following \cite{Lechner:2015}:
\beq 
\label{eq:LHZ}
H^{\mathrm{LHZ}}_{\mathrm{P}} = \sum_{k \in \mc{V}_{\mathrm{LHZ}}} J_k \sigma^z_{k}+ \sum_{c \in \mc{V}_{\mathcal C}} \mathcal C_c\ ,
\eeq
where $\mathcal C_c = - \lambda \sigma^z_{c_u}\sigma^z_{c_d}\sigma^z_{c_l}\sigma^z_{c_r}$ are  four-body interactions between the physical qubits. In addition, three-body interactions appear at the LHZ graph edges, as shown in Fig.~\ref{fig:embs3}. Both the three and four-body interactions can be replaced by two-body interactions, by coupling to ancilla qutrits \cite{Lechner:2015}. $\mc{V}_{\mathrm{LHZ}}$ represents the vertex set of the LHZ physical graph depicted in Fig.~\ref{fig:embs3} by circles, with $K=|\mc{V}_{\mathrm{LHZ}}|$, and where $\mc{V}_{\mathcal C}$ represents the $C=|\mc{V}_{\mathcal C}|=\binom{N-1}{2}$ constraint interactions shown as black dots. Local fields $h_i$ can be included by representing them as couplings to an additional ancilla qubit, via $h_i \sigma^z_i \rightarrow h_i \sigma^z_0 \sigma^z_i$.

\subsection{Leakage errors in embedded quantum annealing}
\label{sec:EDEQA}
%
Due to the fact that in both the ME and LHZ schemes a problem Hamiltonian defined on $N$ logical qubits is embedded into a system with $O(N^2)$ physical qubits, the physical system has a much larger number of physical states ($\sim {2}^{N^2}$) than the original logical system ($\sim {2}^{N}$).  This allows for ``leakage": broken chains in the ME scheme or physical states with unsatisfied constraints in the LHZ scheme, that appear because of thermal and dynamical errors occurring during the annealing process. Such leakage states do not directly correspond to a logical state (see Fig.~\ref{fig:errors} in Appendix~\ref{App:LHZMapping} for typical examples of leakage states one expects in the LHZ scheme for the random weakly correlated errors model and for finite-temperature open-system quantum annealing).

 A leaked physical state must be decoded, i.e., it must be mapped back to a logical state.  This decoding can be considered as partial error correction, in the sense that it recovers a logical state but it does not guarantee the recovery of the logical ground state. Next, we briefly review majority vote decoding; 
 additional decoding schemes for the LHZ scheme will be considered in Sec.~\ref{sec:OP}.

\begin{figure*}[t]
\subfigure[]{\includegraphics[width=0.72\columnwidth]{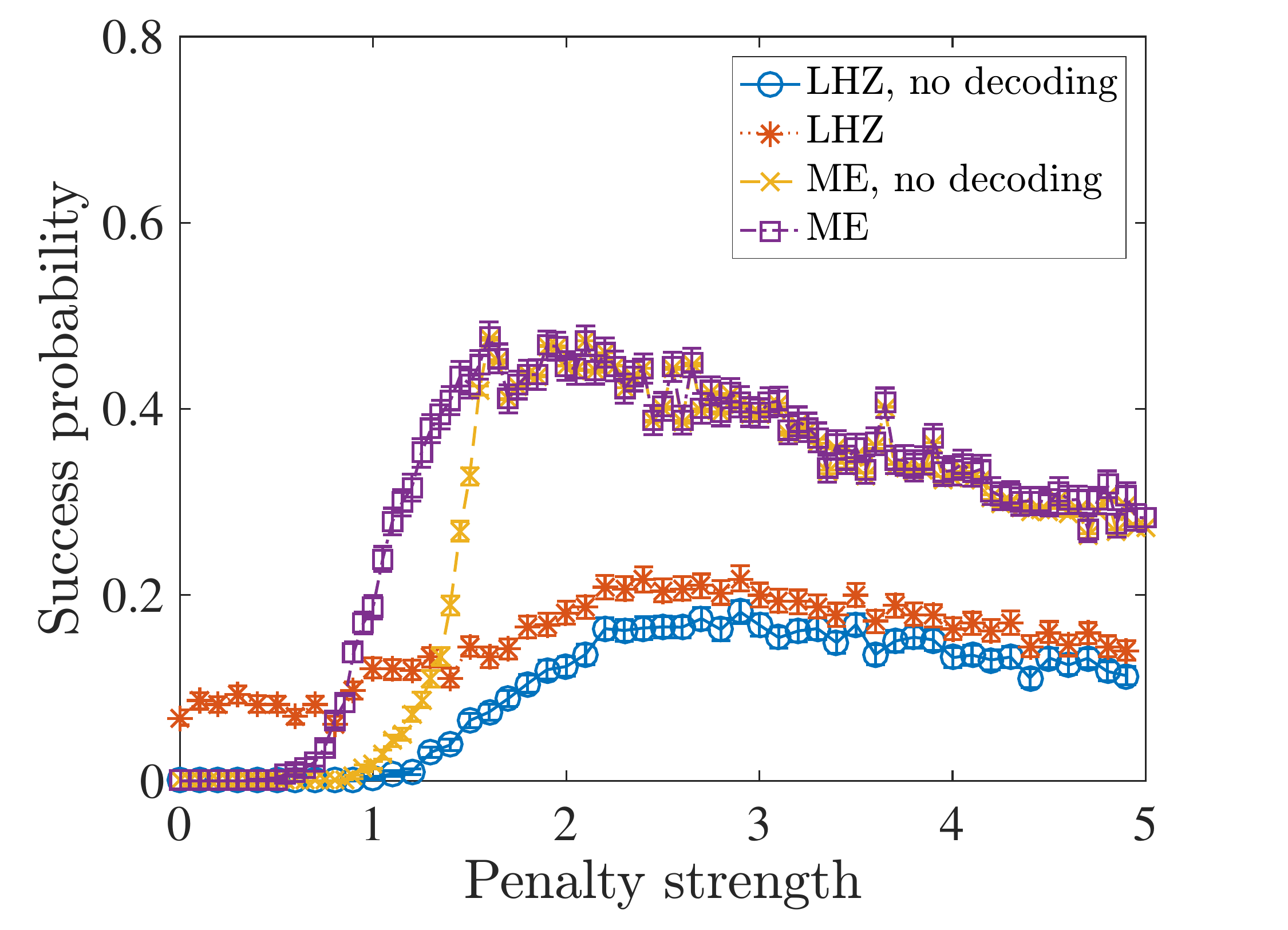} } \hspace{-0.3in}
\subfigure[]{\includegraphics[width=0.72\columnwidth]{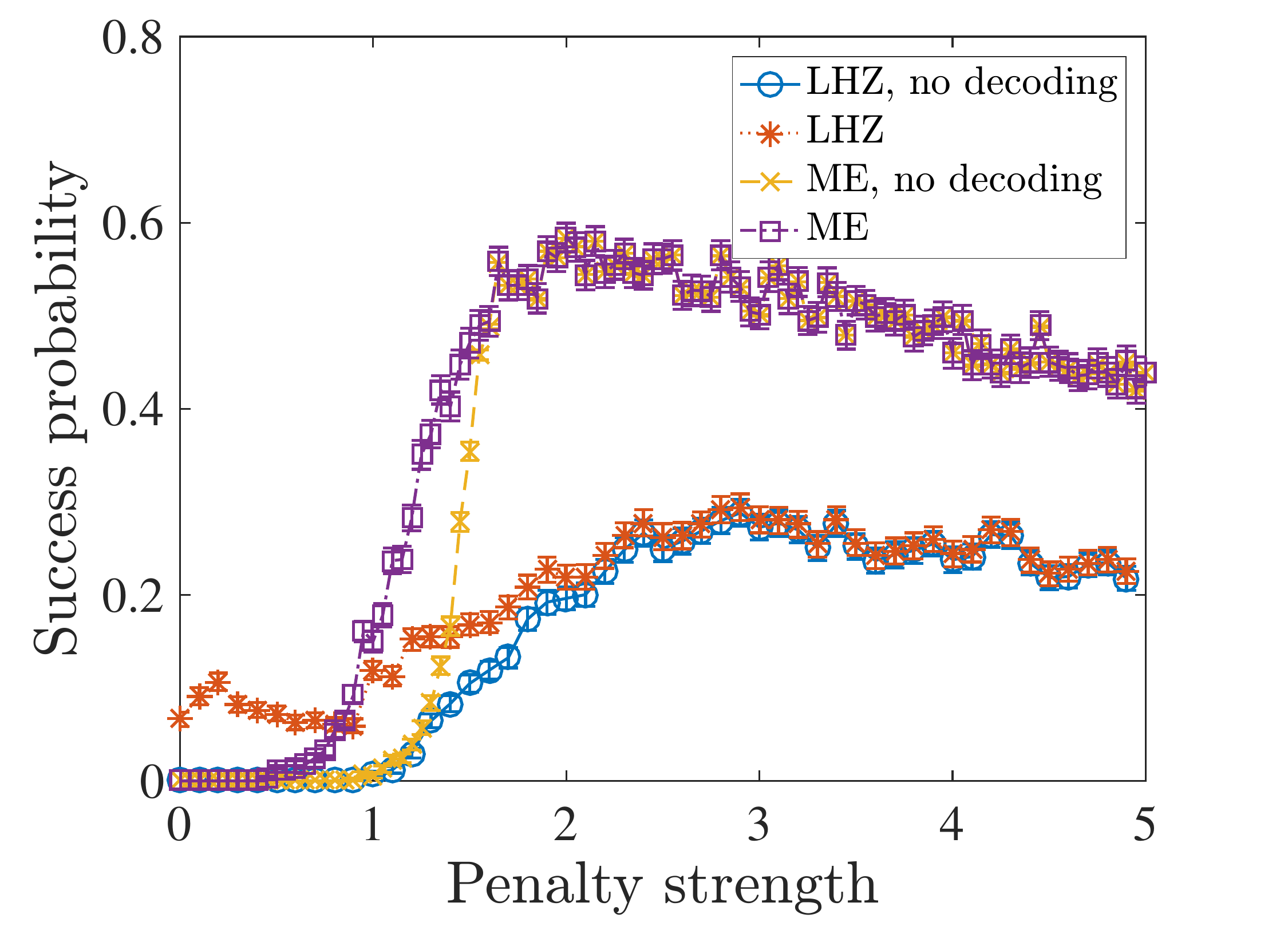}} \hspace{-0.3in}
\subfigure[]{\includegraphics[width=0.72\columnwidth]{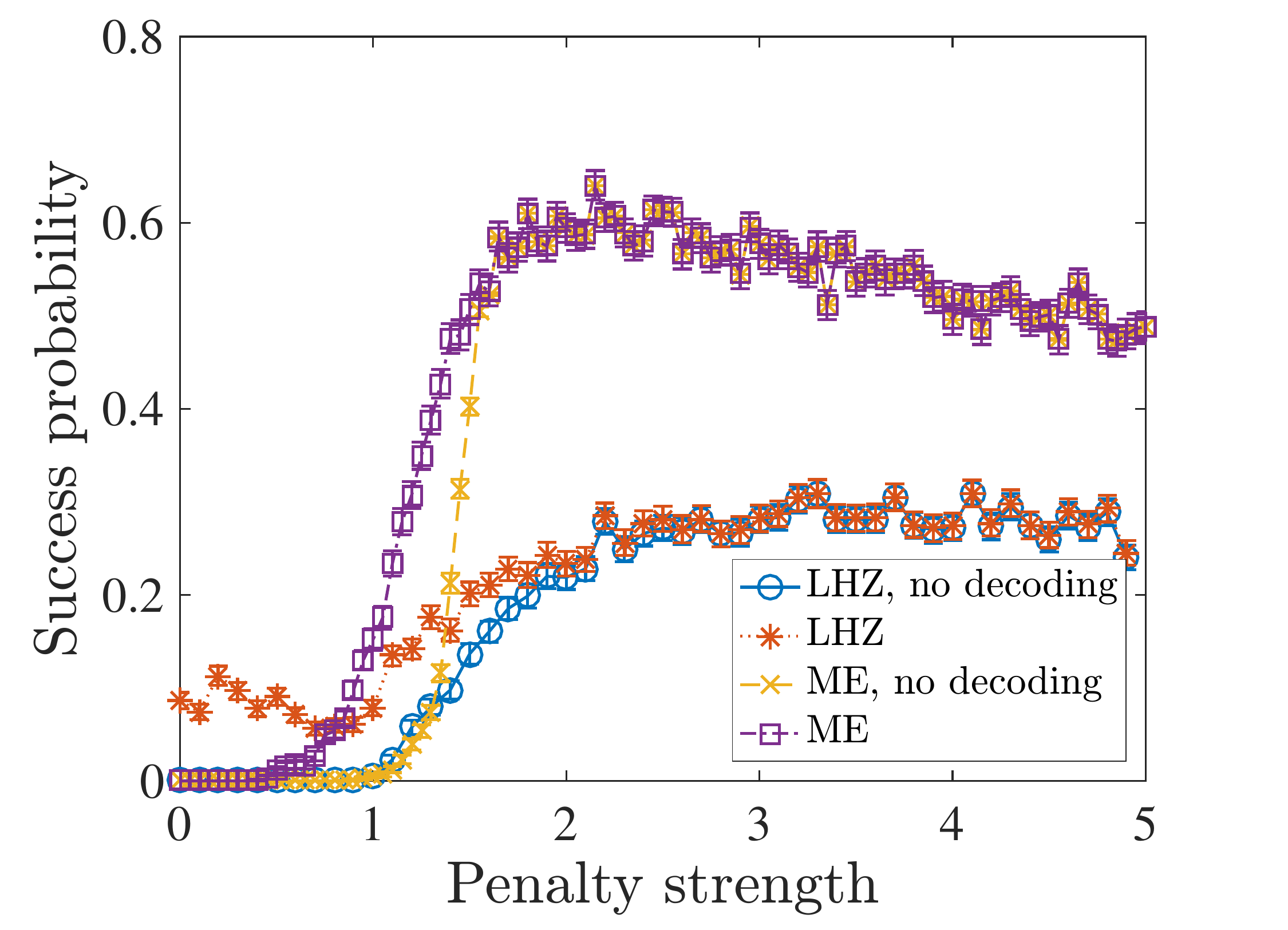} }
\caption{\textbf{Dependence on penalty strength and the number of sweeps.} (Color online) Shown is the success probability of the ME and LHZ schemes for a representative instance generated via SQA, as a function of penalty strength, with a varying number of sweeps: (a) $10^4$ sweeps, (b) $5 \times 10^4$ sweeps, and (c) $10^5$ sweeps. The decoded ME scheme always outperforms the decoded LHZ scheme. 
The optimal penalty strength increases with the number of sweeps.
Performance improves consistently for both ME and LHZ as the number of sweeps increase. The non-vanishing success probability at zero penalty for the decoded LHZ scheme is an artifact of the small size of the logical problems considered, which allows for a non-negligible probability that MVD will retrieve a logical ground state.}
\label{fig:SQA_boosted} 
\end{figure*}

\subsection{Majority vote decoding of the ME Scheme} 
\label{sec:DecodingME}
%
A minor-embedded logical qubit $i$ is encoded into a set of physical qubits $\{a_i\}_{a=1}^{n_A}$. Majority vote decoding (MVD) of logical qubit $i$ consists of computing
\beq
\bar q_{i} = {\rm sign} \sum_{a=1}^{n_A} q_{a_i}\, ,
\eeq
where $ q_{a_i}$ represents the measured value of physical qubit $a_ i$ in the computational basis. In addition to its being used routinely in decoding minor embedded quantum annealing \cite{Venturelli:2014nx,Venturelli:2015pi,Zick:2015cs,humble2014integrated,King:2014uq}, MVD has been successfully used in the context of quantum annealing correction (QAC) \cite{PAL:13,PAL:14,Mishra:2015} and hybrid minor-embedded implementations of QAC \cite{Vinci:2015jt,vinci2015nested}. 

MVD relies on the assumption that the decoded values $\{\bar q_{i}\}$ are the most likely to recover the logical ground state. However, this is not ensured due to the complex way errors are generated in quantum annealing. Alternatives such as energy minimization, which tends to be a better strategy for quantum annealing, have been explored as well \cite{Vinci:2015jt}, but will not be further considered here.

\subsection{Majority vote decoding of the LHZ  Scheme} 
\label{sec:DecodingLHZ}

In the absence of leakage, any spanning tree (i.e., a graph where there is a path connecting all vertices but any two vertices are connected by exactly one path) on the logical graph can be used to reconstruct a logical configuration from the values of the physical qubits. In the presence of leakage, however, different spanning trees may decode the same physical state to different logical states.
A simple decoding technique is to perform a majority vote on $n_T$ logical states decoded from random spanning trees~\cite{Lechner:2015}:
\beq
\bar q_{ i} = {\rm sign} \sum_{t = 1}^{n_T}  \bar{q}_{t,i}\,,
\eeq
where $\bar{q}_{t,i}$ represents the decoded value of logical qubit $i$ retrieved from spanning tree $t$ (see Appendix~\ref{App:LHZMapping} for further details and examples).

\section{Results for majority vote decoding}
\label{sec:NR}

In order to compare the ME and LHZ schemes we first generated a set of $100$ logical random instances on complete graphs $\{K_{8},K_{16}\}$ with couplings ${J}_{ij} \in \left\{ \pm 0.1, \pm 0.2, \pm 0.3, \dots \pm 1 \right\}$ chosen uniformly at random, and all logical local fields set to zero.  For each instance, we constructed the embedded physical Hamiltonians $H^{\mathrm{ME}}_{\mathrm{P}}$ and $H^{\mathrm{LHZ}}_{\mathrm{P}}$.  For each instance and embedding scheme, we ran SQA \cite{sqa1,Heim:2014jf} using the quantum annealing Hamiltonian Eq.~\eqref{eq:adiabatic} with the same annealing schedule, temperature, and number of Monte Carlo sweeps. A single sweep involves applying a Wolff cluster update along the imaginary time direction for all spatial quantum Monte Carlo slices (see Appendix~\ref{sec:NS} for more details on SQA).  

The annealing schedule sets the effective energy scale of the Hamiltonian [Eq.~\eqref{eq:adiabatic}], and the ratio of this scale to the temperature changes during the anneal.  The annealing schedule can play a crucial role in determining the performance of the adiabatic algorithm, and in principle can be optimized separately for the ME and LHZ schemes.  Furthermore, the ME and LHZ schemes may benefit from controlling the constraints (the chains in the ME case, and the four-body terms in the LHZ case) using an independent annealing schedule.  We do not address these issues in this work, and leave the exploration of possible improvements using these strategies for future work.

In order to understand the role of the purely classical Hamiltonians $H^{\mathrm{ME}}_{\mathrm{P}}$ and $H^{\mathrm{LHZ}}_{\mathrm{P}}$ in the success of each scheme, we used parallel tempering (PT) \cite{Geyer:91,katzgraber:06a}. For sufficiently long runtimes, PT samples from the classical thermal (Gibbs) state.  If the temperature is sufficiently low, PT samples predominantly from the ground state, and hence it can be used as a solver to find the ground state of the classical Hamiltonian. PT is  a more efficient algorithm than simulated annealing \cite{kirkpatrick_optimization_1983} for both the sampling and solver tasks (see Appendix~\ref{sec:NS} for more details on PT). 
Here we use PT as a sampler on the ME and LHZ Hamiltonians, whereby we run PT with a long but fixed runtime, while restricting it to single-spin updates in order to keep the algorithm as close to the SQA implementation as possible for both the ME and LHZ schemes\footnote{Cluster updates that take into account the geometry of each embedding scheme are likely to help speed up the convergence of PT to the Gibbs state for both the ME and LHZ schemes. For example, cluster updates associated to the flip of a logical qubit (a chain in the ME case and a cluster consisting of all physical qubits $ \left\{ q_{i,j}, q_{k,i} \right\}_{(j >i, k < i)}$ for logical qubit $\bar q_i$ in the LHZ case).} At sufficiently large penalty values, we expect the leakage states to be completely decoupled from the low-lying levels in both the ME and LHZ schemes. The low energy physical states are then equivalent to the low energy logical states, which are the same for the two schemes, so at sufficiently low temperatures we expect the PT success probabilities for both schemes to coincide. As we increase the temperature but keep the runtime constant, we expect deviations between the two schemes to emerge. This allows us to compare how well each scheme performs in recovering the logical ground state from the low-energy physical states.

In this section we perform the comparison between the ME and LHZ schemes using only majority vote decoding, before considering more sophisticated decoding strategies in Sec.~\ref{sec:OP}.  MVD allows us to equalize the decoding effort between the two schemes.

\begin{figure}[t] 
   \centering
     \includegraphics[width=1\columnwidth]{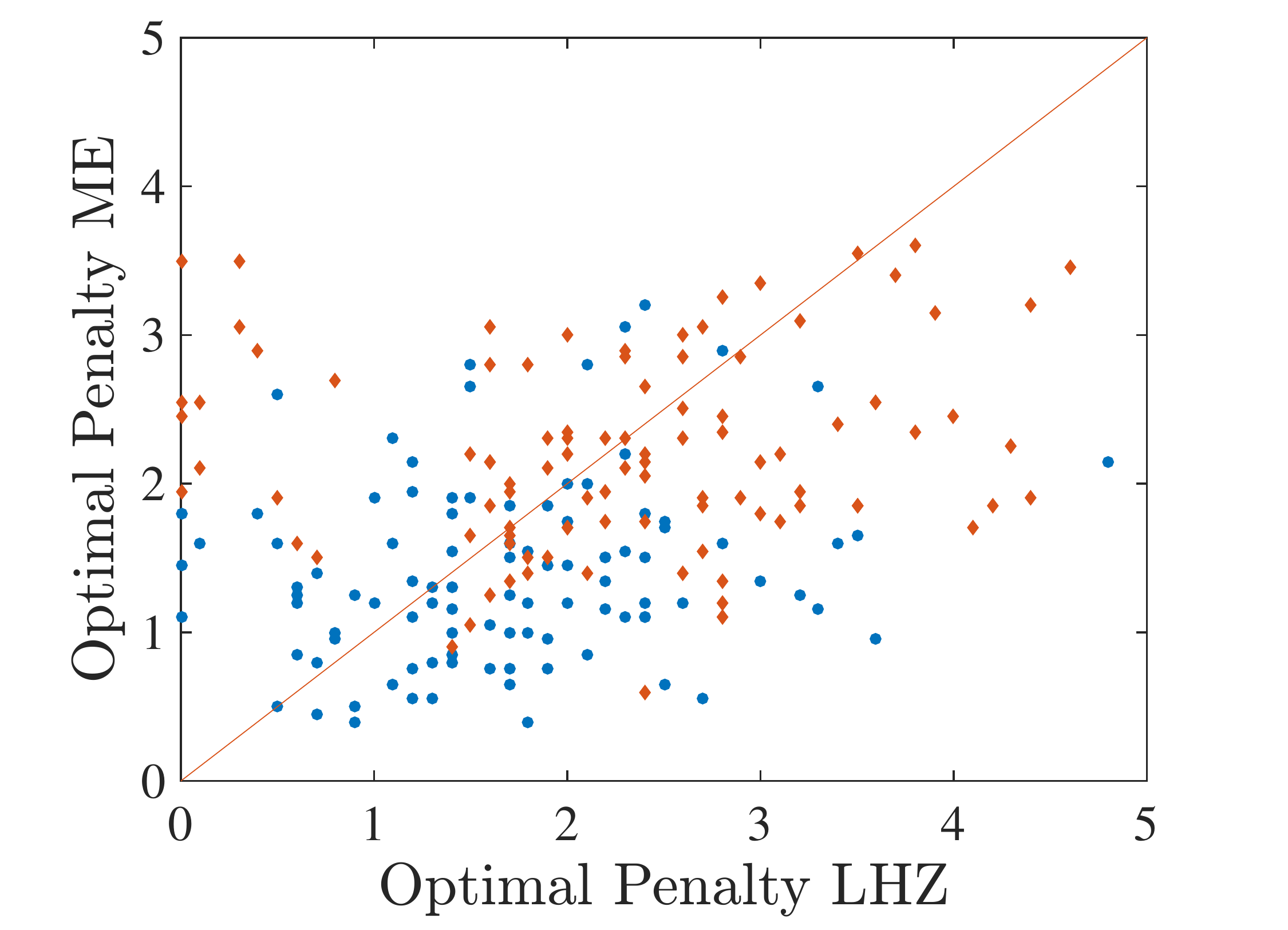}
   \caption{\textbf{Optimal penalty strength comparison.} (Color online) Shown is a comparison of optimal penalties, i.e., the penalty values that maximize the success probability after MVD and using SQA.  These correspond to the penalty strengths used for the results shown in Fig.~\ref{fig:SQAComp}. The optimal penalty is roughly equally distributed on $K_8$ instances (blue dots) but tends to be higher for the LHZ scheme on $K_{16}$ instances (red diamonds). SQA parameters: $10^4$ sweeps, $\beta = 1$.}
\label{fig:example6}
\end{figure}

\begin{figure*}[t]
\subfigure[]{\includegraphics[width=0.72\columnwidth]{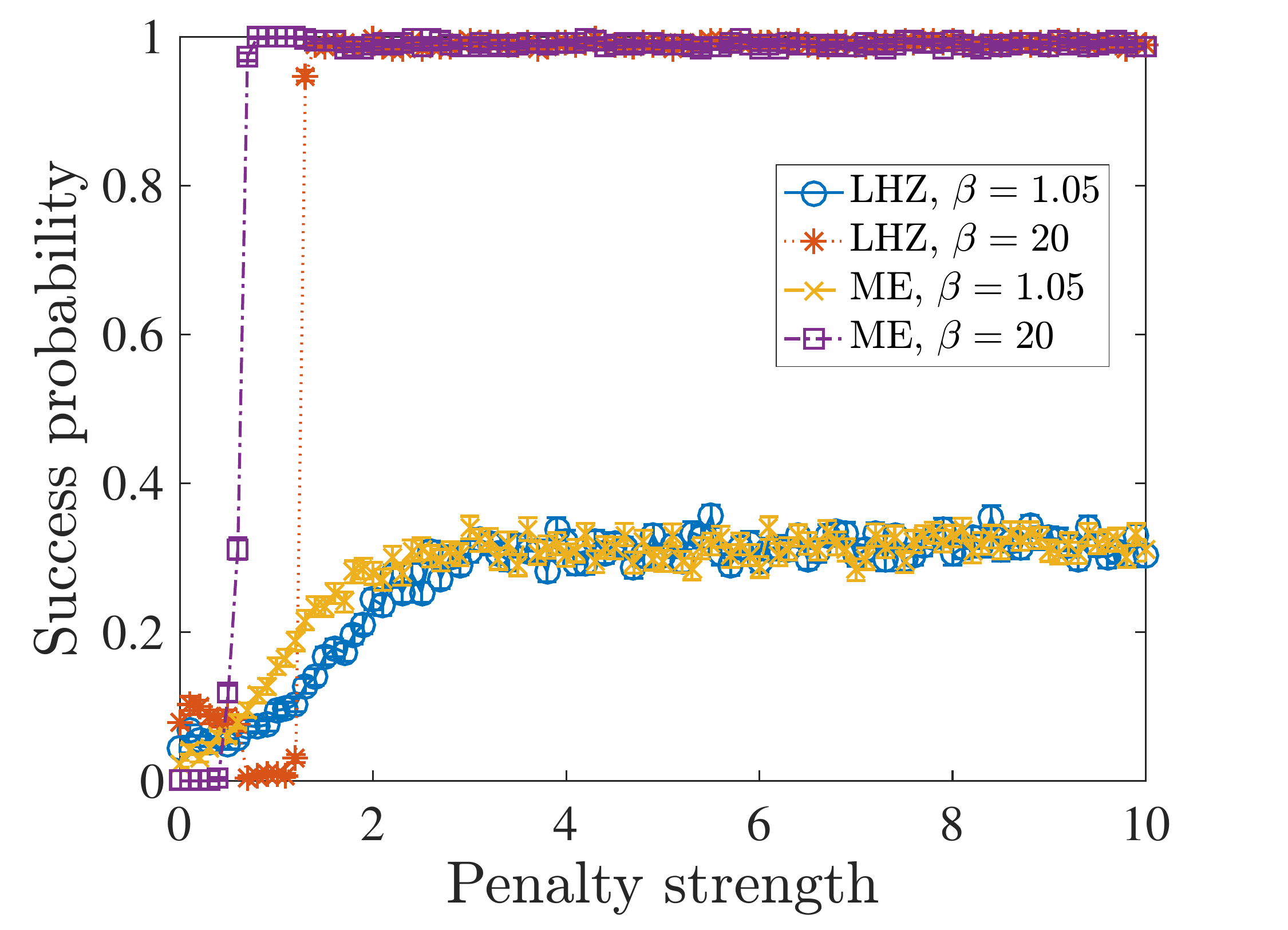} \label{fig:PT1}} \hspace{-0.3in}
\subfigure[]{\includegraphics[width=0.72\columnwidth]{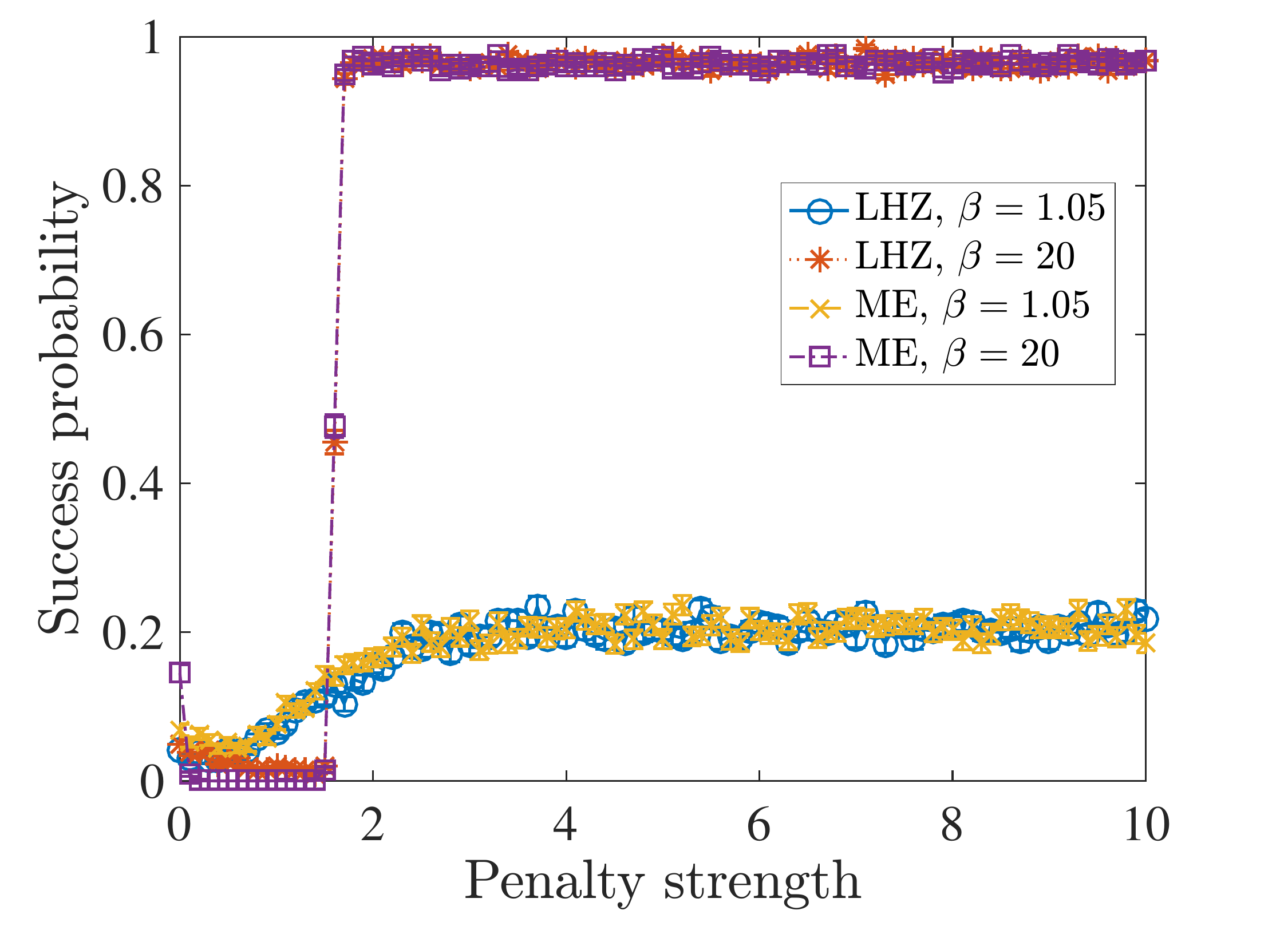} \label{fig:PT2}} \hspace{-0.3in}
\subfigure[]{\includegraphics[width=0.72\columnwidth]{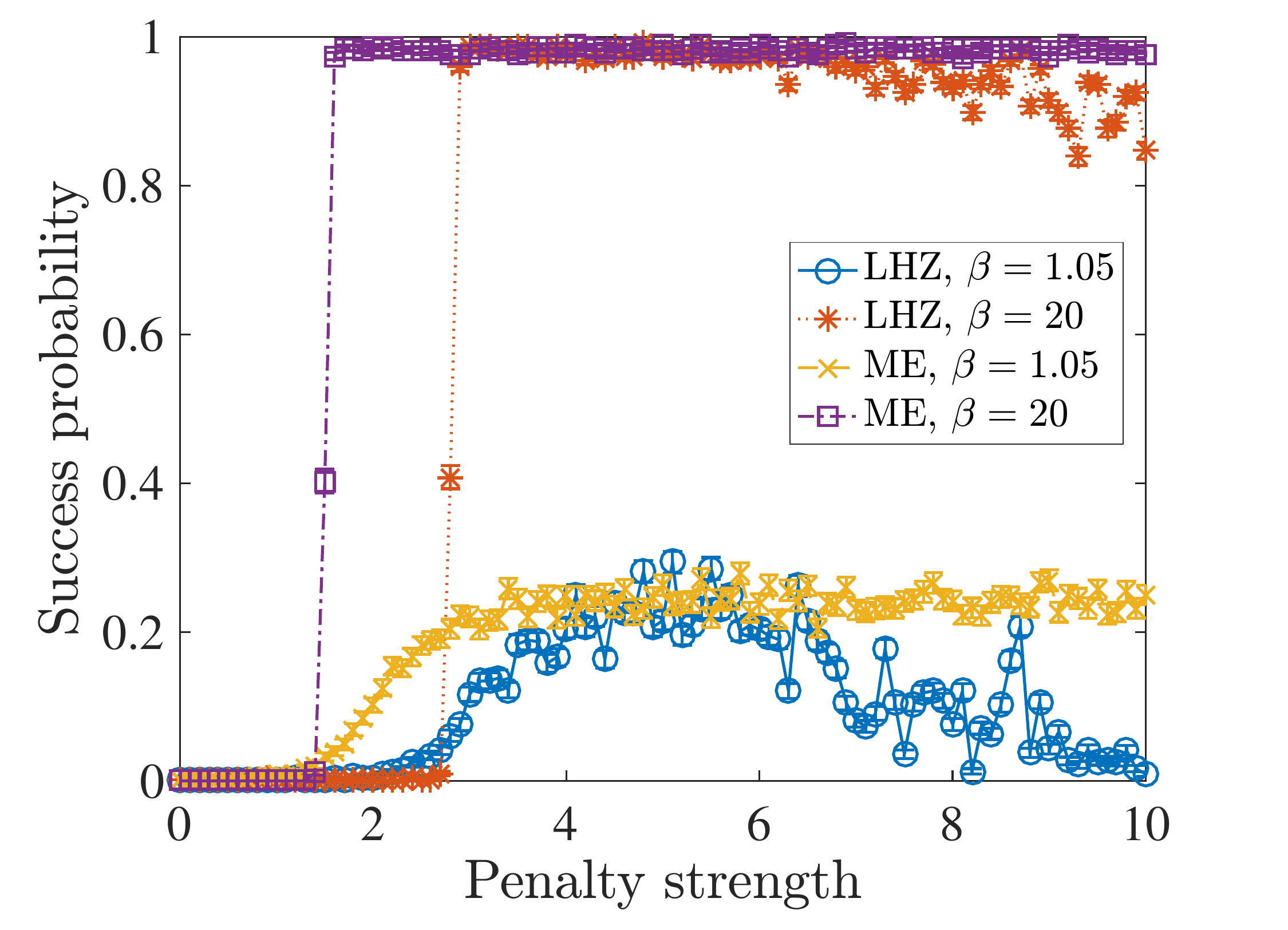} \label{fig:PT3}} 
\caption{\textbf{Performance of the ME and LHZ schemes using PT.} (Color online) Shown are the success probabilities after MV decoding for three representative instances generated via PT on $K_8$ [(a) instance 3 and (b) instance 95] and $K_{16}$ (c) (instance 72) graphs. Results for two temperatures are shown; reducing the temperature boosts the success probabilities, as expected. The ME and LHZ schemes exhibit similar performance for the $K_8$ instances, though the optimal penalty value is again smaller for the ME scheme in (a). Similar results are seen for the $K_{16}$ instances. The drop in success probability at large penalty strength for $\beta=1.05$ in the LHZ case (all but two of our $100$ instances exhibit this signature) can be understood as a signature of incomplete thermalization at the chosen parameter values. It is alleviated by increasing the number of PT swaps (see Appendix~\ref{sec:Add}). PT parameters are discussed in Appendix~\ref{sec:NS}.} 
\label{fig:PT}
\end{figure*}

\subsection{Comparing ME and LHZ via simulated quantum annealing} \label{sec:Compare}
      
\begin{figure}[t]
   \includegraphics[width=1\columnwidth]{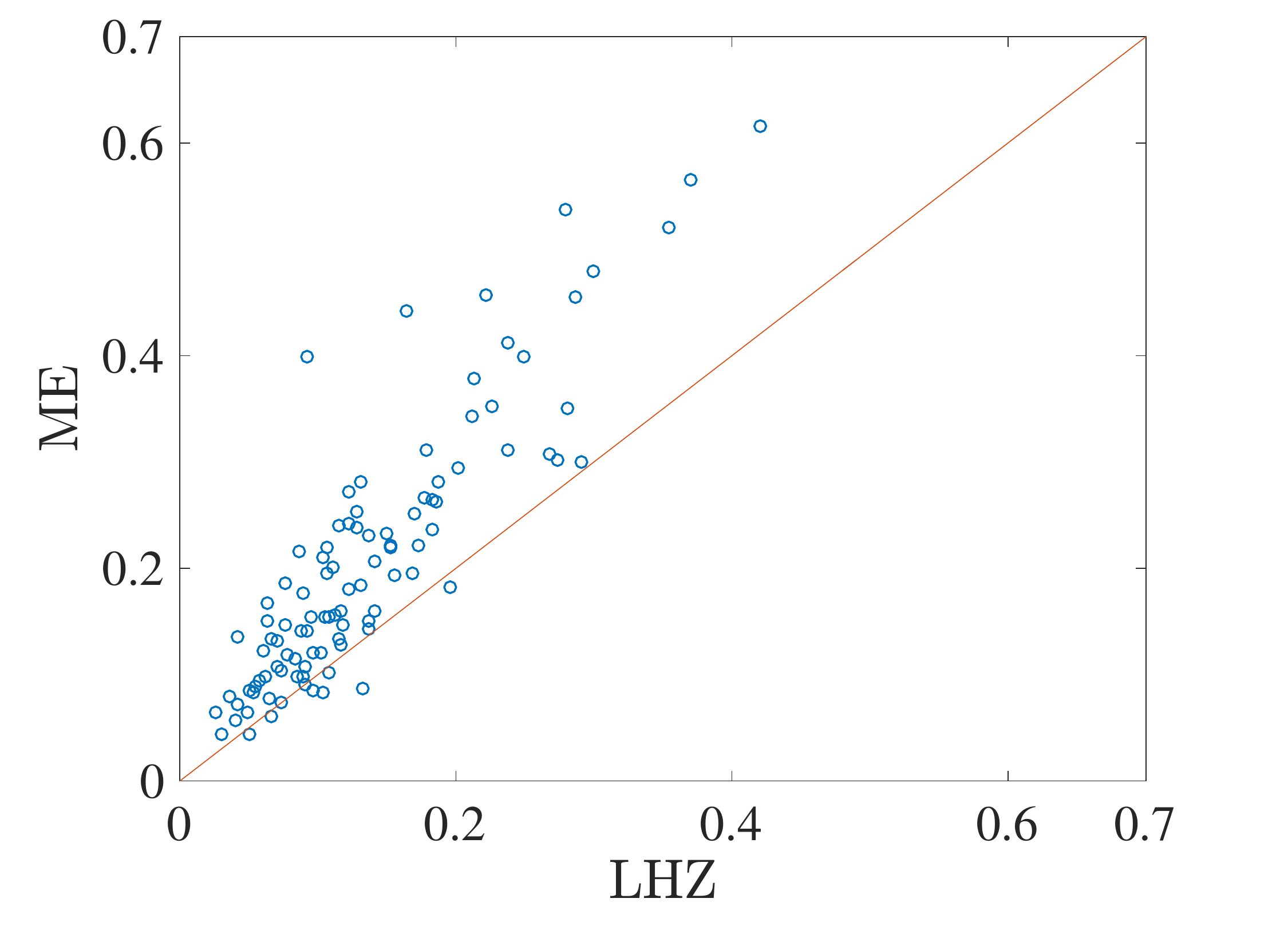} 
   \caption{\textbf{Comparing leakage resiliency of the ME and LHZ schemes.} (Color online) Shown is a scatter plot  for the $K_8$ instances. The success probability on both axes is measured relative to the PT state ($\beta = 1.05$) at a fixed intermediate value of the energy penalty ($\gamma = 1$). Decoding was performed via majority vote.}
\label{fig:PTComp}
\end{figure}
Fig.~\ref{fig:SQAComp} shows scatter plots comparing the ME and LHZ schemes on SQA-generated states. For both the ME and LHZ cases we ran SQA simulations over a wide range of energy penalty values and, for each instance, chose the value that maximizes the success probability, which we refer to as the optimal penalty value.  In both cases we used majority-vote decoding as described in Secs.~\ref{sec:DecodingME} and \ref{sec:DecodingLHZ}. In the ME case, logical values were obtained with a majority vote on $n_A=\lceil N/4 \rceil+1=3$ and $5$ physical qubits, corresponding, respectively, to the length of the chains required to minor-embed $K_8$ and $K_{16}$.
Figure~\ref{fig:SQAComp} 
clearly shows an advantage for the ME scheme over the LHZ scheme using majority vote decoding, with the advantage growing  as the problem size increases from $K_8$ to $K_{16}$.

The number of sweeps is fixed in Fig.~\ref{fig:SQAComp}; to test the dependence on this number, Fig.~\ref{fig:SQA_boosted} shows the success probability for a representative $K_8$ instance as a function of the penalty strength, when the number of sweeps is increased from $10^4$ to $10^5$. The performance of both schemes improves, but we find that the decoded ME scheme's success probability is always larger than LHZ's, except at very small penalty values; we discuss the reason for this below.
In addition, the optimal penalty strength for ME appears smaller; this is studied more systematically in Fig.~\ref{fig:example6}, which shows a roughly equal distribution for $K_8$, but it confirms that the LHZ scheme tends to require higher optimal penalties for $K_{16}$. This could be a disadvantage given practical limitations of quantum annealing devices.

As we demonstrate in the next subsection, the reason for the relatively poorer performance of the LHZ scheme is likely due to its less efficient spin updates in the SQA simulations. Two factors are likely responsible for this difference:  first, the LHZ scheme requires asymptotically twice as many physical qubits as ME [more precisely, $N(N-1)/2 > N(\lceil N/4 \rceil+1)$ for $N\ge 8$]; second, the four-body constraint terms may make updates via time-like cluster spin-flips less effective. It is possible that alternative implementations of the constraints (using for example two-body interactions along with coupling to qutrits, as suggested in Ref.~\cite{Lechner:2015}) may improve the performance of the LHZ scheme under SQA simulation; this is left for a future study.

\subsection{Comparing ME and LHZ via parallel tempering}

We next study the probability with which the logical ground state can be retrieved from the results of PT simulations that use the classical physical Hamiltonians $H^{\mathrm{LHZ}}_{\mathrm{P}}$ and $H^{\mathrm{ME}}_{\mathrm{P}}$. This analysis allows us to directly study the resiliency to leakage due to purely classical thermal errors.  

Fig.~\ref{fig:PT} presents our PT simulation results, as a function of energy penalty, for representative $K_8$ and $K_{16}$ instances, where we also examine the size and temperature-dependence. We confirm in Fig.~\ref{fig:PT} for the $K_8$ case that for sufficiently large penalties and sufficiently low temperatures, the success probability for both schemes is identical and high.
Therefore, the observed difference is attributable to more efficient spin updates in the case of SQA simulations of the ME scheme. 

At smaller energy penalties, thermal excitations are more likely to populate leakage states, leading to differences between the two schemes. This is instance and size-dependent, as can be seen by the differences between the two schemes in Figs.~\ref{fig:PT1} and \ref{fig:PT3}, but the absence of a difference in Fig.~\ref{fig:PT2}. 

The typical behavior is presented in the scatter plot of Fig.~\ref{fig:PTComp}, where we compare the success probabilities of the LHZ and ME schemes at particular values of the inverse temperature and energy penalty, for the same $100$ random $K_8$ instances as in the SQA case of Fig.~\ref{fig:SQAComp1}. At this relatively small penalty value, the ME scheme exhibits a small but consistent advantage in its ability to recover states that have leaked due to thermal excitations. 

Differences between the two schemes are also expected if the number of updates is insufficient to reach the low-energy physical states, and hence leakage states can still be populated.  In this case, even at large penalty values the physical state need not be equivalent to the logical state.  This is what happens in the LHZ case for the $K_{16}$ instance shown in Fig.~\ref{fig:PT3}, where evidently the success probability does not saturate at both the high and low temperature value. We find that increasing the number of steps in the PT simulations eliminates this feature (Fig.~\ref{fig:MoreSwaps} in Appendix~\ref{sec:Add}).  Moreover, we find that for most of the instances and penalty values, the SQA states  are closer to the PT states for ME than for LHZ (Fig.~\ref{fig:PTSQADistance} in Appendix~\ref{sec:Add}). Thus, we see once again that the LHZ scheme has greater difficulty with single spin updates than ME, in agreement with our earlier SQA results.  

\begin{figure}[t] 
   \centering
   \includegraphics[width=1\columnwidth]{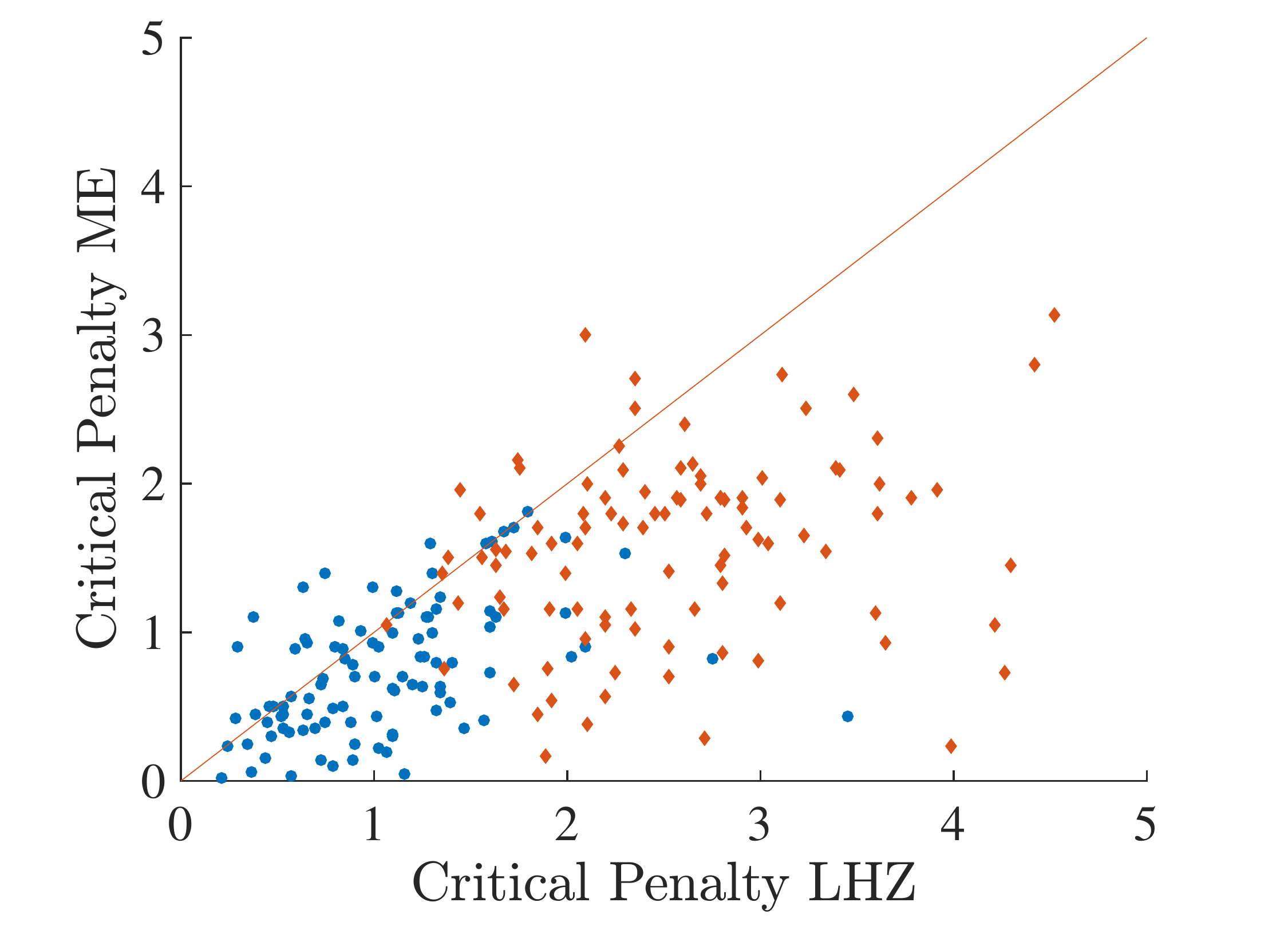}
   \caption{\textbf{Critical penalty strength comparison.}  (Color online) Shown is the penalty strength where the PT success probability after MVD is $0.5$ for $\beta = 20$. There is a systematic increase in the critical penalty strength for both schemes as the problem size increases from $K_8$ (blue dots) to $K_{16}$ (red diamonds). Additionally, the ME critical penalty strength is systematically lower for the $K_{16}$ instances.}
   \label{fig:criticalPenalty}
\end{figure}

The different response to small and large energy penalties explains the striking step-like behavior of the success probability observed in Fig.~\ref{fig:PT} for the low temperature simulations, where PT overwhelmingly samples the physical ground state. This step-like behavior appears at the smallest value of the energy penalty such that the physical ground state is correctly mapped to the logical ground state. A scatter plot of the corresponding critical energy penalty values is shown in Fig.~\ref{fig:criticalPenalty}, where it is seen to be larger for the LHZ scheme,  in agreement with the SQA optimal penalty results seen in Fig.~\ref{fig:example6}.
However, a lower bound argument shows that the value of the optimal penalty should scale at least with $N$ in both schemes (see Appendix~\ref{sec:SEP}). 

%
\section{Minimum Weight, Maximum Likelihood, and Belief Propagation Decoding of the LHZ Scheme}
\label{sec:OP}
In an attempt to improve the performance of the LHZ scheme, in this section we present and study a decoding strategy that extends the approaches of Refs.~\cite{Lechner:2015,Pastawski:2015}, and is designed to address independent or weakly correlated physical spin-flip errors. As we have already stressed, this is not an accurate model of noise in quantum annealing. However, adopting this perspective allows us to develop well-defined decoding strategies, which we in turn test on relevant distributions generated via SQA and PT. We shall find that this leads to substantial improvements.

\subsection{Minimum Weight Decoding}
\label{sec:MWD}
Let us denote the final states of the $K$ physical qubits  at the end of a quantum annealing run by $\Psi_{t_f} = \{s_k\}_{k=1}^K$ (with $s_k = \pm 1$). This readout amounts to a syndrome measurement $\mathcal S$ defined by the list of values of the four-local constraints: $\mathcal S = \{ \zeta_c\}_{c=1}^C = \{ - \lambda s^z_{c_u}s^z_{c_d}s^z_{c_l}s^z_{c_r}\}_{c=1}^C$, with $\zeta_c = \pm 1$.  
Given a physical state with violated constraints, the goal of minimum-weight decoding (MWD) is to find  the nearest Hamming distance constraint-satisfying physical state.  

Consider the ground state $\Phi =  \{e_k\}_{k=1}^K$ ($e_k = \pm 1$) of the following Hamiltonian, defined on the same LHZ graph as the original optimization problem: 
 \beq 
\label{eq:LHZ_dec}
H_{\mathrm{MWD}} = - \sum_{k \in \mc{V}_{\mathrm{LHZ}}} \sigma^z_{k} -  \sum_{c \in \mc{V}_{\mathcal C}} \zeta_c C_c\ ,
\eeq
where it is assumed that the four-body interactions $\mathcal C_c = - \lambda \sigma^z_{c_u}\sigma^z_{c_d}\sigma^z_{c_l}\sigma^z_{c_r}$  are sufficiently strong such that $\Phi $ minimizes (satisfies) all constraint terms. 
The MWD decoded state is then given by:
\beq
\Psi_{\mathrm{MWD}} = \{ e_k s_k\}_{k=1}^K\, ,
\eeq
for the following reason:
in the absence of any constraint-violating terms $\Phi  =  \{+1\}_{k=1}^K$, i.e., the original physical state need not be changed.  When the physical state violates a constraint, a sequence of spin-flips must be performed to correct this violation.  The second term in Eq.~\eqref{eq:LHZ_dec} forces the state $\Phi$ to undo the measured syndrome, and the first term minimizes the number of spin-flips (i.e., $e_k=-1$) in $G$.  

\begin{figure*}[t]
\subfigure[]{\includegraphics[width=0.72\columnwidth]{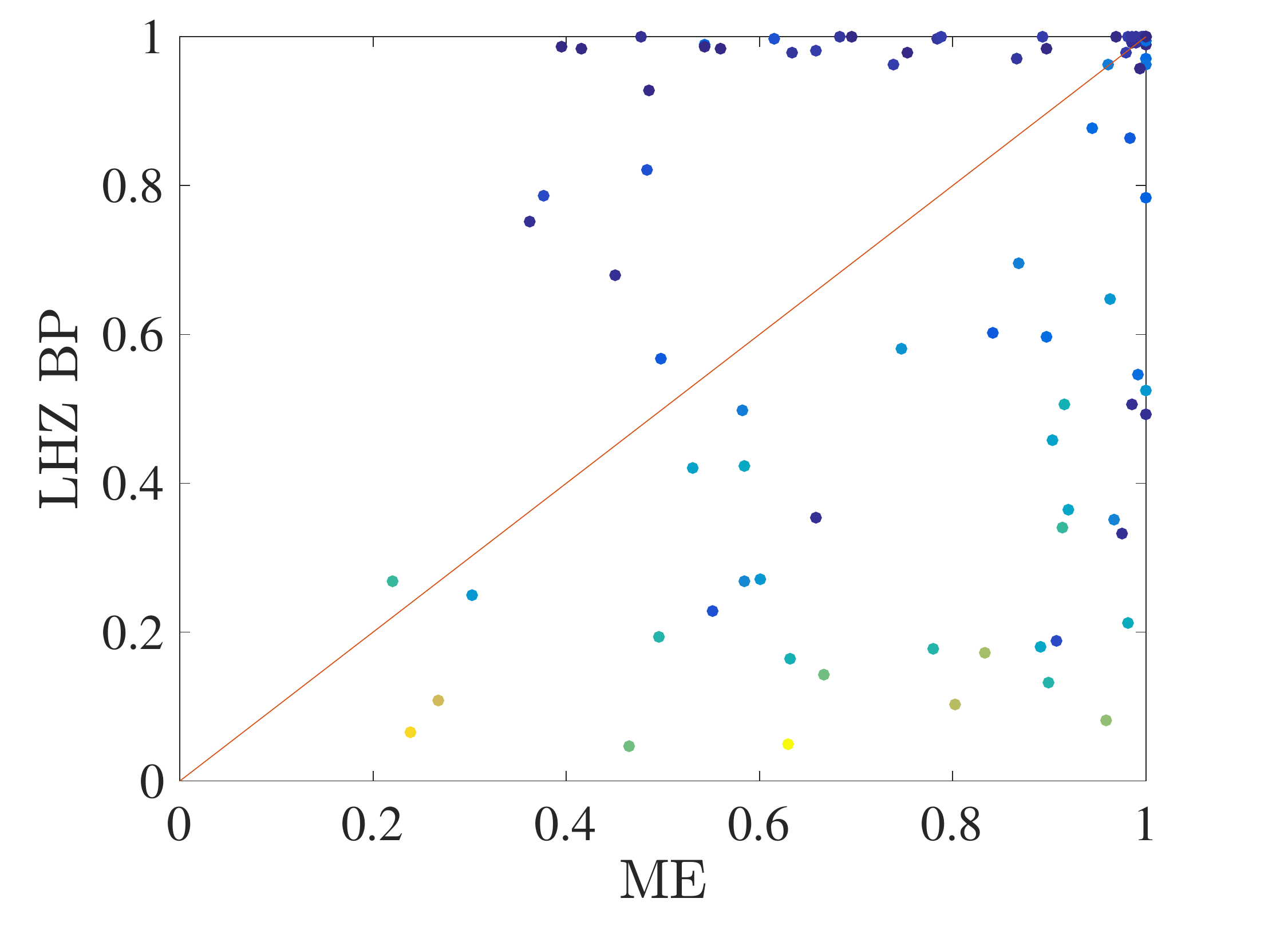}\label{fig:comparison13} }\hspace{-0.3in}
\subfigure[]{\includegraphics[width=0.72\columnwidth]{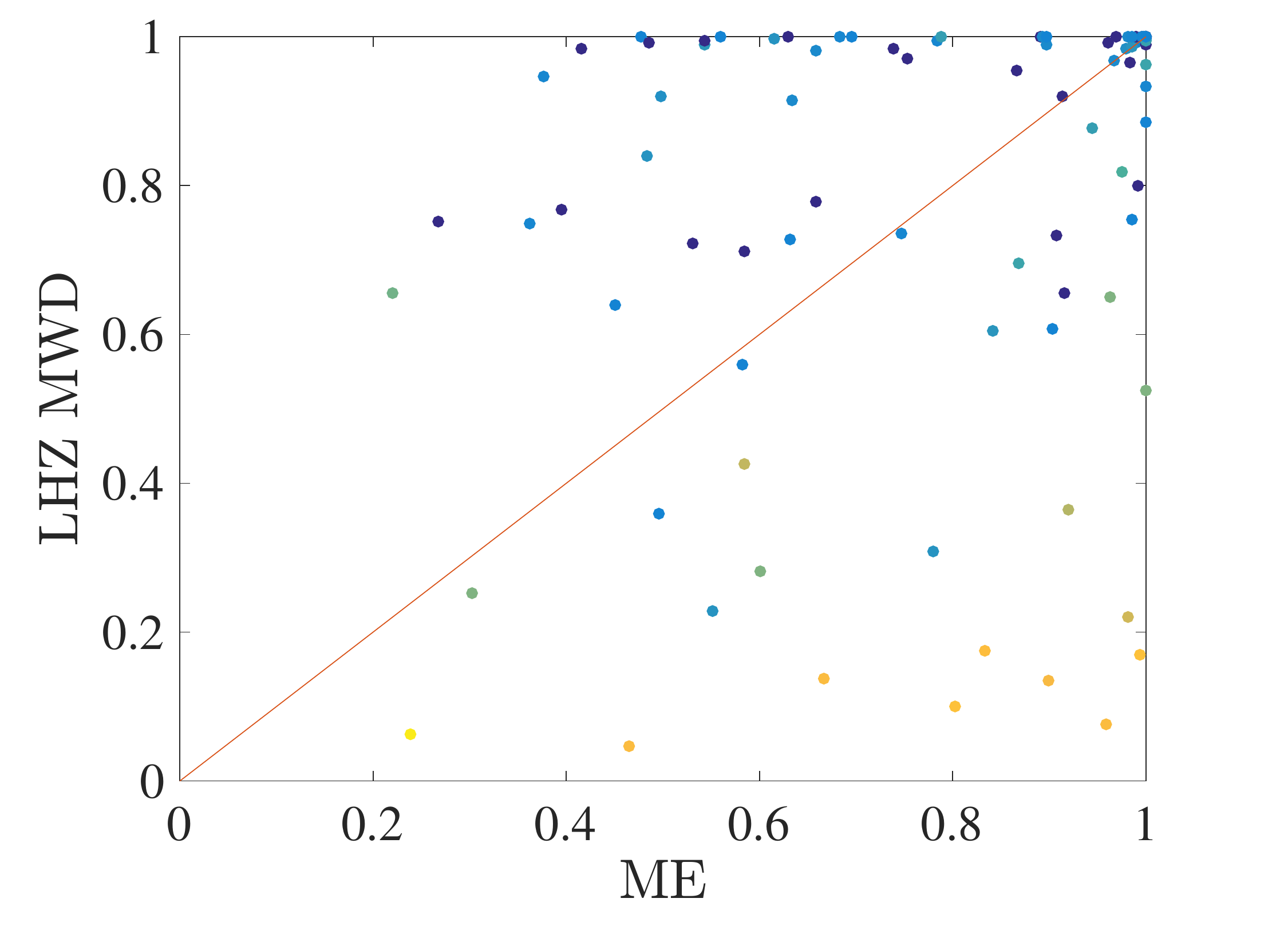}\label{fig:comparison11} }\hspace{-0.3in}
\subfigure[]{\includegraphics[width=0.72\columnwidth]{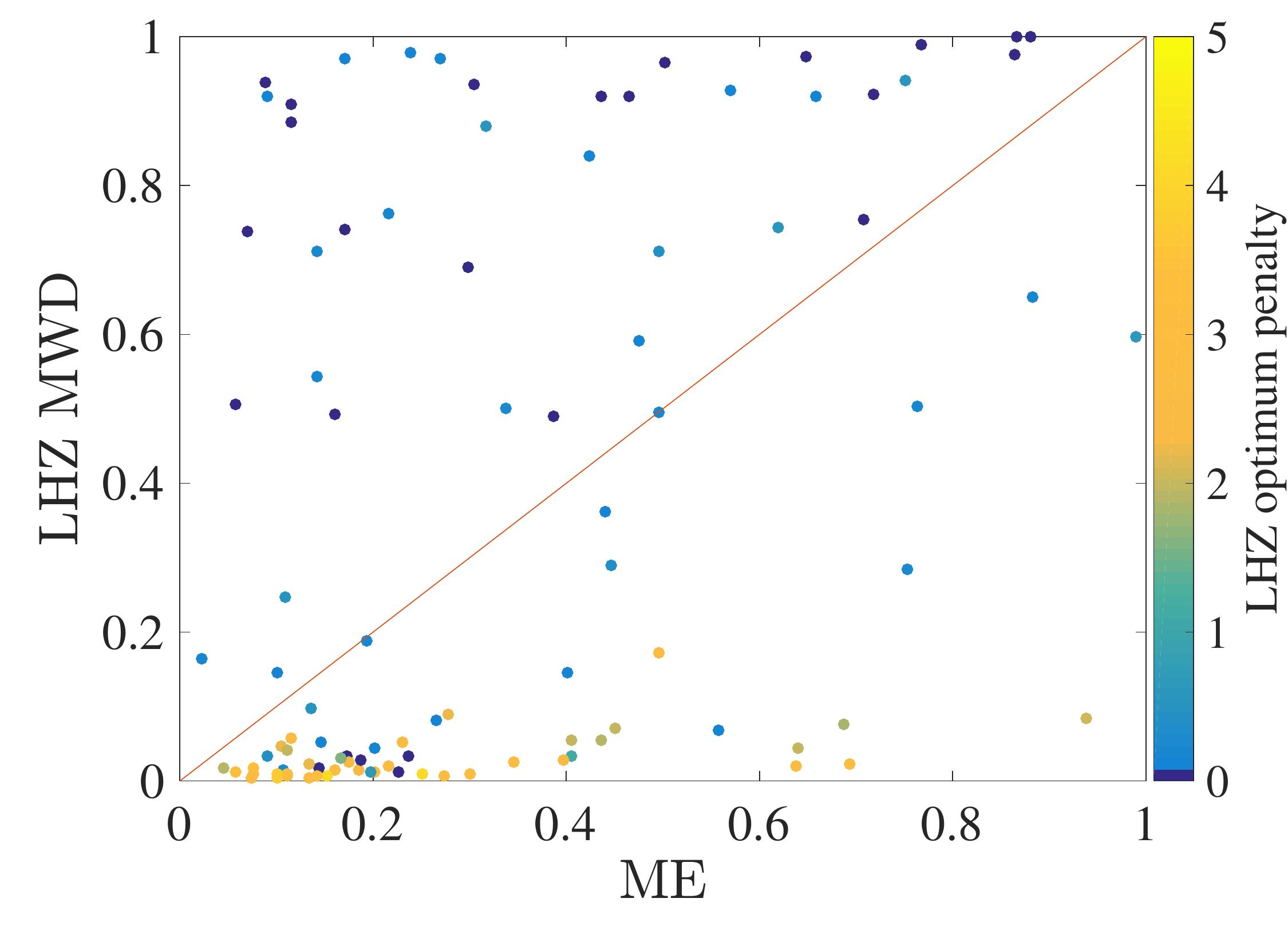}\label{fig:comparison12}}
\caption{\textbf{Performance of different decoding strategies for the LHZ scheme.} (Color online) Shown are scatter plots comparing success probabilities obtained at the optimal penalty strength for (a) the LHZ scheme with BP decoding (using an error rate of 0.2 and 10 iterations) and the ME scheme on the $K_8$ set; (b) the LHZ scheme with MWD decoding and the ME scheme on the $K_8$ set; (c) same as (b) but for the $K_{16}$ set. Data was obtained with SQA simulations ($10^4$ sweeps, $\beta = 1$). Color denotes the energy penalty values for the LHZ scheme. Note that LHZ outperforms ME only for optimal penalty strength $\ll 1$ (blue dots).}
 \label{fig:comparison1}
\end{figure*}

To gain additional insight into the MWD problem, and to connect it to maximum likelihood decoding of Sourlas codes \cite{Sourlas:1989,Sourlas:1994}, we first note that the MWD Hamiltonian~\eqref{eq:LHZ_dec} can be brought back to the LHZ form~\eqref{eq:LHZ}. Let us assume that we can find a set of spin-flips $G =  \{g_k\}_{k=1}^K$ with $g_k = \pm 1$ ($-1$ means  a spin-flip) that removes all unsatisfied constraints, i.e., that replaces every $\zeta_c$ by $+1$. It is easy to find at least one such $G$ state, by performing a sequence of spin-flips that changes the sign of each constraint one by one [this is not necessarily the MWD state since it need not minimize the number of spin-flips; an example is shown in Appendix~\ref{sec:SEP}, Fig.~\ref{fig:flip2}]. Alternatively, one can think of $G$ as a gauge transformation $\sigma^z_{k}\mapsto g_k \sigma^z_{k}$ such that $\zeta_c \mapsto 1$, which allow Eq.~\eqref{eq:LHZ_dec} to be rewritten as
\beq 
\label{eq:LHZ_dec_gauged}
H^g_{\mathrm{MWD}} = - \sum_{k \in \mc{V}_{\mathrm{LHZ}}} g_k \sigma^z_{k}+ \sum_{c \in \mc{V}_{\mathcal C}}{\mathcal{C}}_c\ .
\eeq
Equation~\eqref{eq:LHZ_dec_gauged} is now in the form of Eq.~\eqref{eq:LHZ}, and thus can be mapped back to an optimization over a complete (logical) graph $K_N$ with couplings $g_{ij} = g_k$:
\beq 
\label{eq:K_NDec}
H^{\mathrm{spin}}_{\mathrm{MWD}} =- \sum_{i<j } g_{ij}\sigma^z_i\sigma^z_j\ .
\eeq
We can extract a vector $E^g =  \{e^g_k\}_{k=1}^K$ from the ground state $\{s_i^g\}_{i=1}^N$ of this optimization problem, where  $e^g_k = g_{ij} s^g_i s^g_j $ [i.e., $e^g_k$ is positive or negative if the  coupling corresponding to the pair $(i,j)$ is  satisfies or unsatisfied, respectively]. $E^g$  is the minimum-weight corrected state. Taking into account the gauge transformation we have:
\beq \label{eqt:stateG}
\Psi_{\mathrm{MWD}} = \{ e^g_k g_k s_k\}_{k=1}^K\,.
\eeq

Note that a syndrome measurement does not uniquely identify an error.  
In fact, there are $2^{{N}}$ possible $G$ states, related by gauge transformations on the logical graph. Two different choices of $G$ will thus define two optimization problems over a $K_N$  that only differ by a gauge transformation, i.e., the two corresponding decoding processes are equivalent.

\subsection{Maximum Likelihood Decoding}
\label{sec:MLD}
In Eq.~\eqref{eq:K_NDec} we have formulated the decoding process in terms of a spin system with ferromagnetic or antiferromagnetic unit interactions $g_{ij} = \pm 1$. In this formulation, the spin flip error probability $P_\epsilon$ corresponds to the probability of an antiferromagnetic interaction $g_{ij} = -1$. Errors thus corrupt the sign of the couplings of the spin system Eq.~\eqref{eq:K_NDec}.
This formulation of the decoding of the LHZ scheme corresponds precisely to the decoding of a Sourlas code for error correction of classical signals transmitted over a noisy channel, wherein the transmitted physical bits correspond to the couplings, while the logical bits are encoded in the ground state of an Ising model \cite{Sourlas:1989,Sourlas:1994}. 
As was shown in Refs.~\cite{Rujan,NishTemp}, optimal, maximum-likelihood decoding (MLD) decoding of  a Sourlas code corresponds to performing a thermal average at the Nishimori inverse temperature $\beta_N = \frac{1}{2} \ln [(1-P_{\epsilon})/P_{\epsilon}]$:
\beq
(e^g_k)_{\mathrm{MLD}} = g^g_{ij} {\rm sign}(\langle s^g_i \rangle_{\beta_{N}}\langle s^g_j \rangle_{\beta_{N}})\,.
\label{eq:MLD}
\eeq 
An intuitive explanation for the need to perform a finite temperature decoding for optimal performance is that for sufficiently large error probabilities $P_\epsilon$, the Ising Hamiltonian~\eqref{eq:K_NDec} is corrupted to such a degree that the correct state is most likely encoded in one of the excited states rather than in the ground state. These excited states are optimally explored by thermal sampling, if performed at the Nishimori temperature.


\subsection{Hardness of MWD and MLD}

\begin{figure*}[t]
\begin{center}
\subfigure[]{\includegraphics[width=0.72\columnwidth]{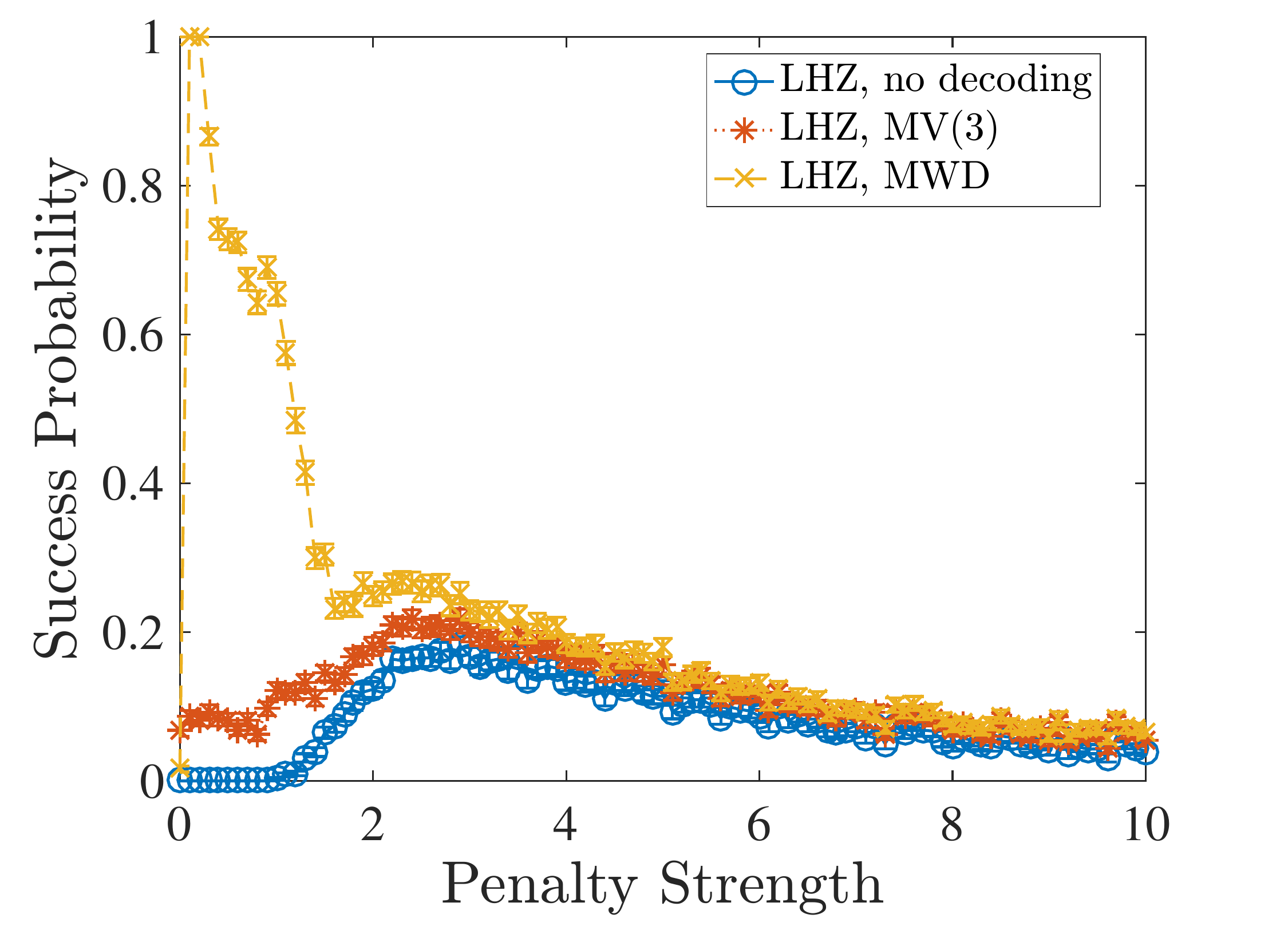} \label{fig:LechnerInstance11}} \hspace{-0.3in}
\subfigure[]{\includegraphics[width=0.72\columnwidth]{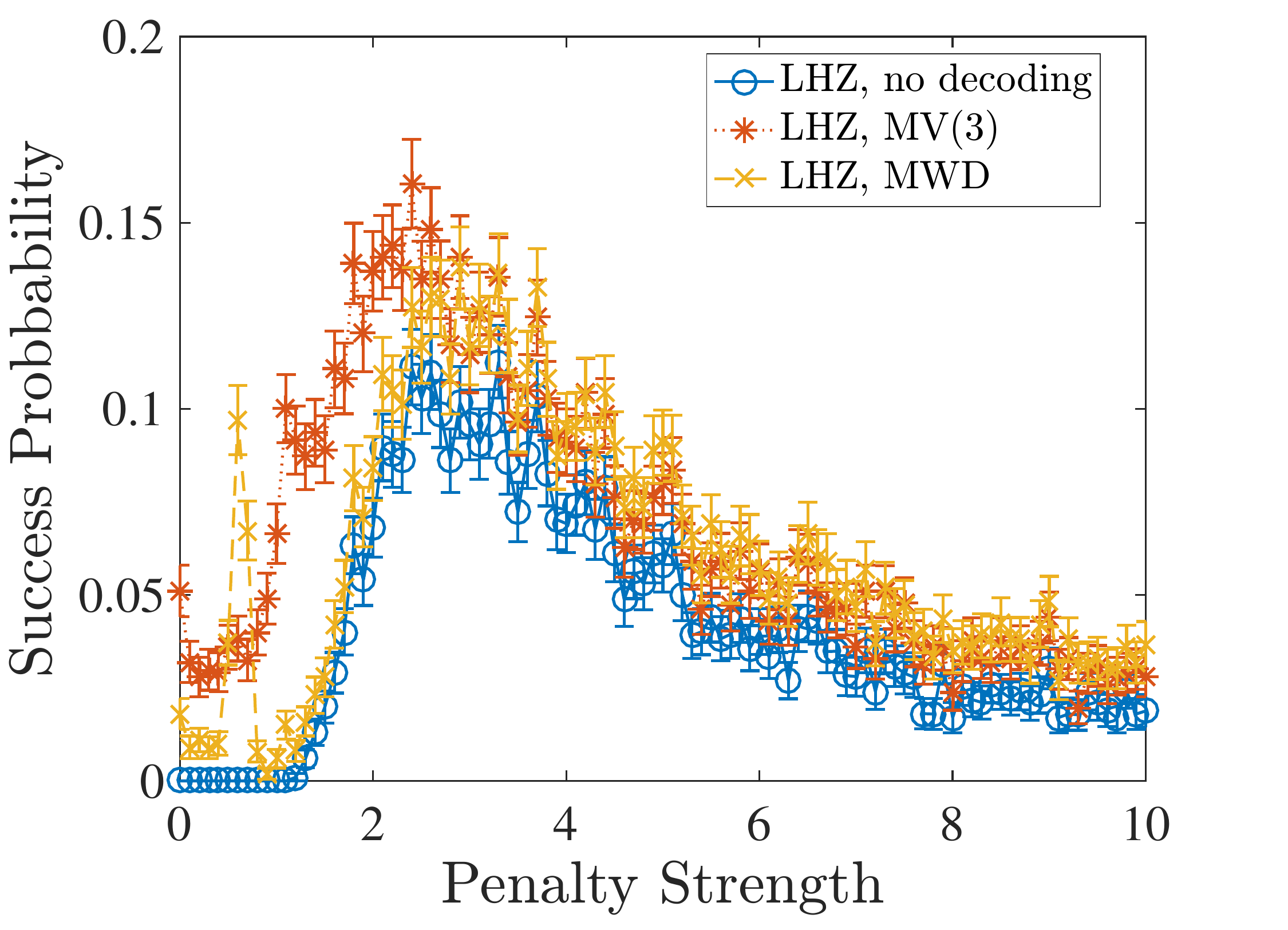}\label{fig:LechnerInstance12} }\hspace{-0.3in}
\subfigure[]{\includegraphics[width=0.72\columnwidth]{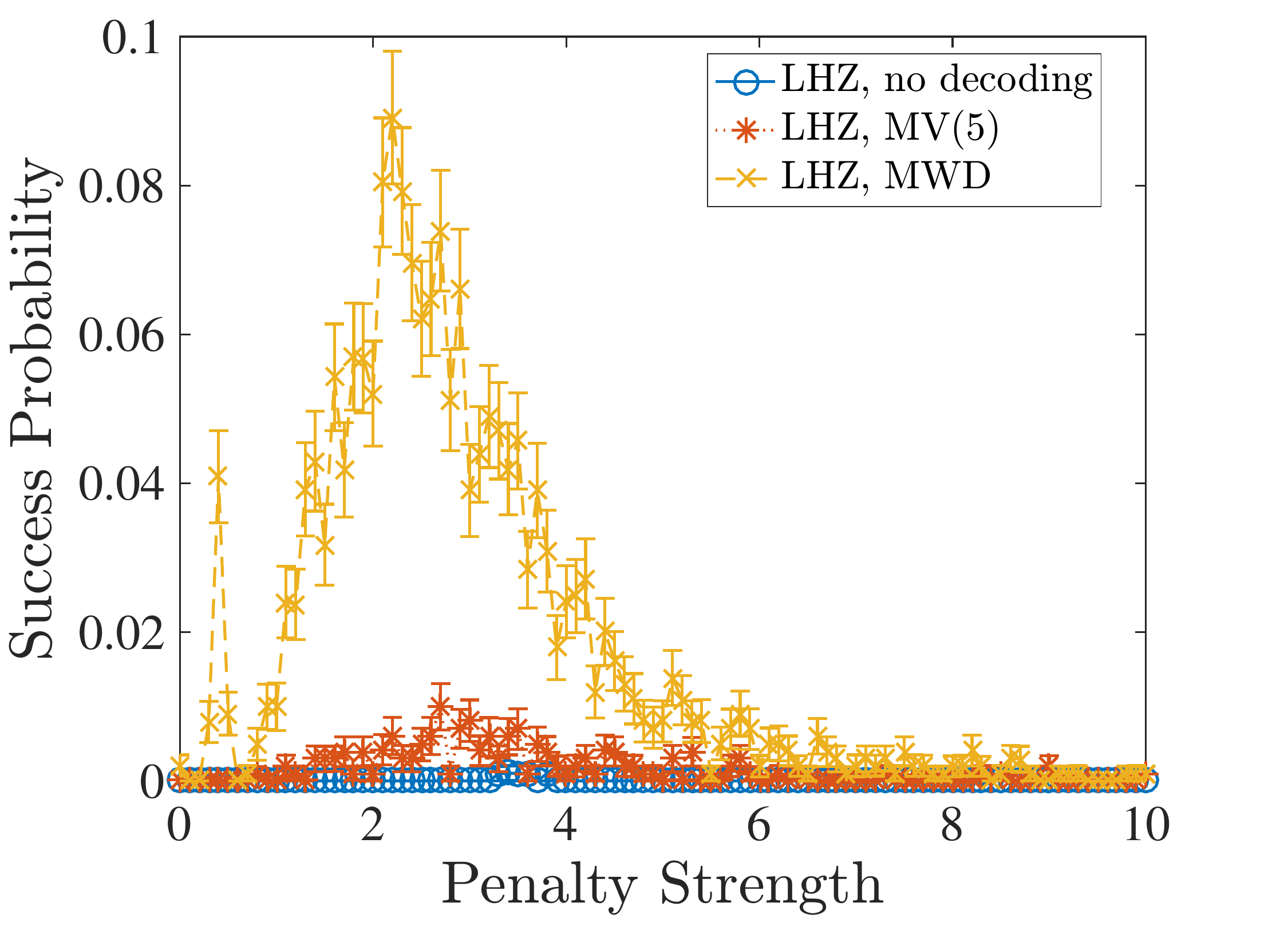} \label{fig:LechnerInstance13}}
\caption{\textbf{Comparison of MWD and MVD for the LHZ scheme as a function of energy penalty.} (Color online) Shown are results for two representative $K_8$ instances (instance 3 and 5) and a representative $K_{16}$ instance (instance 5). (a) MWD wins by peaking at a very small energy penalty value. (b) A rare example where MVD performs better than MWD on a $K_8$ instance. (c) A typical $K_{16}$ instance where MWD beats MVD by a large margin.} 
\label{fig:LechnerInstance1}
\end{center}
\end{figure*}

MWD requires the minimization of the Hamiltonian defined in Eq.~\eqref{eq:K_NDec}. This is equivalent to solving an instance of MAX-2-SAT, which is NP-hard; MLD requires thermal sampling, which is $\#$P-hard and thus even harder \cite{Papadimitriou:book,Valiant:1979,Dahllof:2005}.    Nevertheless, we expect the typical instance to be easy in the small $P_\epsilon$ regime. This observation can be made more precise thanks to the connection between Sourlas decoding and statistical mechanics.  Specifically, the spin system defined in Eq.~\eqref{eq:K_NDec} will be in a ferromagnetic phase for sufficiently small error probabilities $P_\epsilon$, where decoding is expected to be easy. On the other hand, a disordered phase will arise for sufficiently large $P_\epsilon$. A phase transition between the decodable (ordered) and undecodable (disordered) regimes is thus expected at a threshold error probability {$P_{\mathrm{th}}$}.  
A mean field analysis of the problem in Eq.~\eqref{eq:K_NDec}, valid in the large-${N}$ limit, reveals the presence of a phase transition between ferromagnetic and spin-glass phases at $P_{\mathrm{th}} = 1/2$~\cite{Sherrington75,Kirkpatrick:1978dn}. Moreover, in the ferromagnetic phase the thermal average in Eq.~\eqref{eq:MLD} recovers the uncorrupted state perfectly at any finite temperature. In this sense, in the large-${N}$ limit MWD is equivalent to MLD.\footnote{At finite ${N}$, MWD decoding is equivalent to MLD decoding  in the $P_\epsilon \to 0$ ($\beta_{N}\to \infty$) limit only.} On the other hand, for $P_\epsilon \ge P_{\mathrm{th}} = 1/2$ we are in the undecodable regime, in which $(e^g_k)_{\mathrm{MLD}}$ does not recover the uncorrupted state. Note that the perfect decoding achievable with MWD/MLD for $P_\epsilon \le1/2$ is a consequence of the fault tolerance of the LHZ scheme when the only source of errors is random spin-flips \cite{Lechner:2015,Pastawski:2015}. 

\subsection{Weight-$3$ Parity Check with Belief Propagation}
\label{sec:BP}
Given a classical error correcting code, instead of performing MLD, which is hard, one can use belief propagation (BP) as a suboptimal, but more manageable decoding technique, where decoded values are iteratively updated based on the expected error rate. It turns out that BP works very well with low-density parity check (LDPC) codes.   

The relative alignment of a pair of logical spins $\bar{q}_i$ and $\bar{q}_j$ can be recovered in multiple ways.  One way is to use the value of the physical spin $q_{i,j}$, but using weight-$3$ parity checks there are $N-2$ additional independent ways, specifically, $q_{i,k} q_{k,j}$, with $k \neq i,j$.  This particular approach was proposed by Pastawski and Preskill (PP) \cite{Pastawski:2015} and can be viewed as a classical LDPC code, which can be decoded with belief propagation. We follow the same implementation of BP as used in Ref.~\cite{Pastawski:2015}.  For the case of weakly correlated errors, the probability of a decoding error is exponentially small in $N$ \cite{Pastawski:2015}. PP noted that including higher-weight parity checks can yield further improvements. Indeed, to approach the performance of the MLD described in section~\ref{sec:MLD}, one should include all parity checks and use finite temperature decoding. We do not consider these extensions here; they are an interesting possibility for future work.

\subsection{Decoding Results}

To understand whether the advantage that ME has over LHZ originates in the decoding step, we studied the efficacy of several other decoding strategies for the LHZ scheme. We included MWD, BP of Ref.~\cite{Pastawski:2015}, MVD over a large number of trees ($100$). We did not include MLD since it gives a well-defined decoding procedure only in the case of random spin-flips, where the error probability $P_{\epsilon}$ and the corresponding Nishimori temperature are known. 

Different decoding strategies using majority vote give very similar results, as shown in more detail in Appendix~\ref{sec:dec}.  This suggests that majority voting schemes that attempt to exploit a large number of decoding trees are not beneficial, as a consequence of the fact that the noise generated via SQA simulations does not correspond to uncorrelated random bit flips. (We explicitly show how the SQA results differ from uncorrelated noise in Appendix~\ref{sec:corr}.)  In the latter case we would expect the decoding errors to decay exponentially with the number of trees (below a decoding threshold as discussed in the previous subsection). 

We focus our attention here on MWD, which gives a substantial improvement over MVD and slightly improved performance over BP as seen in Figs.~\ref{fig:comparison13} and \ref{fig:comparison11}. Figures~\ref{fig:comparison11} and \ref{fig:comparison12} show that after MWD the LHZ scheme becomes competitive with ME on the $K_8$ instances, although its  performance remains worse on the $K_{16}$ instances. We can thus not conclude that MWD (which requires substantial effort beyond MVD) is sufficient to enable the LHZ scheme to outperform ME.

A closer examination shows that when the LHZ scheme with MWD is superior to ME, it typically does so by having a peak in the success probability at almost vanishing penalty strengths [represented by the color scale of Figs.~\ref{fig:comparison11} and \ref{fig:comparison12}]. Fig.~\ref{fig:LechnerInstance11} shows the success probability as a function of the penalty strength for an instance with the aforementioned behavior.  In this situation, success arises entirely from the decoding process. This is related to the fact that MWD at zero penalty strength corresponds to finding the ground state of the following approximation of the original logical problem: $J_{ij} \rightarrow {\rm sign }(J_{ij})$. To see this, note that when the energy penalties vanish, the physical ground state is trivially given by the physical configuration with all spins aligned to their local fields. This configuration is typically a leaked state. Following the steps described in Sec.~\ref{sec:MWD} it is easy to check that $g_{ij} = - {\rm sign }(J_{ij})$ is a possible choice for a gauge transformation which allows one to write the MWD problem Eq.~\eqref{eq:K_NDec}, i.e., the approximation of the logical problem mentioned above. It turns out that for our set of $K_8$ instances, the ground state of this approximate problem typically corresponds to the ground state of the original problem, resulting in the observed peak in success probability. Even when this does not happen, however [as in the examples of Figs.~\ref{fig:LechnerInstance12} and \ref{fig:LechnerInstance13}], MWD gives comparable or better results than MVD.

\section{Conclusions}
\label{sec:Conc}

The path toward the achievement of ``quantum supremacy" \cite{Preskill:2012dp} using quantum annealing is fraught with many challenges, among which is the embedding problem we have focused on here. Lechner, Hauke and Zoller have proposed an embedding scheme which they claimed exhibits intrinsic fault tolerance \cite{Lechner:2015}. 
Unfortunately, our findings do not support the fault tolerance claim in the context of a realistic error model for quantum annealing: we find that the performance of the LHZ scheme suffers under the errors generated using simulated quantum annealing, as well as (to a smaller degree) under classical thermalization. This is perhaps unsurprising, since the fault tolerance claim was based on a model of weakly correlated spin-flip errors, which does not accurately describe the errors that arise due to dynamical and thermal excitations during the course of a quantum annealing evolution (see Appendix~\ref{sec:corr}). 

Since the earlier minor embedding scheme due to Choi \cite{Choi1,Choi2} has already found widespread use, it is important to compare the performance of the two schemes. A particularly attractive feature of the LHZ scheme is that it separates the problem of controlling local fields and couplings.
In principle this appears to allow for a drastic design simplification. Still, it remains to be seen how flexible the LHZ scheme is to the occurrence of faulty qubits or couplers, a problem that has prompted the development of improvements to Choi's original ME scheme, showing that ME is easily adaptable to architectures with faults~\cite{klymko_adiabatic_2012,Boothby2015a}. Another attractive feature of the LHZ scheme is that it is amenable to a host of different decoding techniques, beyond the majority vote decoding strategy that is typically used for ME. Indeed, we have derived optimal minimum weight and maximum likelihood decoding strategies for the random bit-flip error model, and demonstrated that the former boosts the performance of the LHZ scheme for SQA. It is possible that the LHZ scheme may rival or surpass the ME scheme with even better decoding strategies, tailored to correlated spin-flip errors.  

However, in several other important respects the ME scheme appears to be more attractive. First, the LHZ scheme requires a factor of two more physical qubits than ME for a given number of logical qubits. Second, the LHZ scheme requires four-body interactions (or two-body interactions with qutrits; it is an open question whether this would change our results), which can be harder to implement, while the ME scheme only requires standard, realizable two-body interactions between qubits. Third, we found that the optimal penalty strength tends to be lower for the ME scheme, which can be important given practical constraints on the highest energy scales achievable, especially in superconducting flux qubit based quantum annealing devices. Given that the penalties are implemented differently in the two schemes, it is possible that a suitable physical implementation of the LHZ scheme achieves a higher maximum penalty strength, making this point minor.  However, both schemes suffer from requiring energy penalties to grow with problem size, an issue that needs to be addressed in both schemes for scalability (see Appendix~\ref{sec:SEP}). 
Finally, and perhaps most importantly, we found that the ME scheme seems to outperforms the LHZ scheme under a broad set of conditions. Namely, subject to identical simulation parameters and decoding effort, the success probabilities of the ME scheme are nearly always higher than those of the LHZ scheme for randomly generated Ising instances over complete graphs, up to the largest sizes we were able to test. We have explained this finding in terms of better spin- update properties of the ME scheme under SQA simulations. 

We note several caveats regarding our results. First, the LHZ scheme's spin-update bottleneck under SQA simulations does not necessarily correspond to a performance bottleneck for an actual quantum annealing device: we used a discrete-time quantum Monte Carlo version of SQA with only time-like cluster updates, and such a model is ultimately not a complete model of a true quantum annealer. Only experiments or more detailed open-system quantum simulations can definitely address whether one scheme has a true implementation advantage over the other. However, the past success of SQA simulations in reproducing experimental quantum annealing data does suggest that our results have predictive power. 

Second, our study is not comprehensive, and our conclusions are obviously limited to the set of instances we have considered.  Thus, while we have observed a drop in performance for the LHZ scheme relative to the ME scheme when going from $K_8$ to $K_{16}$, our study does not suffice to ascertain whether the observed relative performance between the two embedding schemes persists for larger problem sizes.  

Third, we focused on simulations with identical parameters and annealing schedule, with the choices based on present values for  quantum annealing devices \cite{q-sig2}.  We expect that using lower temperatures and longer anneals will improve the performance of both schemes, but whether it changes the relative performance is unclear.  Furthermore, we did not explore the possibility of separately optimizing all the parameters of both schemes to improve their respective performance.  For example, one possibility is to optimize the annealing schedule for each scheme separately, as well as using a third independent annealing schedule for the constraint strength.

Finally, in our tests, we did not include the effects of errors in controlling the local fields and couplings.  Given that the LHZ only needs to control the local fields precisely to implement the logical problem, while the ME scheme requires control of both local fields and couplings, it is conceivable that under inclusion of noise on the local fields and couplings, the LHZ scheme may be robust to such hardware implementation errors.  This in itself would be very advantageous since such errors can dominate the performance of quantum annealing devices \cite{Martin-Mayor:2015dq,Zhu:2015pd}.

In conclusion, there is no question that the field of quantum annealing will benefit from continued research into improved embedding methods. 
An important consideration that should guide such efforts is the integration of quantum error correction to achieve scalability. Our work highlights the importance of testing new embedding methods using errors models that directly match quantum annealing.

\vspace{1cm}
\section{Acknowledgements}

The authors would like to thank Itay Hen, Philipp Hauke, Wolfgang Lechner, Hidetoshi Nishimori, Fernando Pastawski, John Preskill, Federico Spedalieri, and Peter Zoller for useful discussions and comments on the manuscript.  Part of the computing resources were provided by the USC Center for High Performance Computing and Communications. This work was supported under ARO grant number W911NF- 12-1-0523, ARO MURI Grant Nos. W911NF-11-1-0268 and W911NF-15-1-0582, and NSF grant number INSPIRE- 1551064.

\appendix

\begin{figure}[b]
\begin{center}
\subfigure[\, ]{\includegraphics[width=0.49\columnwidth]{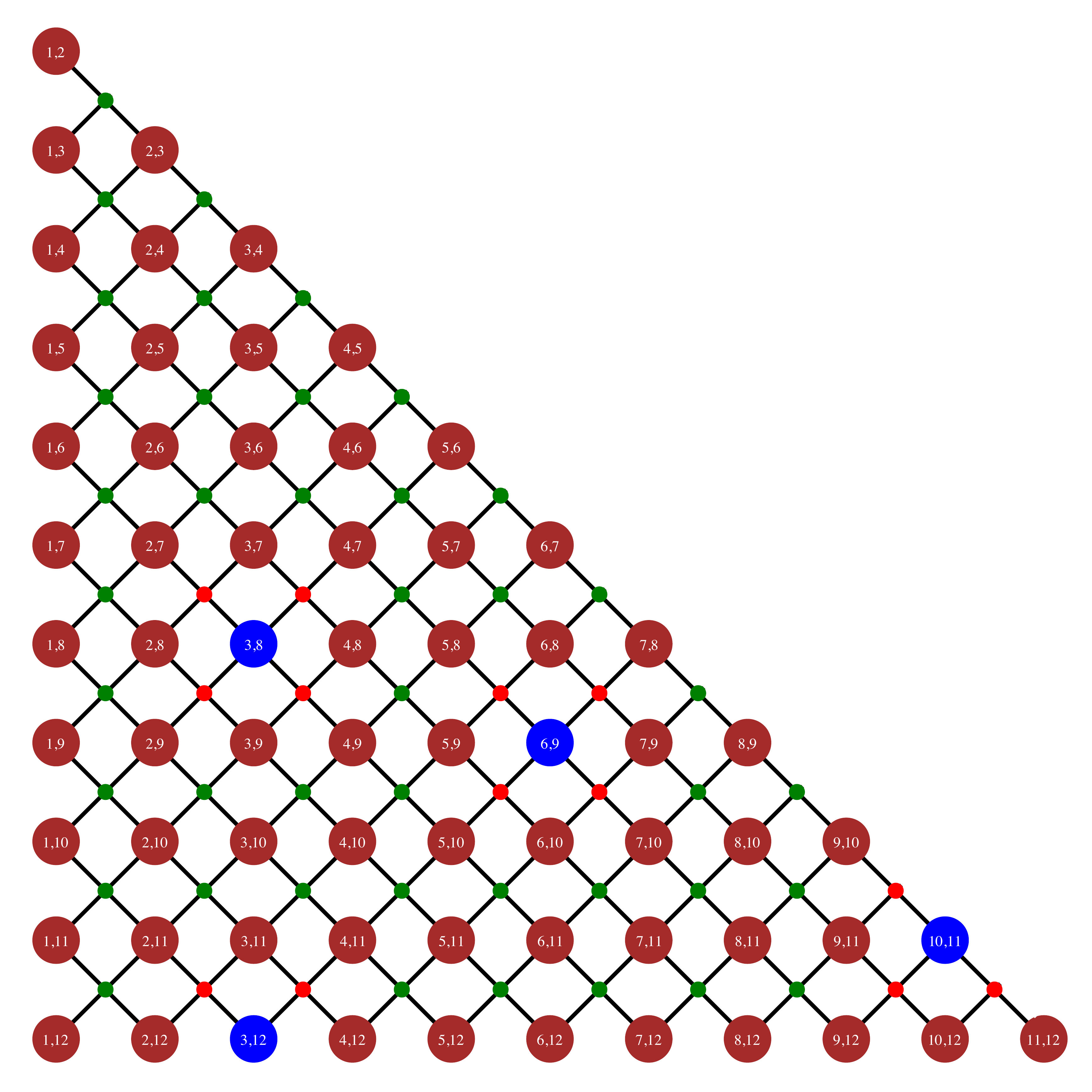}\label{fig:errors1}}
\subfigure[\, ]{\includegraphics[width=0.49\columnwidth]{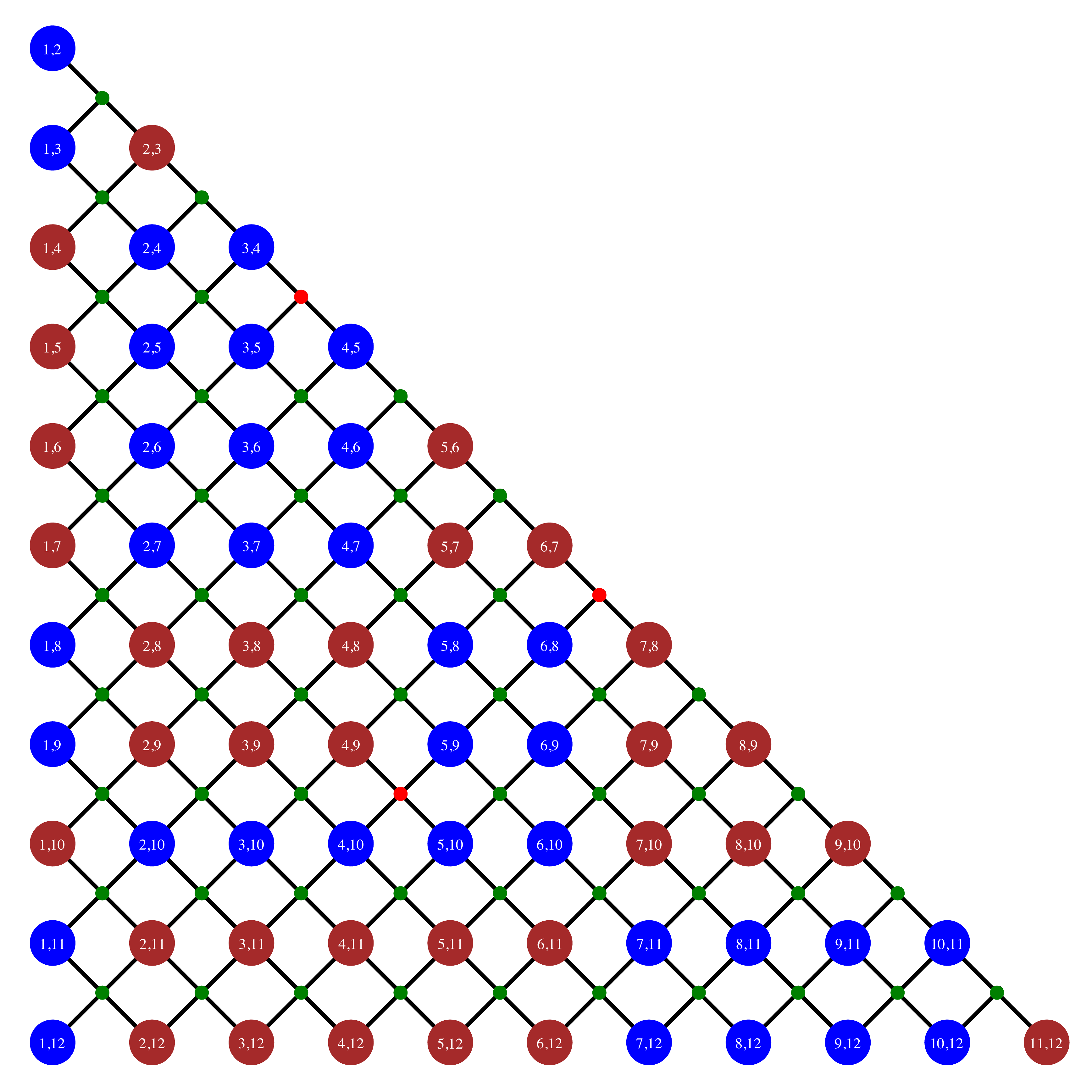}\label{fig:errors2}}
\caption{\textbf{Examples of leakage errors in the LHZ scheme.} (Color online) Blue circles represent logical qubit errors, i.e., the corresponding physical qubits have been flipped from the physical state representing the logical ground state (all red circles). Because of the presence of unsatisfied constraints (red dots) the two configurations shown are leakage states, i.e., do not represent a logical state and need to be decoded. Panel (a): a small number of spin-flips, as expected for a model of weakly correlated, random spin-flips. Panel (b): a small number of constraints is unsatisfied, but a large number of spin flips have occurred, as expected for a realistic model of open-system quantum annealing.} 
\label{fig:errors}
\end{center}
\end{figure}

\section{Mapping Physical States to Logical States for LHZ} 
\label{App:LHZMapping}
Examples of leakage errors are given in Fig.~\ref{fig:errors}. The absence of unsatisfied penalties (all four-local constraints are satisfied) ensures that the physical configuration describes a legitimate logical configuration. As described in Ref.~\cite{Lechner:2015}, a logical qubit configuration can be reconstructed from the values of an appropriately chosen group of independent physical qubits. One possibility is to choose a group of physical qubits that corresponds to a spanning tree of the logical graph. For example, the group of physical qubits corresponding to the upper diagonal of Fig.~\ref{fig:errors1} [or Fig.~\ref{fig:errors2}] defines a spanning chain $ q_{1,2}, q_{2,3},\dots, q_{N-2, N-1},q_{N-1, N}$. To see how to perform decoding, recall that the values of the physical qubits in the LHZ scheme specify the alignment of the corresponding logical pairs. This implies that each physical configuration corresponds to a logical state only up to an overall flip of all logical qubits. We may thus choose, by convention, to fix  the value of one logical qubit: $\bar q_{1} = +1$. To decode the value of logical qubit $\bar q_{ i}$, we simply read out the relative alignments of the logical pairs following the chain induced by the spanning tree which connects $\bar q_{i}$ to $\bar q_{1}$. Using the upper diagonal chain mentioned above we can decode as follows:
\beq
\bar q_{ i} = \prod_{m=1 }^{ i-1}  q_{m,m+1}\,.
\label{eq:chain}
\eeq
Similarly, the first column of Fig.~\ref{fig:errors1} defines a star-like spanning tree where each  qubit is only connected to $\bar q_{1}$: $ q_{1,2}, q_{1,3},\dots, q_{1, N-1},q_{1, N}$, thus we have:
\beq
\bar q_{ i} =   q_{1,i}\,.
\label{eq:star}
\eeq
In the examples given in Fig.~\ref{fig:errors1}, decoding using the chain of Eq.~\eqref{eq:chain} gives the wrong state (one error hit), while decoding using the star of Eq.~\eqref{eq:star} gives the correct state (no errors hit).

\begin{figure}[t] 
   \centering
   \includegraphics[width=1\columnwidth]{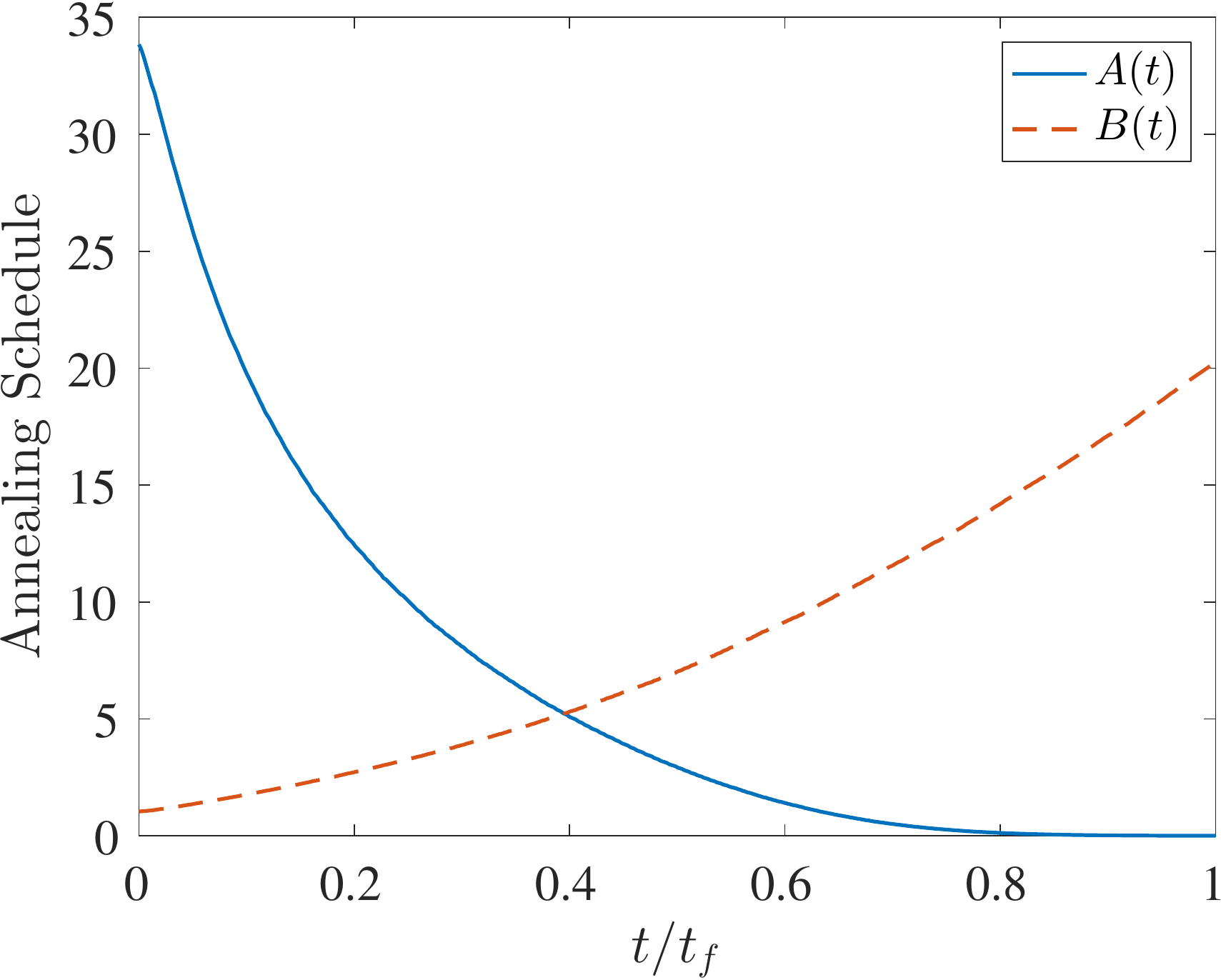} 
   \caption{\textbf{Annealing schedule used in our SQA simulations.} (Color online) The vertical axis units are arbitrary but correspond to the energy scale of our Hamiltonians.}
   \label{fig:AnnealingSchedule}
\end{figure}

\begin{figure}[t]
\subfigure[]{\includegraphics[width=0.5\columnwidth]{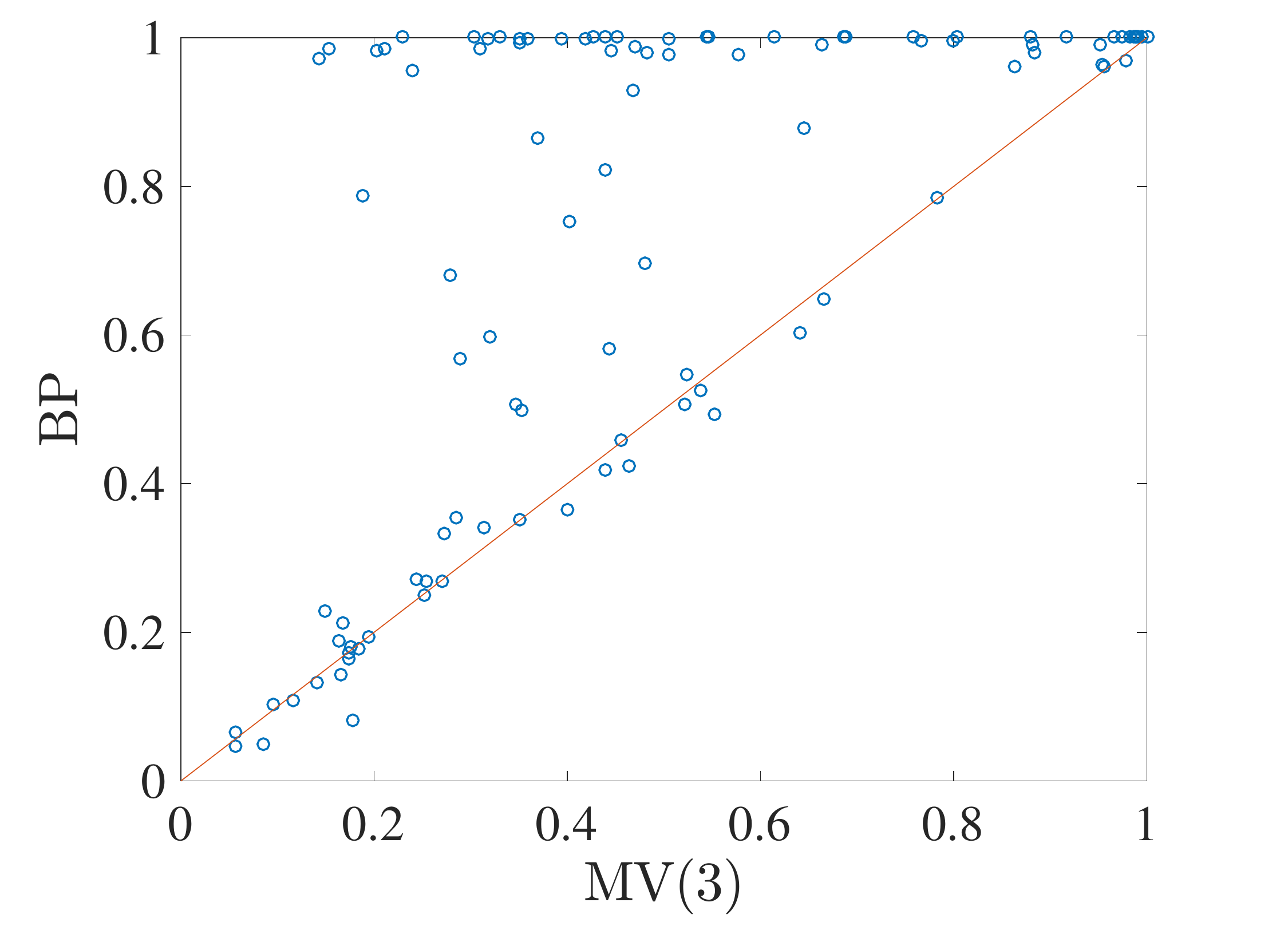}\label{fig:comparison11a} }
 \hspace{-0.2in}
\subfigure[]{\includegraphics[width=0.5\columnwidth]{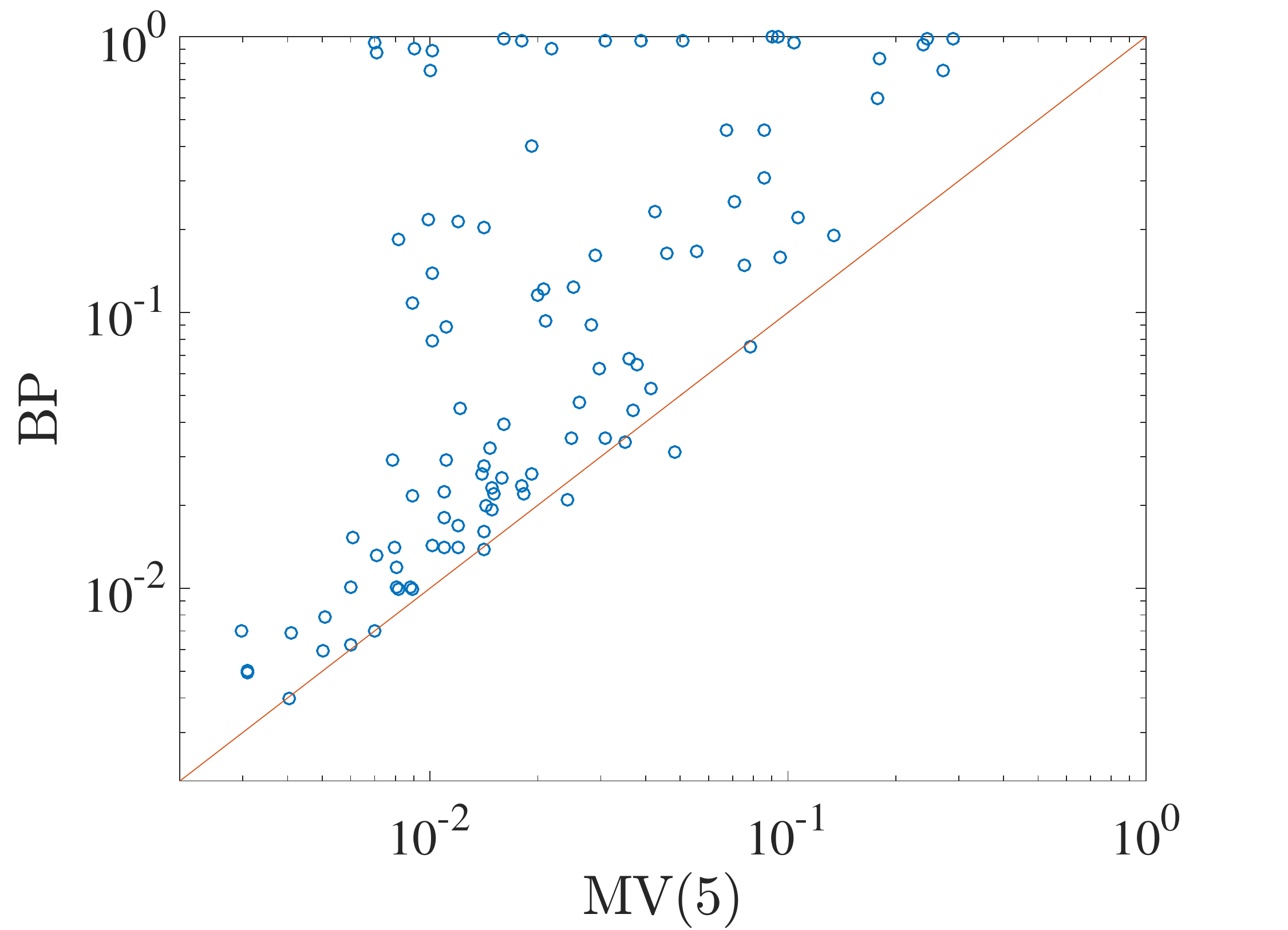} }\\
\subfigure[]{\includegraphics[width=0.5\columnwidth]{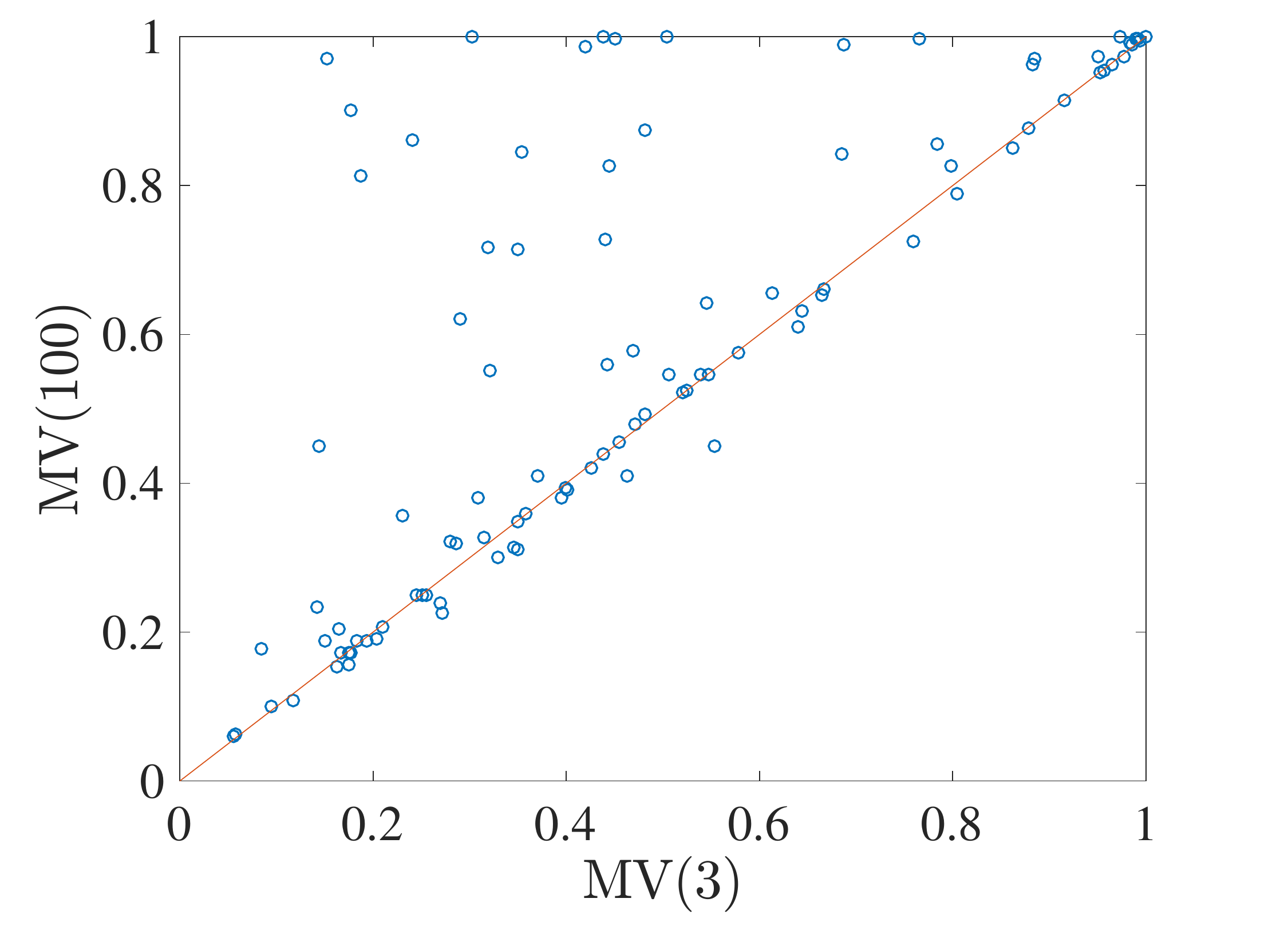}\label{fig:comparison12b}}
 \hspace{-0.2in}
\subfigure[]{\includegraphics[width=0.5\columnwidth]{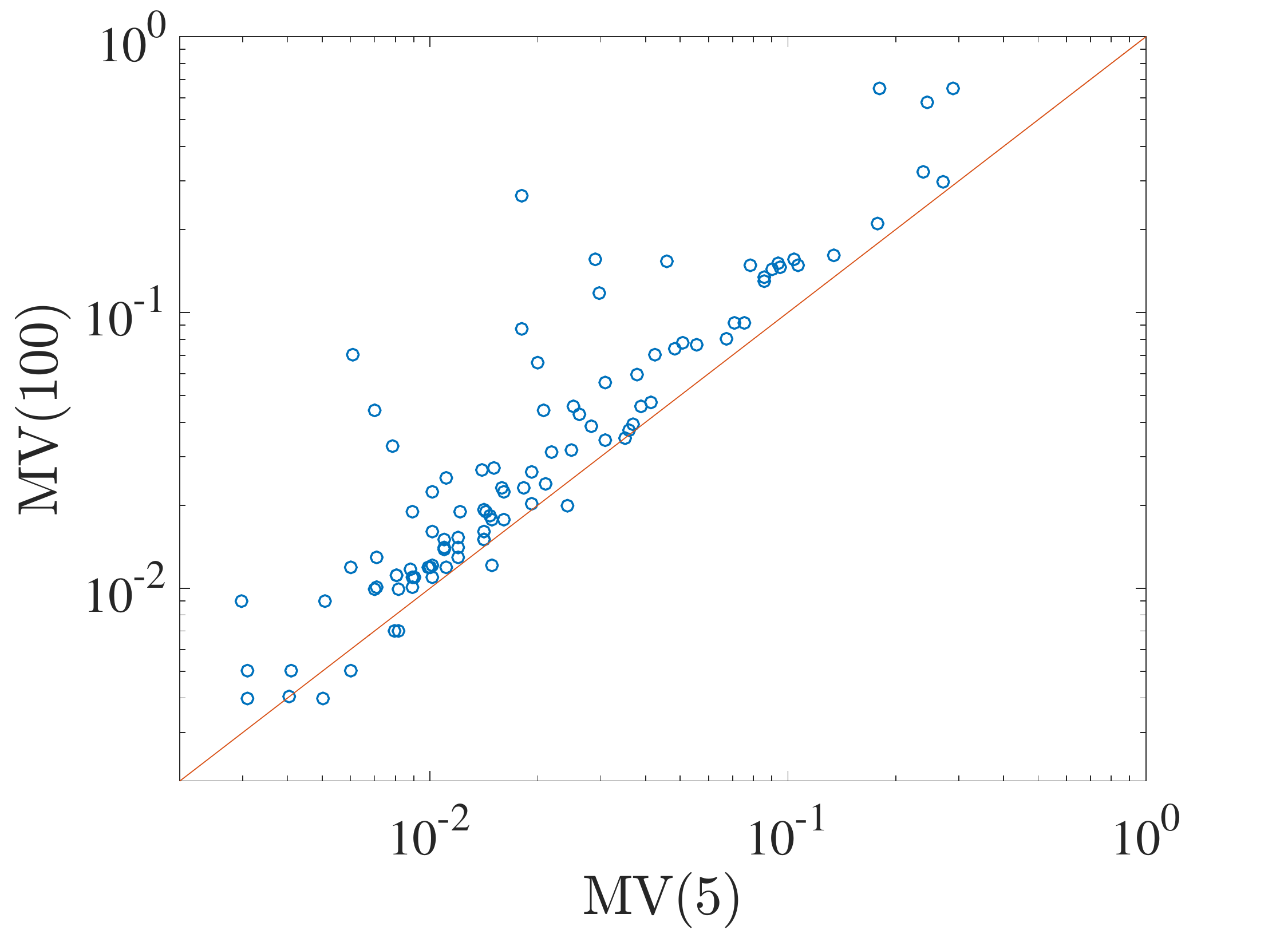}}\\
\subfigure[]{\includegraphics[width=0.5\columnwidth]{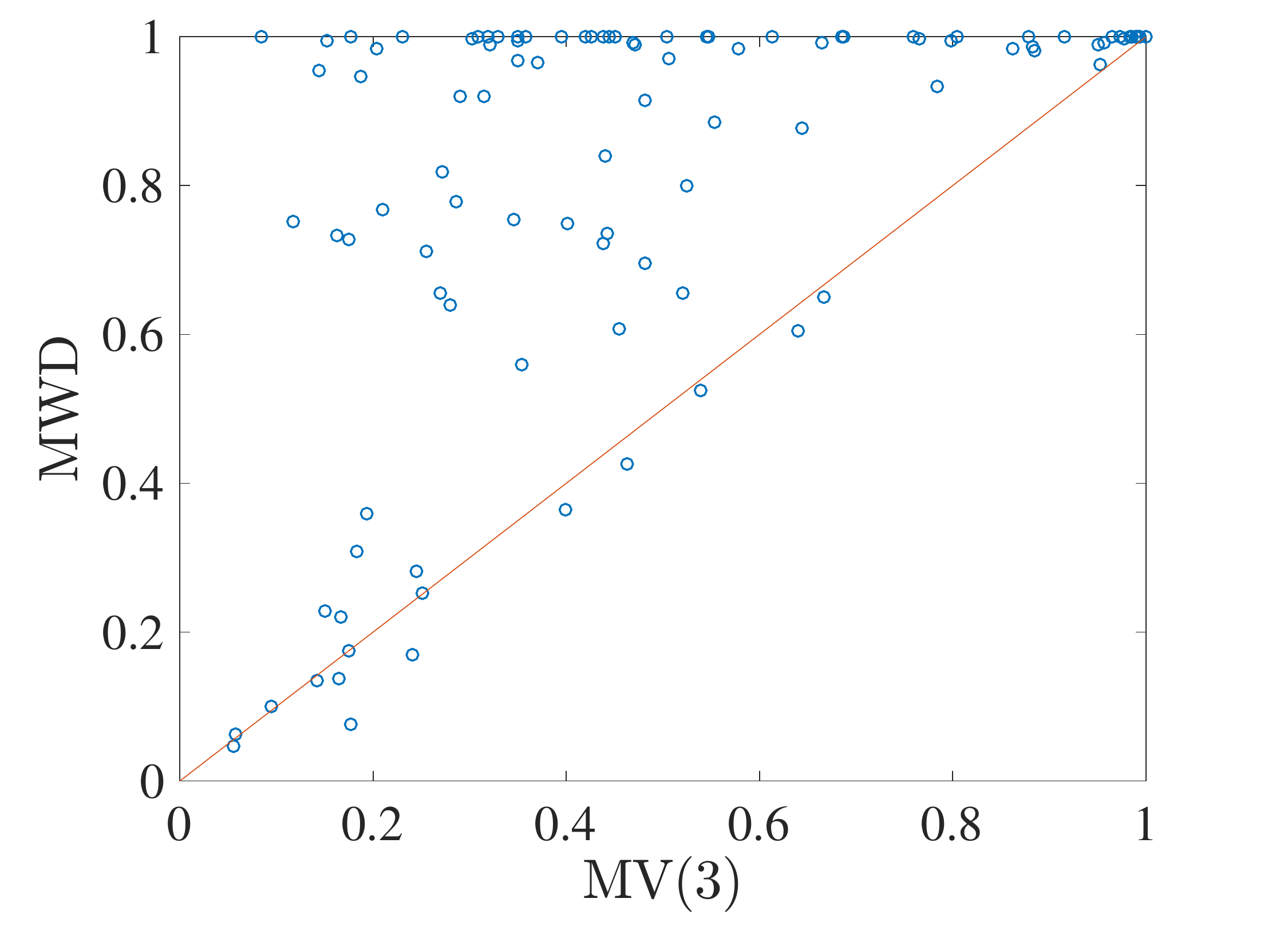}\label{fig:comparison13c} }
 \hspace{-0.2in}
\subfigure[]{\includegraphics[width=0.5\columnwidth]{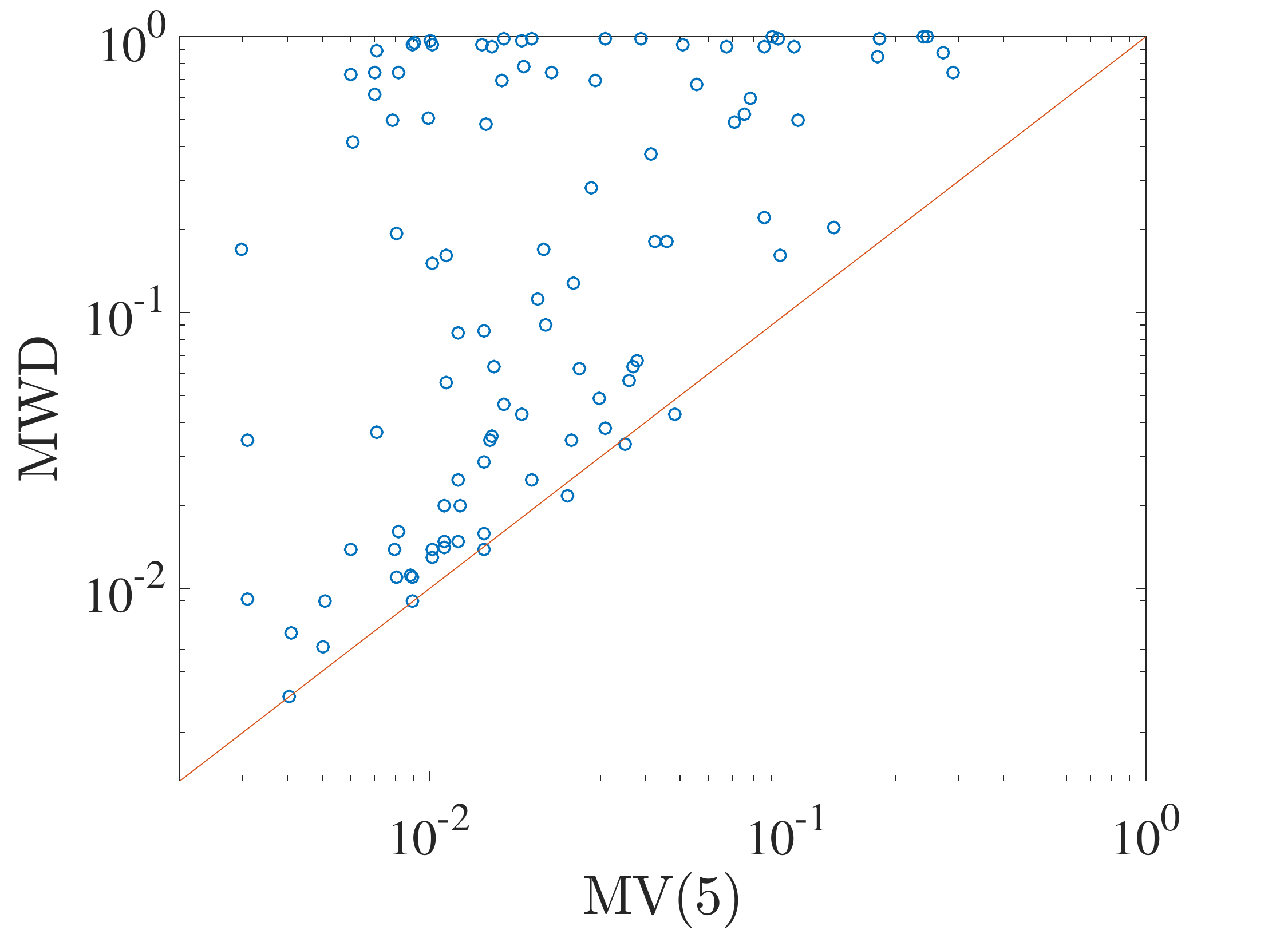}}
\caption{\textbf{Comparison of different decoding strategies for the LHZ scheme.}  (Color online) Shown are scatter plots comparing the success probabilities computed via SQA for (a,b) belief propagation of Ref.~\cite{Pastawski:2015}, (c,d) majority vote over $100$ spanning trees, (e,f) MWD, all relative to majority vote decoding over three and five spanning trees for the $K_8$ (a,c,e) and $K_{16}$ (b,d,f) instances respectively.  The optimal penalty was used for each instance.  SQA parameters: $10^4$ sweeps for $K_8$ and  $5\times 10^4$ sweeps for the $K_{16}$ instances, $\beta = 1$. }
\label{fig:dec_comp}
\end{figure}

\begin{figure}[t] %
   \centering
   \includegraphics[width=1\columnwidth]{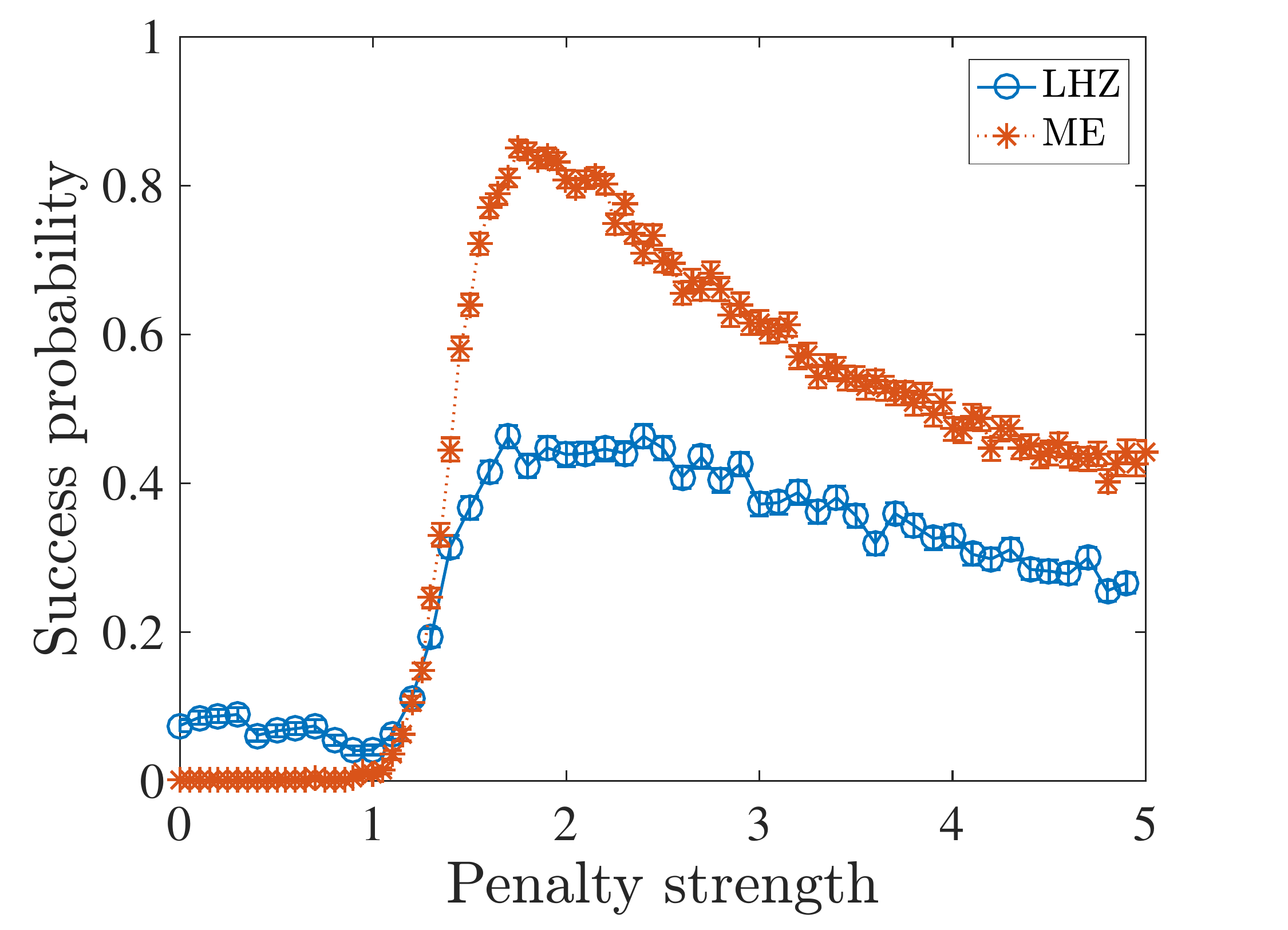} 
   \caption{\textbf{Effect of increasing the number of sweeps and lowering the temperature.} (Color online) A comparison of the SQA results using majority vote for the $K_{8}$ instance shown in Fig.~\ref{fig:SQA_boosted} for an increasing number of sweeps ($10^6$) and lowering the temperature ($\beta = 5$).}
   \label{fig:MoreSweeps}
\end{figure}

\begin{figure}[t] %
   \centering
   \includegraphics[width=1\columnwidth]{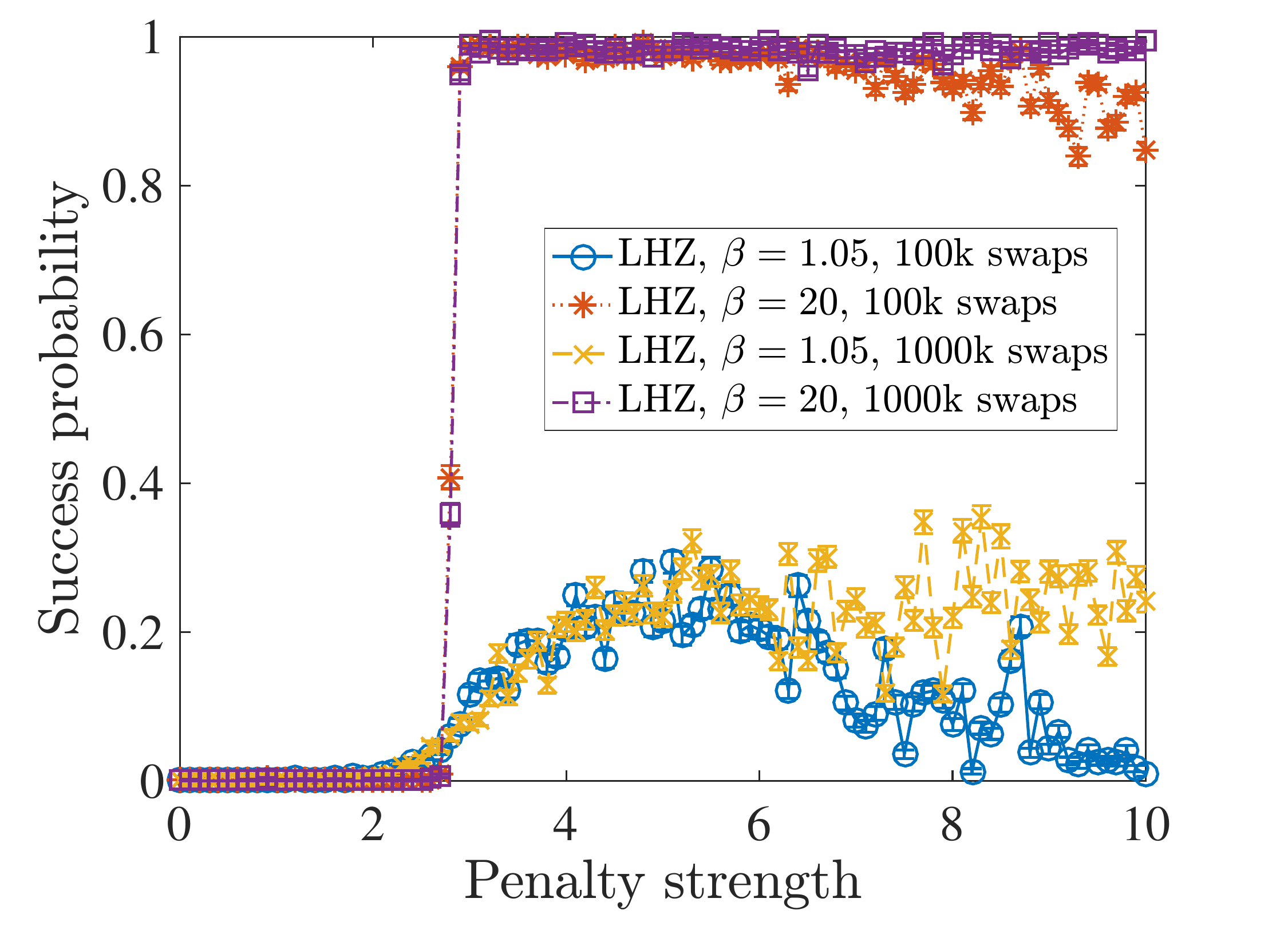} 
   \caption{\textbf{Effect of increasing the number of swaps.} (Color online) A comparison of the PT results for the $K_{16}$ instance shown in Fig.~\ref{fig:PT} for different numbers of swaps. Increasing the number of swaps keeps the success probability more fixed at high penalty values.}
   \label{fig:MoreSwaps}
\end{figure}

\section{Numerical Simulations}
\label{sec:NS}

%
{\it Simulated Quantum Annealing.} Our simulated quantum annealing simulations are based on a discrete-time quantum Monte Carlo algorithm \cite{sqa1,Santoro}.  Details of the implementation of this algorithm have been described elsewhere \cite{Muthukrishnan:2015jt}.  Here we use a fixed Trotter slicing of $N_{\tau} = 64$.  The annealing schedule used is shown in Fig.~\ref{fig:AnnealingSchedule}.
%

{\it Parallel Tempering}. Parallel tempering (also known as exchange Monte Carlo) \cite{Hukushima:1996} is a standard method used to thermalize spin systems.  In PT simulations, replicas of the spin system at different temperatures are evolved independently for a fixed number of Metropolis updates, followed by a `swap' operation.  A swap involves comparing the energies of neighboring replicas, and swapping their temperatures with a probability $p_{\mathrm{swap}}$ given by:
\beq
p_{\mathrm{swap}} = \min \{ 1 , \exp \left[ \left(E_{i+1} - E_i \right) \left(\beta_{i+1} - \beta_{i} \right) \right] \}
\eeq
where $\left(\beta_{i+1} - \beta_{i} \right)  < 0$, $\forall i$.  For our simulations we perform $10$ sweeps of single spin Metropolis updates per swap, and we perform a total of $10^5$ swaps.  We use $64$ different temperatures distributed according to: 
\beq
\beta_i = \left( \frac{0.1}{20} \right)^{(i-1)/63} \beta_1
\eeq
with $\beta_1 = 20$ and $i = 1, \dots, 64$.

\section{Decoding Strategies}
\label{sec:dec}

Here we provide a more direct comparison of the various decoding strategies we have tried: MWD, BP as in Ref.~\cite{Pastawski:2015}, and majority vote on a large number ($100$) of random spanning trees. As remarked in the main text, we did not implement MLD. 

To perform MWD, we ran simulated annealing with the Hamiltonian in Eq.~\eqref{eq:K_NDec} $100$ times with a linear schedule for $\beta \in [0.1, 5]$ and $1000$ sweeps.  We found this to be sufficient to find the ground state of the decoding Hamiltonian Eq.~\eqref{eq:K_NDec} for our relatively small problem sizes. In the BP decoding approach of Ref.~\cite{Pastawski:2015} the alignment of a given pair of logical qubits $(\bar q_{ i},\bar q_{ j})$ is determined by iteratively updating its value using a majority vote on $ N-2$ parity checks $(q_{il},q_{lj})$ ($l\neq i,j$).  We then use a single random spanning tree to read out the full logical state.

The various panels of Fig.~\ref{fig:dec_comp} present our SQA success probability results for the same $100$ random $K_8$ and $K_{16}$ instances considered earlier. It can be seen that the MVD scheme give comparable results, with the BP and MWD scheme giving much better results, as discussed in the main text. 

\section{Additional results for the LHZ scheme} 
\label{sec:Add}
%
\subsection{SQA simulations for at colder temperatures and larger number of sweeps}
In the main text, our SQA simulations for the $K_8$ instances were limited to $\beta = 1$ and $10^4$ sweeps.  Here we show for the same instance in Fig.~\ref{fig:SQA_boosted} that increasing the number of sweeps and lowering the temperature does not necessarily improve the performance of LHZ relative to ME.  We show in Fig.~\ref{fig:MoreSweeps} the performance of each when using $\beta = 5$ and $10^6$ sweeps.  Both the ME and LHZ performance is improved, but relative performance is unchanged with ME continuing to show better results.

\subsection{PT simulations for LHZ at higher swaps}
In Sec.~\ref{sec:Compare} of the main text we noted that the fixed number of updates performed in the PT simulations on the LHZ embeddings was insufficient to properly thermalize the PT replicas, which led to the drop in success probability at large penalty strength in Fig.~\ref{fig:PT}.  In order to validate this conclusion, we increase the total number of swaps (the number of sweeps per swap remains fixed) by an order of magnitude and show the corresponding results in Fig.~\ref{fig:MoreSwaps}.  With this increased number of swaps, the performance saturates for large penalties as expected, indicating that the behavior observed in Fig.~\ref{fig:PT} is indeed a consequence of an insufficient number of updates. This behavior is to be expected, since single spin-flip thermalization is completely frozen in the limit of very large penalty values; in such a limit, only cluster updates of physical qubits corresponding to a logical spin-flips would be efficient update moves (these are moves we do not perform in our simulations).

\subsection{Distance between the SQA and PT states}
\begin{figure}[t] %
   \centering
   \subfigure[]{\includegraphics[width=0.5\columnwidth]{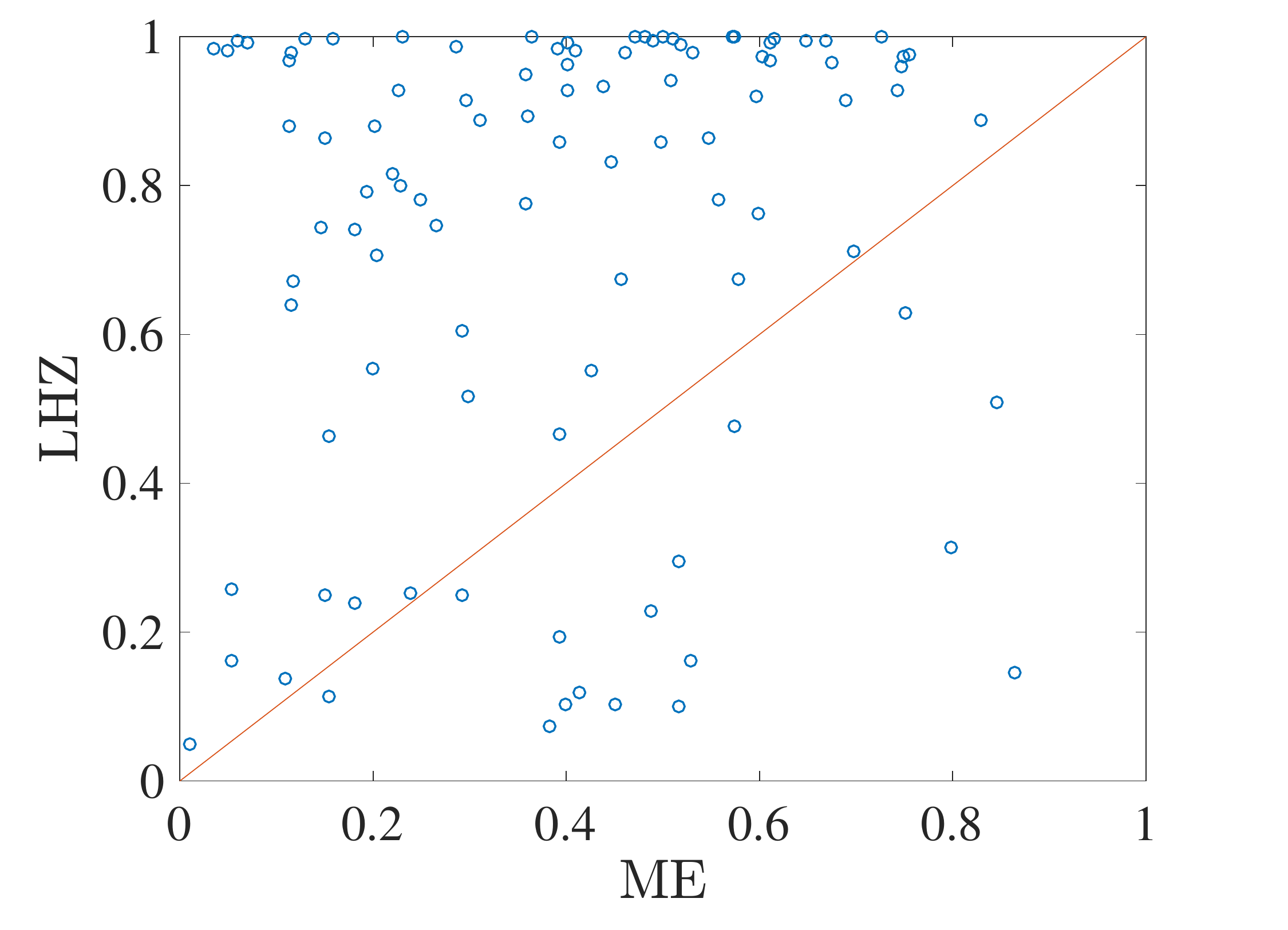}}
   \hspace{-0.2in}
    \subfigure[]{\includegraphics[width=0.5\columnwidth]{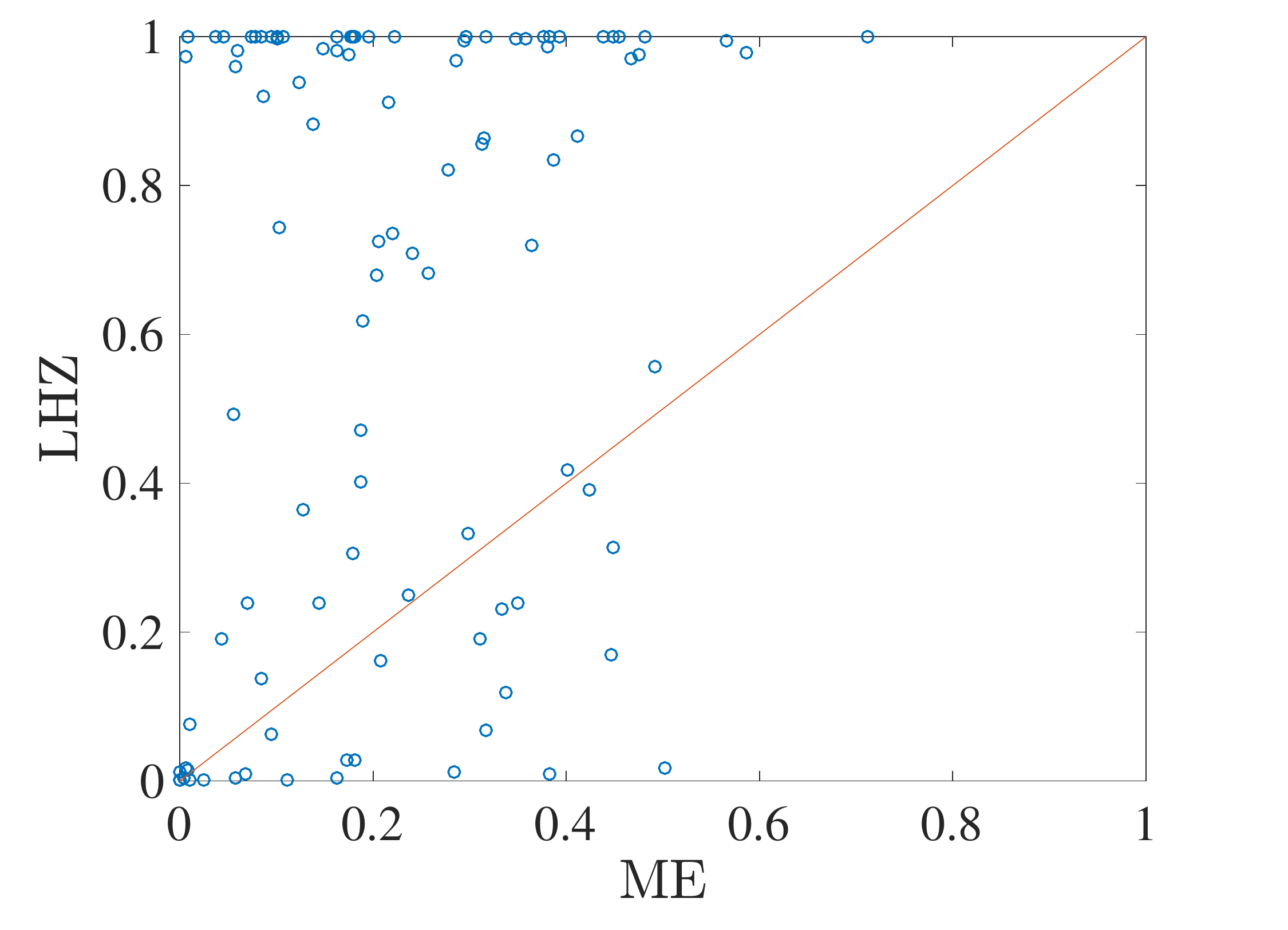} }
   \caption{\textbf{Distance test between PT and SQA.} (Color online) Scatter plot of the distance [Eq.~\eqref{eq:dist}] of the LHZ and ME states obtained via SQA from their respective PT simulation states. Panel (a): higher temperature, $\beta = 2.1$. Panel (b) lower temperature, $\beta = 4.0$. For most instances the ME scheme distance is smaller, indicating better thermalization. Results are shown for all $100$ $K_8$ instances.}
   \label{fig:PTSQADistance}
\end{figure}

In order to quantify how close the ME and LHZ states are to the PT states, we use the following distance measure \cite{Albash:2014if}.  For each instance $i$ we define the probability distribution function $p_i(E)$ of finding a state with physical energy $E$:
\beq
p_i(E) = \frac{1}{N_E}\sum_{n=0}^{N_E} \delta_{E_{n},E}\ ,
\eeq 
where $E_n$ is the energy of the $n^{\rm th}$ excited state and $N_E$ is the total number of  energy levels observed for the given instance.  We then compute the total variation distance 
\begin{equation} 
\mathcal{D} \left(p,q \right) = \frac{1}{2} \sum_{x} \left| p(x) - q(x) \right|
\label{eq:dist}
\end{equation}
for a given instance $i$ between the probability distributions for SQA and PT, i.e., we let $p = p_i^{\mathrm{SQA}}$ and $q=p_i^{\mathrm{PT}}$.  We resort to this distance measure because of the relatively small number of states ($10^3$) we have for each scheme, which prevents us from reliably computing, e.g., the trace-norm distance between states.  Figure~\ref{fig:PTSQADistance} is a scatter plot of the distances comparing the ME and LHZ schemes, evaluated at the penalty that maximizes the success probability (after MV decoding) for each respectively, at two inverse temperatures of PT.  For the majority of the instances, the ME states tend to be closer to the PT states than the LHZ scheme, in support of our assertion that the ME scheme thermalizes more easily than the LHZ scheme.

\begin{figure}[t] %
   \centering
   \subfigure[]{\includegraphics[width=0.5\columnwidth]{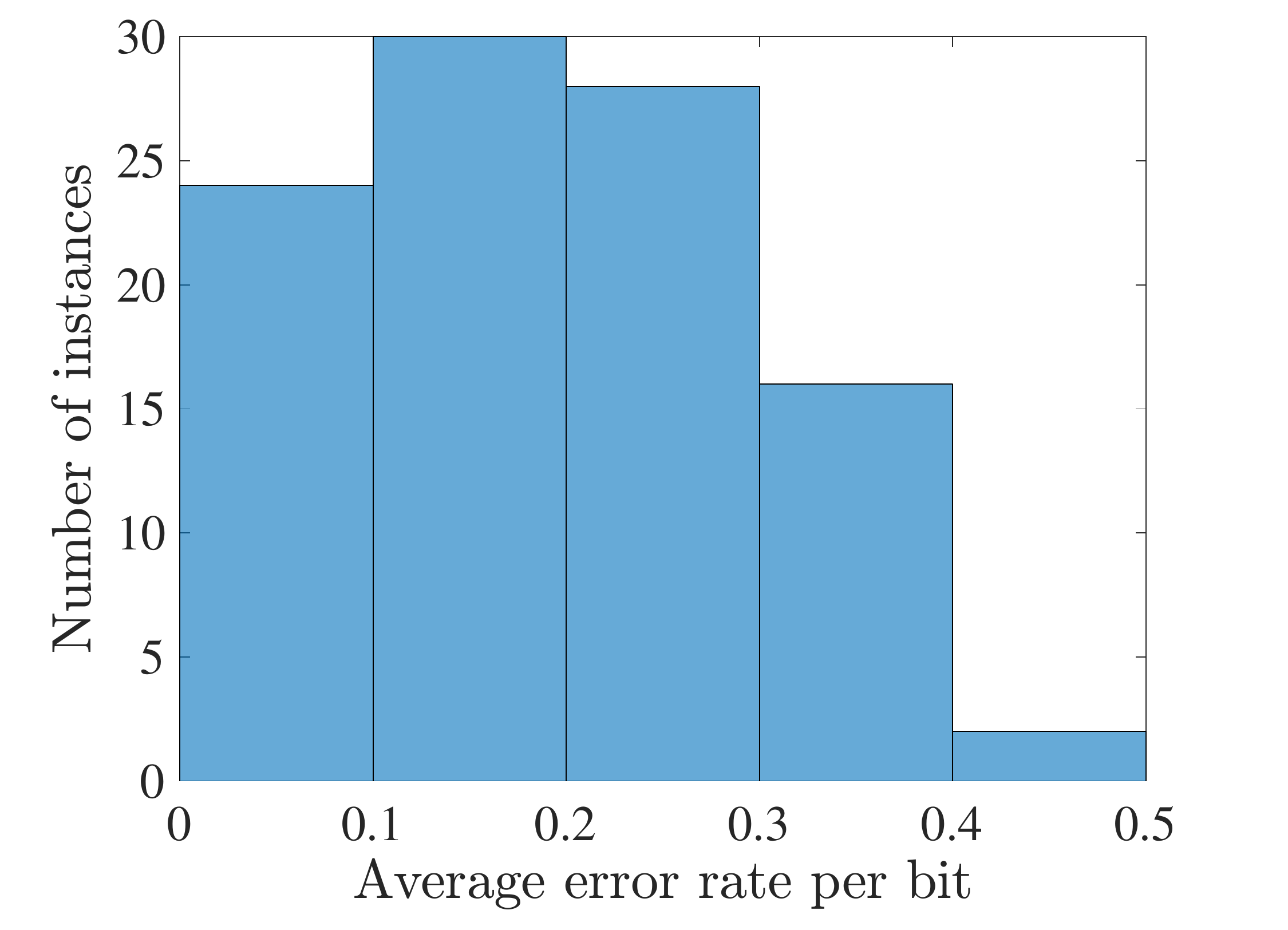} }\hspace{-0.2in}
    \subfigure[]{\includegraphics[width=0.5\columnwidth]{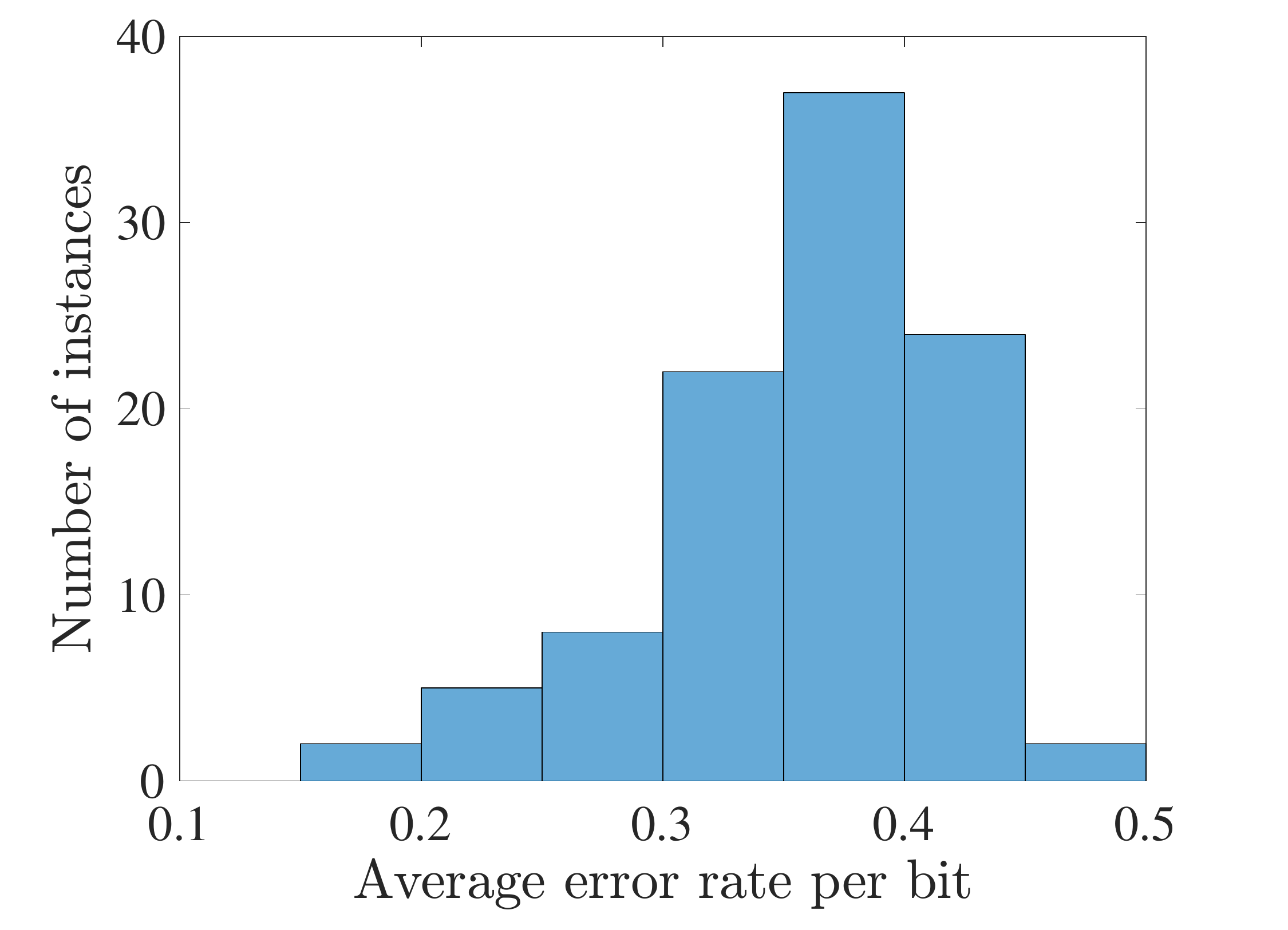} }
   \caption{\textbf{Average error rate.} Shown is a histogram of the average error rate for the $100$ instances for the (a) $K_8$ and (b) $K_{16}$ instances for SQA with $\beta = 1$ and $10^4$ sweeps.  The error rate is calculated as follows.  For the $1000$ states generated for each instance and each penalty value, we find the average minimum distance in terms of spin flips to the nearest degenerate ground state.  The distance divided by the total number of spins gives the average error rate for a given instance and penalty value.  The data in the histogram uses the error rate at the optimal penalty from the data shown in Fig.~\ref{fig:SQAComp}.}
   \label{fig:AverageErrorRate}
\end{figure}

\begin{figure}[t] %
   \centering
   \subfigure[]{\includegraphics[width=.5\columnwidth]{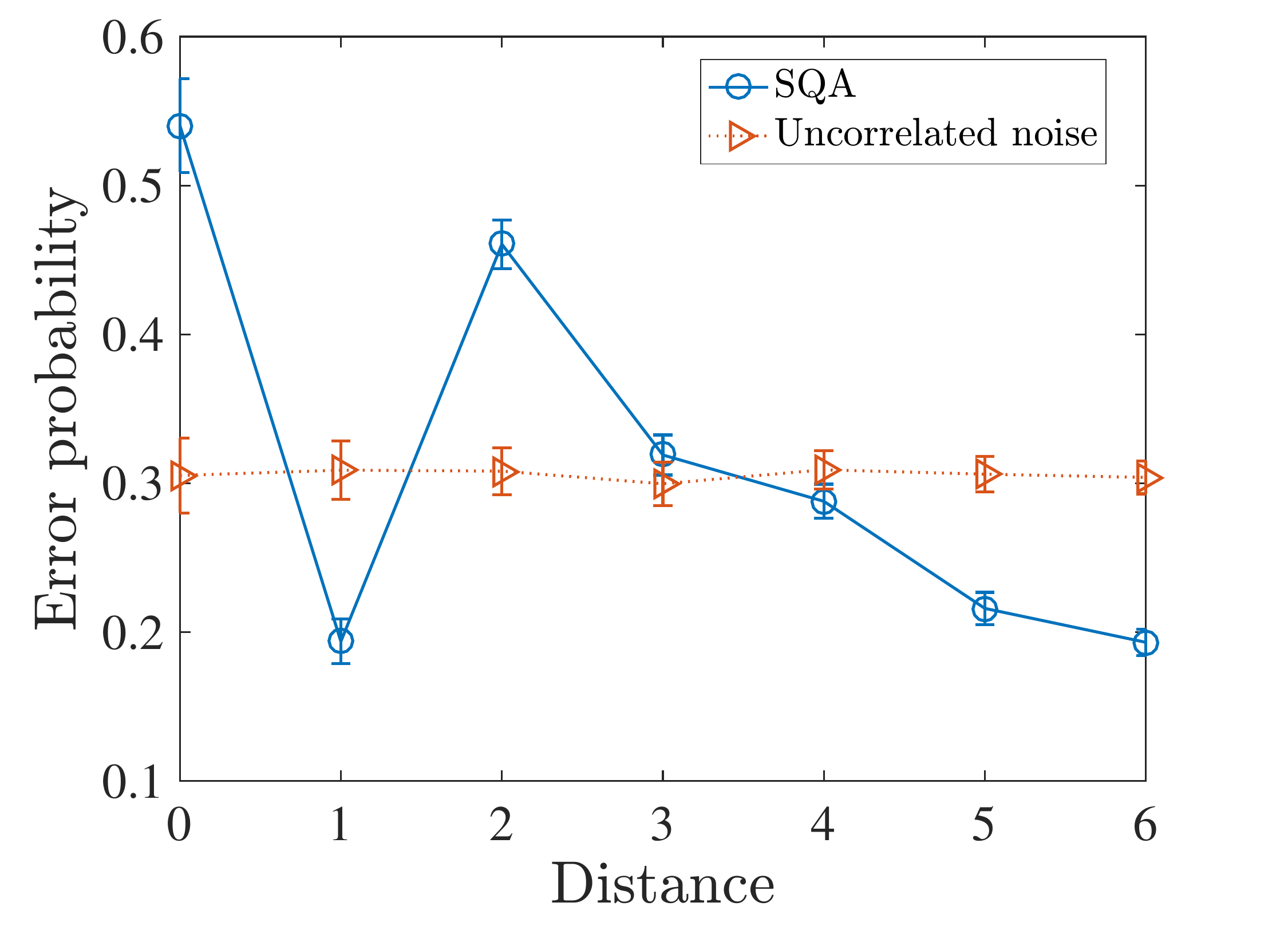} \hspace{-0.2in}
\label{fig:SQACorrelationA} }
   \subfigure[]{\includegraphics[width=.5\columnwidth]{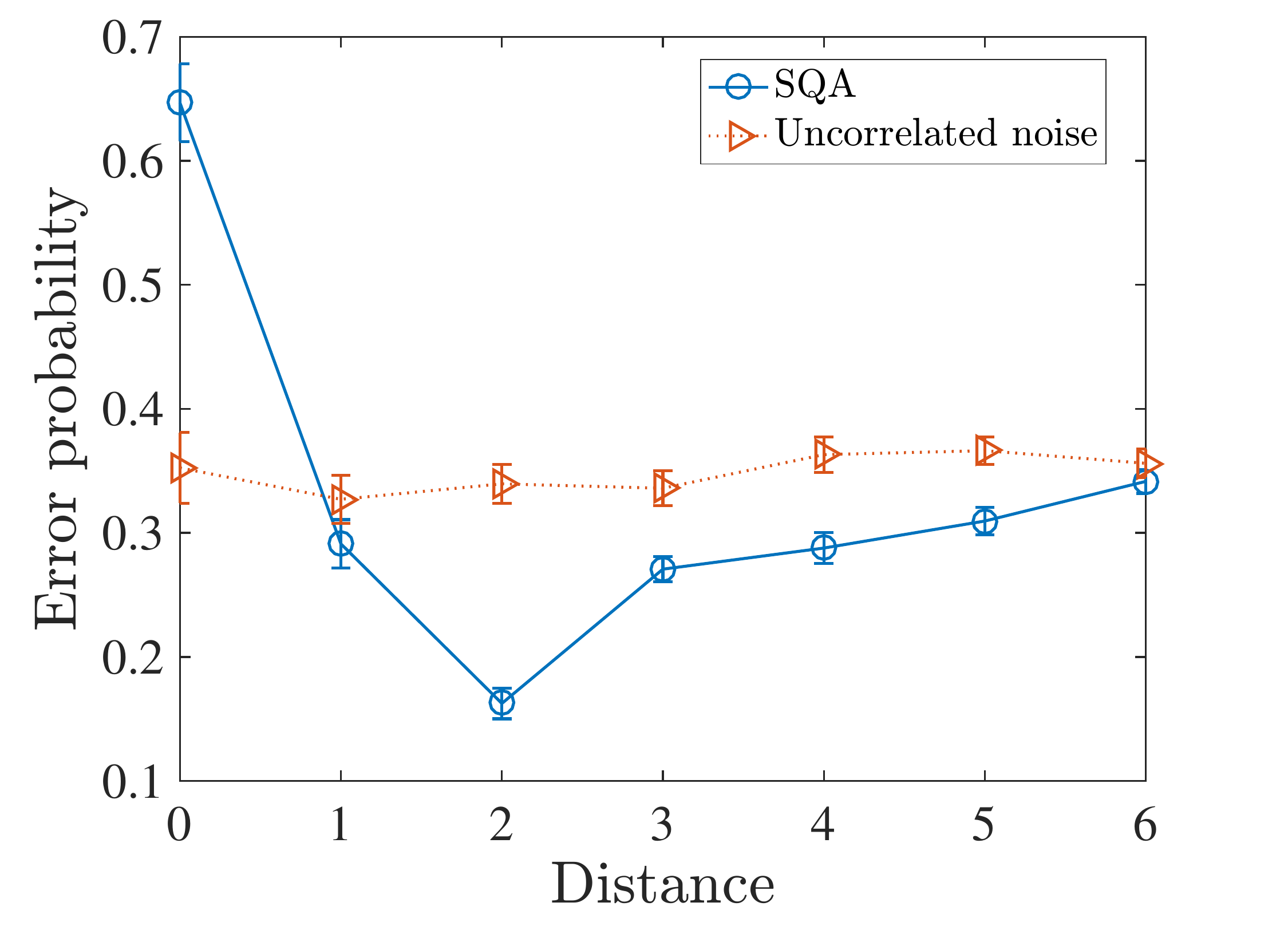} 
\label{fig:SQACorrelationB} }
   \caption{\textbf{Error probability as a function of distance from the corner of the LHZ graph.} (Color online) Shown is the probability that a spin at distance $d$ from the top corner of the LHZ graph flips subject to uncorrelated noise and SQA-generated noise, for two $K_8$ instances. (a) $K_8$: instance 3, $\gamma= 2.9$.  (b) $K_8$: instance 95, $\gamma= 2.4$.  The uncorrelated case is flat as expected, while the SQA case has a non-trivial distance dependence.}
   \label{fig:SQACorrelation}
\end{figure}

\begin{figure}[t] %
   \centering
{\includegraphics[width=\columnwidth]{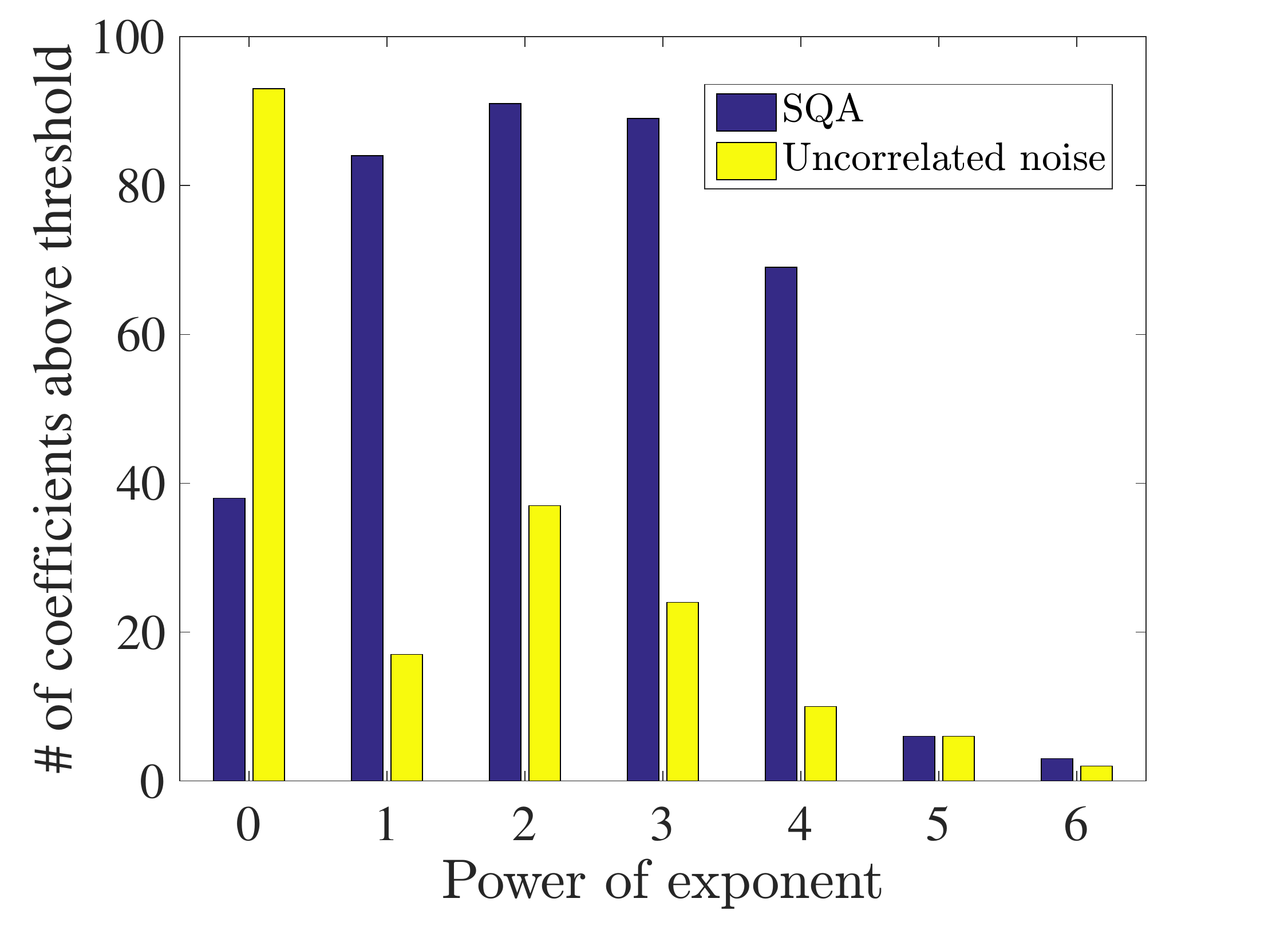}  \label{fig:SQACorrelationC} }
   \caption{\textbf{Distance dependence of error probability across all $K_8$ instances.} (Color online) Shown is a histogram of the values of the coefficients of a degree-$6$ polynomial fit to the curves generated as in Fig.~\ref{fig:SQACorrelation}, but for all $100$ $K_8$ instances. We define a cutoff based on the average (over distance) error rate minus twice the standard deviation of the error rate of the uncorrelated noise.  For each coefficient of the polynomial, if the coefficient is above this threshold, we count it, otherwise we do not. The histogram shown is the result of this binning process.}
   \label{fig:SQACorrelationHisto}
\end{figure}

\section{SQA noise is not uncorrelated}
\label{sec:corr}
Our motivation for using SQA as a noise model is that it mimics thermal noise effects we expect from a finite temperature quantum annealer.  Here we demonstrate how SQA noise differs from uncorrelated noise on the instances we have studied.  First, Fig.~\ref{fig:AverageErrorRate} shows the average error rate for our instances in the SQA simulations.  For all instances, the average error rates are below $0.5$. Thus, our SQA simulations generate noise in a regime that is favorable from the perspective of decoding the LHZ scheme if in addition the noise were uncorrelated \cite{Pastawski:2015}.
  
However, SQA generates noise that is not uncorrelated. To show this, we contrast it with uncorrelated noise and perform the following test. Starting at the top corner of the LHZ graph for the $K_8$ case [the spin labeled $1,2$ in Fig.~\ref{fig:embs3}], we calculate the average error rate for the spins at distances $0$ to $6$ (the largest for the $K_8$ case) from this corner spin; see the caption of Fig.~\ref{fig:AverageErrorRate} for how the error rate is calculated.  Note that there are $d+1$ spins at distance $d$.  For uncorrelated noise we expect there not to be any dependence on distance.  We show in Fig.~\ref{fig:SQACorrelation} the behavior of SQA-generated noise for the two instances shown in Figs.~\ref{fig:PT1} and \ref{fig:PT2}, relative to uncorrelated noise, generated by randomly flipping spins with a probability equal to the averaged SQA error rate per spin (over all distances), to generate $10^3$ uncorrelated error states on the physical state associated with the corresponding logical ground state of the LHZ scheme.

Clearly, the two cases behave very differently: as expected, the uncorrelated noise is effectively flat, while the SQA-generated noise has a non-trivial dependence on distance.  This distance behavior is strongly instance and penalty strength dependent. To demonstrate that this difference occurs across all instances, we fit both the uncorrelated and SQA results to a degree-$6$ polynomial, as shown in Fig.~\ref{fig:SQACorrelationHisto}.  Ideally, the uncorrelated noise case will have its entire weight on the constant ($x^0$) term. We see that, as expected, the uncorrelated noise has most of its weight on the low powers of the polynomial, whereas the SQA-generated noise has most of its weight on the higher powers.

\section{Scaling of the Strength of Energy Penalties}
\label{sec:SEP}

\begin{figure}[t]
\begin{center}
\subfigure[]{\includegraphics[width=0.49\columnwidth]{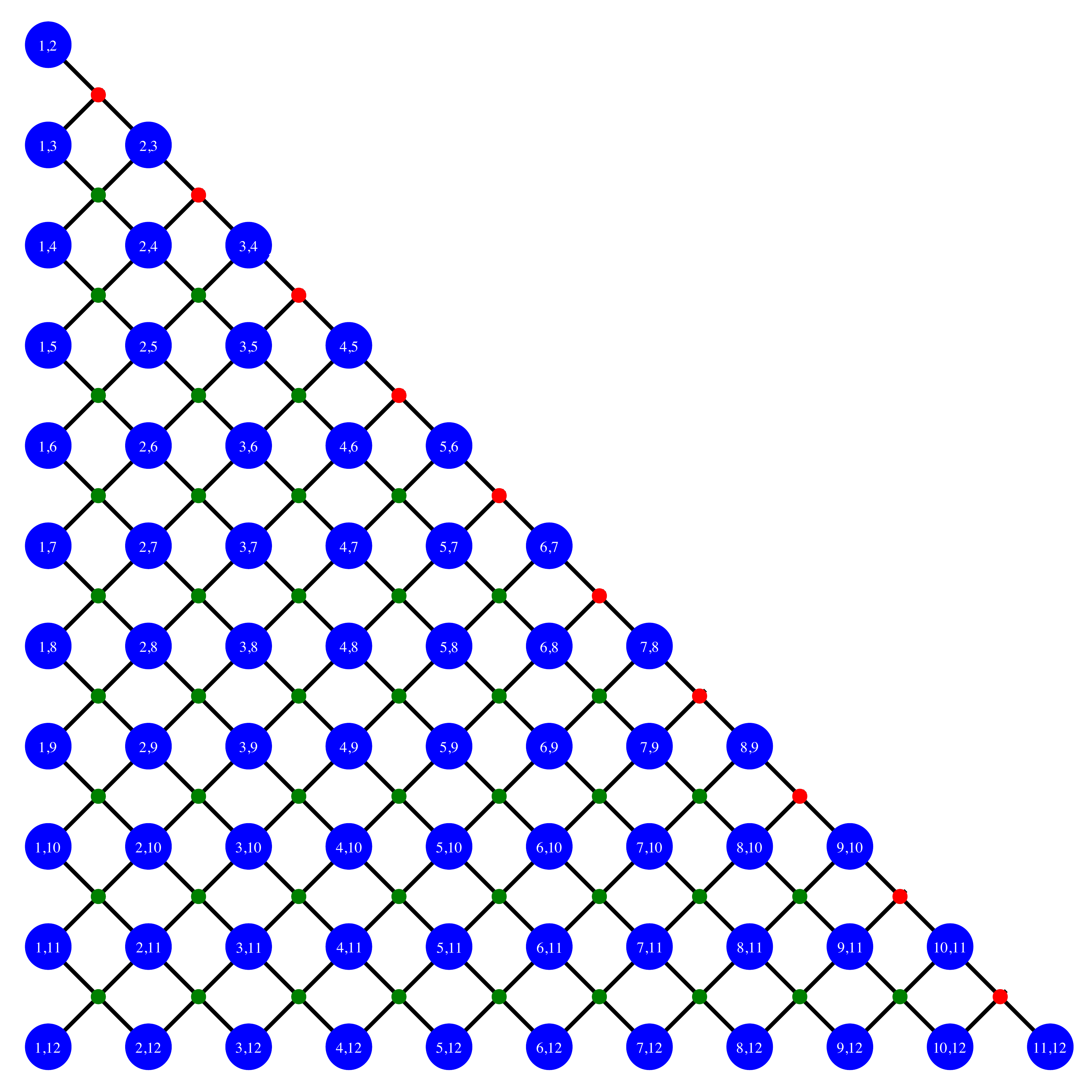}\label{fig:flip1}}
\subfigure[]{\includegraphics[width=0.49\columnwidth]{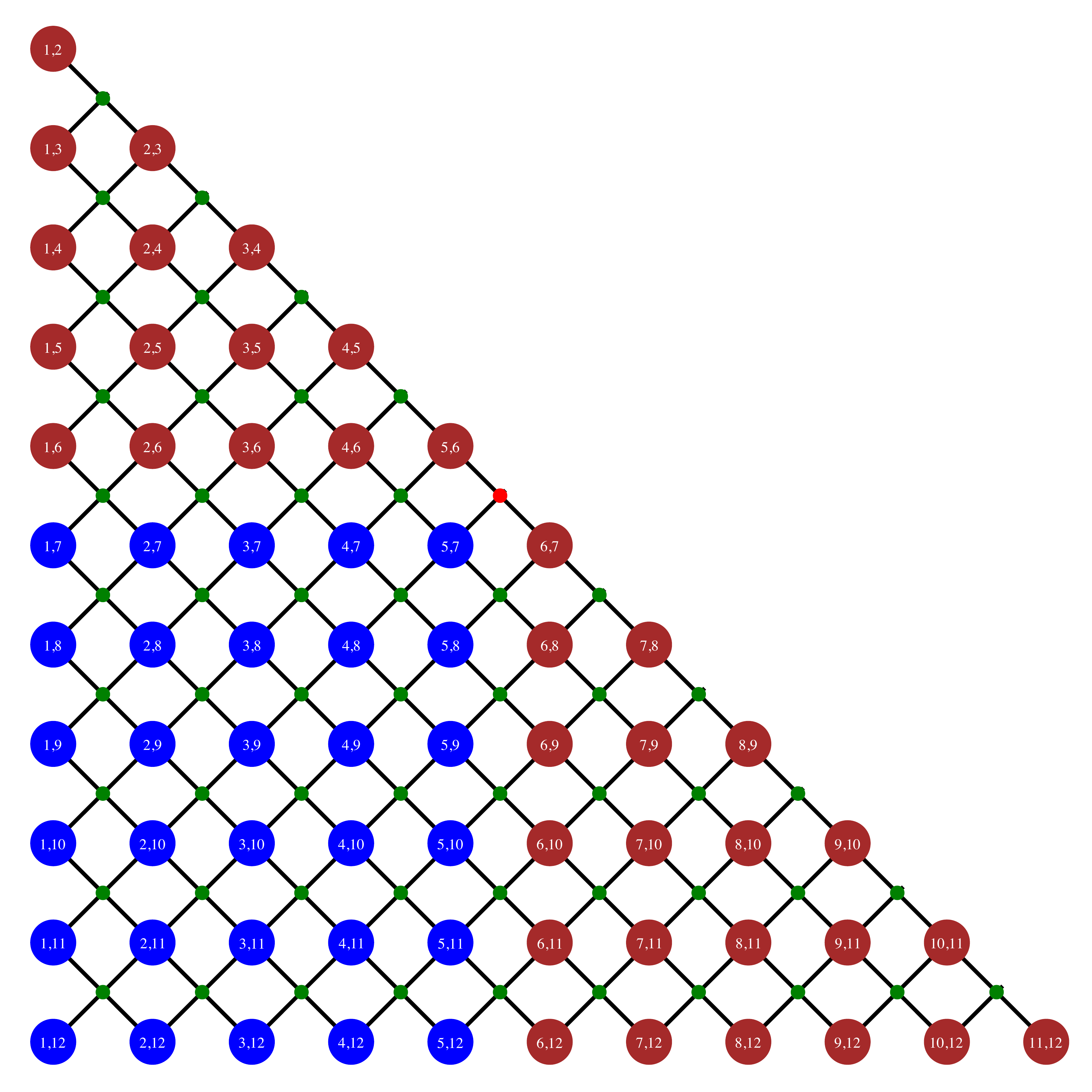}\label{fig:flip2}}
\caption{\textbf{Configurations used to derive a lower bound on the scaling of the energy penalty strength.}  (Color online) (a) Antiferromagnetic case with $J_{ij} = 1$. Colors represent alignment to the local fields: all the $\sim  N^2$ can be aligned (blue) to the physical local field by violating only $N$ constraints. (b) Random couplings case with $J_{ij} = \pm 1$. Colors represent physical flips (blue) from the logical ground state: a domain of $\sim N^2$ physical qubits can be flipped by violating only  one constraint. To prevent these two configurations from being the physical ground state, the strength of the energy penalties should scale in both cases with $N$.}
\label{fig:flip}
\end{center}
\end{figure}

An important limitation of the embedded approach is that the strength of the energy penalties grows with problem size. This is true for the ME scheme~\cite{Venturelli:2014nx}, but also for the LHZ scheme. We do not have a general lower bound for the strength of the penalties in the LHZ scheme, but simple arguments suggest that in both schemes, the strength of the energy penalties must grow linearly with the number of logical qubits $ N$. To see this in the case of the LHZ scheme, consider, e.g., a completely antiferromagnetic $K_{ N}$. Roughly half of the couplings are frustrated in the logical ground state.  This should be mapped to roughly half the physical qubits in the LHZ scheme being down and the other half being up. However, for sufficiently weak constraints, the LHZ ground state is a physical configuration like that shown in Fig.~\ref{fig:flip1} where all physical qubits are pointing down, i.e., aligned with the physical local field. The energy from aligning with the local fields is of order ${N}^2$, while the number of violated constraints is of order ${N}$. This implies that the strength of the constraints must grow at least with ${N}$  in order for the physical ground state to faithfully represent the logical ground state. Consider now a $K_{ N}$ with random couplings $J_{ij} = \pm 1$. In this case we expect the logical ground state to be highly frustrated, with roughly half of the corresponding physical qubits not aligned to the local fields. Flipping a domain of order ${N}^2$ physical qubits like that shown in Fig.~\ref{fig:flip2} will typically result in a change in the energy of the physical configuration of the order of $\sqrt{{N}^2} = N$. To avoid the possibility of such domain-flips to lower the energy of the physical configuration, the constraints must grow again at least with ${N}$.


%

\end{document}